\pdfoutput=1
\documentclass[a4paper,11pt]{article}
\usepackage[T1]{fontenc}
\usepackage[utf8]{inputenc}
\usepackage{lmodern}
\usepackage[margin=25mm]{geometry}
\usepackage{amsmath,amssymb,amsthm,mathtools}
\usepackage{microtype}
\usepackage{graphicx}
\usepackage{booktabs}
\usepackage{array}
\usepackage{xcolor}
\usepackage{enumitem}
\usepackage[section]{placeins}
\usepackage[title]{appendix}
\usepackage[numbers,sort&compress]{natbib}
\usepackage[hidelinks]{hyperref}
\usepackage{bookmark}
\graphicspath{{figures/}}
\newcommand*{\doi}[1]{\href{https://doi.org/\detokenize{#1}}{doi:\detokenize{#1}}}

\allowdisplaybreaks[2]
\setlist{itemsep=2pt,topsep=4pt}

\newtheorem{theorem}{Theorem}[section]
\newtheorem{proposition}[theorem]{Proposition}
\newtheorem{corollary}[theorem]{Corollary}
\newtheorem{lemma}[theorem]{Lemma}
\theoremstyle{definition}

\theoremstyle{remark}
\newtheorem{remark}[theorem]{Remark}

\newcommand{\dd}{\mathrm{d}}
\newcommand{\B}{\mathcal{B}}
\newcommand{\Y}{\mathbf{Y}}
\newcommand{\Lop}{\boldsymbol{\mathcal{L}}}

\title{Separability defects in artificial satellite theory:\\
intermediaries, zonal harmonics and secular separatrices}
\author{Vladimir Martinusi\thanks{Corresponding author. E-mail:
\href{mailto:vmartinusi@technion.ac.il}{vmartinusi@technion.ac.il}.
ORCID: \href{https://orcid.org/0000-0001-9689-7657}{0000-0001-9689-7657}.}\\[1ex]
\small Stephen B.\ Klein Faculty of Aerospace Engineering,\\
\small Technion -- Israel Institute of Technology, Technion City, Haifa 3200003, Israel}
\date{}

\begin{document}
\maketitle

\begin{abstract}
A five-input Poisson-bracket operator, which reproduces the Levi-Civita
separability expressions and is cubic in the Hamiltonian, is used as a diagnostic of artificial satellite theory, in a modified
polar-nodal chart regular at prograde equatorial orbits. For the general
intermediary proposed by the author in 2020, covering those of Sterne,
Garfinkel, Brouwer, Aksnes, Cid and Lahulla and Deprit, the linearised defects
vanish if and only if the integrability ansatz holds, the condition for
additive first-order Hamilton--Jacobi separation in polar-nodal variables.
Second-order defects follow in closed form, and the coupling with the
nodal pair comes from the dependence on the physical latitude, shared by the
zonal term, Aksnes's intermediary and the originals of Sterne and
Garfinkel. For every axisymmetric field the radial defect is linear, with
a torque form, and the first-order zonal defect is the operator response to
the first-order Lie generator, which also gives the second-order mean radial
defect per revolution. In its gauge the
long-period second-order average is common to intermediaries without
latitude terms. Oblate structures with a displaced centre separate the zonal
field exactly if and only if its coefficients satisfy the recurrence of
Vinti's translated spheroidal potential, and along orbits propagated over $300$
periods in the field truncated at $J_4$ they conserve the approximate third integral better than the centred structure, by
median factors of $23$ to $77$. On secular separatrices the integrated defect is the derivative of
the Melnikov function, tested in the eccentric Kozai--Lidov problem, and transverse zero sets of the operator define event
surfaces.
\end{abstract}

\medskip
\noindent\textbf{Keywords:} Satellite theory; Intermediary orbits; Separability; Lie transforms; Critical inclination; Melnikov method

\section{Introduction}
\label{sec:intro}

The main problem of artificial satellite theory, the motion of a particle in
the gravitational field of an oblate planet truncated after the second zonal
harmonic, has no additional analytic first integral \citep{IrigoyenSimo1993} and has
regions of chaotic motion \citep{CellettiNegrini1995}, and
its analytical treatment rests on an integrable approximation of the zonal
Hamiltonian, the intermediary, and on a near-identity canonical
transformation that connects the variables of the intermediary with those of
the actual motion \citep{Brouwer1959,Kozai1959,Deprit1969,Lara2021}. \citet{Brouwer1959} takes
the average of the perturbation over the mean anomaly as the intermediary,
whereas \citet{Sterne1958}, \citet{Garfinkel1958}, \citet{Aksnes1965},
\citet{CidLahulla1969} and \citet{Deprit1981} retain part of the short-period
dependence and solve the resulting problem in closed form. \citet{Vinti1961}
chose a potential that separates in oblate spheroidal coordinates and
reproduces the second zonal harmonic exactly, and \citet{AlfriendCoffey1984},
\citet{GurfilLara2014} and \citet{Lara2021} developed the radial
intermediaries that follow from the elimination of the parallax
\citep{Deprit1981}.

A general form containing these intermediaries as special cases was given by
\citet{Martinusi2020}, after an earlier systematic attempt by
\citet{FerrandizFloria1991}. In polar-nodal variables it adds to the Kepler
Hamiltonian four potentials with the radial weights $r^0,\dots,r^{-3}$, whose
coefficients depend on the angular momentum and its polar component and may
contain terms in twice the argument of latitude, and it selects the
coefficients by integrability, imposed through a St\"ackel-type ansatz, and by the efficiency condition of \citet{Deprit1981}. In that construction,
integrability enters as an assumption on the form of the Hamiltonian, and the
departure of a member of the family, or of the zonal problem itself, from
separation in the chosen variables, is not quantified.

The five-input operator introduced by \citet{MartinusiPD} gives such a quantitative
measure. It is built from nested Poisson brackets of a Hamiltonian $H$ and four probe
functions, reproduces the separability expressions of \citet{LeviCivita} on
two canonical pairs, and has off-diagonal values that are intrinsic on a
symplectic product structure. Since it is cubic in $H$, its expansion along a
line $H_0+\varepsilon V$ terminates after the cubic term, and the first-order
coefficient, the linearised operator $\Lop_{H_0}$, is related to the Melnikov
integrand by an exact identity for Hamiltonians with a cyclic angle
\citep[Theorem~4.7]{MartinusiPD}.

The present paper applies that operator to satellite theory and follows one
scalar through the problem: evaluated on the factors of a chosen product
structure and expanded in the perturbation, the same defect gives the exact
defects of the intermediaries, the torque form of the zonal defect, the mean
radial defect along propagated orbits, the long-period part of the
second-order average and the derivative of the Melnikov function on secular
separatrices. Several classical statements, obtained historically by separate
arguments, appear in this way as properties of one quantity: the
integrability ansatz of the general intermediary is the vanishing of its
linearised defects, the torque equation of the Gauss form is the radial
defect of every axisymmetric field, the first-order Lie generator is the
transformation of the product structure whose response cancels the zonal
defect, Vinti's focal radius and the translated spheroidal potential are the
coordinate responses of the oblate family, and on a secular separatrix the
derivative of the Melnikov function is an integrated defect. The five-input
construction, the Melnikov--defect identity and the polar-nodal torque
bracket are taken from \citet[Proposition~4.2 and Theorem~4.7]{MartinusiPD}.

The polar-nodal variables used in most of the literature on intermediaries
\citep{Deprit1981,Lara2021} are singular at equatorial orbits, where the
product structure itself degenerates (Section~\ref{sec:charts}). The
statements below are formulated in a canonical modified polar-nodal chart, in
which the sum of the argument of latitude and the node replaces the argument
of latitude, as in the nonsingular elements of \citet{Lara2015}, and a
Poincar\'e-type pair built from the node and $\Theta-N$ replaces the nodal
pair \citep{Lyddane1963,MurrayDermott2000}. Since the Kepler Hamiltonian does
not depend on this pair, vanishing linearised defects in the modified chart do
not imply first-order separation, and the separation statements are made in
the polar-nodal chart at fixed polar momentum
(Proposition~\ref{prop:general-form} and Remark~\ref{rem:degenerate}).

The results are the following, in the order of the argument.
\begin{enumerate}[label=(\roman*)]
 \item \emph{Operator and modified chart} (Section~\ref{sec:charts}). The
 defects that involve the nodal Poincar\'e pair vanish at first order for every
 perturbation of the Kepler problem (Proposition~\ref{prop:structure}), the
 interaction terms of a sum of perturbations are polarisations of the cubic
 form of the operator (Section~\ref{sec:sums}), and two kernel families
 have identically vanishing defects (Lemma~\ref{lem:kernel}).
 \item \emph{Exact defects of intermediaries} (Section~\ref{sec:intermediaries}).
 The linearised defects of the general form vanish identically if and only
 if the argument of latitude appears only in the potential with weight
 $r^{-2}$, which is the integrability ansatz and the condition for additive
 separation of the first-order Hamilton--Jacobi equation in polar-nodal
 variables (Proposition~\ref{prop:general-form}). The second-order defects are
 explicit (Propositions~\ref{prop:radial-angular} and \ref{prop:general}), and
 the coupling with the Poincar\'e pair is produced by the dependence on the
 physical latitude, shared by the zonal term, by the intermediary of
 \citet{Aksnes1965} and by the original potentials of \citet{Sterne1958} and
 \citet{Garfinkel1958}, which separate exactly in the polar-nodal chart
 (Table~\ref{tab:intermediaries}).
 \item \emph{Zonal perturbations and the first-order Lie generator}
 (Section~\ref{sec:generator}). For every axisymmetric potential the radial
 defect is exactly linear and has a torque form, and the zonal harmonics
 interact only through the total latitudinal force
 (Proposition~\ref{prop:axisym}). Along propagated orbits the mean radial
 defect over each revolution follows closed-form laws at first and second
 order in $J_2$ (Proposition~\ref{prop:second-mean}), and the first-order defect
 is the operator response to the first-order Lie generator
 (Proposition~\ref{prop:generator}).
 \item \emph{Second-order efficiency and the critical inclination}
 (Section~\ref{sec:second}). The intermediaries with vanishing defects have
 averages independent of the argument of pericentre, so that, in the gauge of
 the first-order generator, the efficiency condition can be imposed at second
 order only on the secular part, and the long-period part is common to all
 efficient intermediaries without terms in the argument of latitude
 (Proposition~\ref{prop:second}). Its $J_2^2$ term is the coefficient obtained
 by the Lie calculation of \citet{Lara2021redux}, and at the critical
 inclination the resonant coefficient, which no regular axially symmetric
 change of gauge alters, is proportional to $J_4+J_2^2$
 (Corollary~\ref{cor:critical}).
 \item \emph{Oblate spheroidal coordinates} (Section~\ref{sec:vinti}). The
 oblate defect factorises through the residuals of \citet{An2013}. The
 coordinate response selects Vinti's focal radius at first order, and the
 second-order residual vanishes if and only if $J_4=-J_2^2$
 (Proposition~\ref{prop:vinti-response}). Displaced oblate structures separate
 the zonal potential exactly if and only if $J_{n+2}=2\delta J_{n+1}-aJ_n$ for all
 $n$, the recurrence of the translated spheroidal potential of
 \citet{Vinti1966inv}, and the response places the centre at
 $\frac12R_eJ_3J_2^{-1}$ (Lemma~\ref{lem:ladder} and
 Proposition~\ref{prop:shifted}). Along propagated orbits of the Earth and of
 Mars, the displaced structures reduce the excursion of the approximate third
 integral over $300$ periods by median factors between $23$ and $77$.
 \item \emph{Secular separatrices} (Section~\ref{sec:separatrices}). Coboundary
 parts of a perturbation have zero Melnikov function and zero integrated
 defect (Proposition~\ref{prop:coboundary}). In the eccentric Kozai--Lidov
 (EKL) problem \citep{Naoz2016,LithwickNaoz2011} the rotor is slower than the
 separatrix exponent at the saddle, and the changes of the axial action over
 finite Kozai cycles follow the first-order kick, with errors between $0.3$
 and $0.45$ times the distance of the cycle endpoints from the saddle.
 \item \emph{Event surfaces} (Section~\ref{sec:event}). Transverse zero sets of
 the operator define event surfaces. In the resonant canonical variables of
 \citet{Martinusi2023}, the defect of a planar Kepler problem with a uniform
 force and a radial potential is a single product to all orders, and for
 solar radiation pressure with the equatorial $J_2$ term its transverse zeros
 are the turning points of the parabolic coordinates aligned with the Sun
 (Proposition~\ref{prop:rcv}).
\end{enumerate}

Several of these statements are known, and they are identified where they
occur: the Levi-Civita conditions, the St\"ackel-type form of the general
intermediary, the torque equation and the classical event surfaces, the
$J_2^2$ long-period coefficient \citep{Lara2021redux}, the role of $J_4+J_2^2$ at
the critical inclination \citep{Brouwer1959,Izsak1963,Lara2026}, the residuals
and the separability condition of translated spheroidal coordinates
\citep{An2013}, the translated spheroidal potential and Vinti's reference
potential with the third harmonic \citep{Vinti1966inv,Vinti1966} and the
separability of the Stark problem in parabolic coordinates. The operator
recovers them with the same definitions, while the methods by which they were
first obtained remain more direct. The operator does not prove the
non-integrability of the main problem
\citep{IrigoyenSimo1993,CellettiNegrini1995}, whose stochastic zones are
negligible for Earth satellites \citep{Simo1991}, and the dynamical statements
are restricted to the regimes in which the first-order theory applies.
Section~\ref{sec:limitations} states the limitations, and
Appendix~\ref{app:numerics} collects the numerical procedures and diagnostics,
all of which are reproduced by the codes archived in \citet{CodeCMDA}.

\section{The separability operator and the modified polar-nodal chart}
\label{sec:charts}

\subsection{The five-input operator}

Let $M$ be a symplectic manifold of dimension $2n$ with Poisson bracket
$\{F,G\}=\sum_i(F_{q_i}G_{p_i}-F_{p_i}G_{q_i})$ in canonical coordinates,
subscripts denoting partial derivatives \citep{Arnold}. Following
\citet{MartinusiPD}, for functions $a,b,c,d,H$ the minor and the derived field
are
\begin{equation}
 \mathcal D_{a,b}(F,G)=\{a,F\}\{b,G\}-\{a,G\}\{b,F\},\qquad
 \Y^H_{a,b}(F)=\mathcal D_{a,b}(F,H),
\end{equation}
and with the response rates $u=\{H,c\}$ and $v=\{H,d\}$ the five-input
operator is
\begin{equation}
 \B_{a,b;c,d}[H]=u\,\Y^H_{a,b}(v)-v\,\Y^H_{a,b}(u).
 \label{eq:B}
\end{equation}
On two canonical pairs the operator reproduces the expression of
\citet{LeviCivita} \citep[Proposition~3.1]{MartinusiPD},
\begin{equation}
 b_{q_iq_j}[H]:=\B_{q_i,p_i;q_j,p_j}[H]
 \begin{aligned}[t]
 &=-H_{p_i}H_{p_j}H_{q_iq_j}+H_{q_i}H_{p_j}H_{p_iq_j}\\
 &\quad+H_{p_i}H_{q_j}H_{q_ip_j}-H_{q_i}H_{q_j}H_{p_ip_j},
 \end{aligned}
 \label{eq:LC}
\end{equation}
and the off-diagonal values $b_{q_iq_j}$, $i\ne j$, are called the defects of
$H$ on the product structure defined by the pairs. They are intrinsic on the
product structure \citep[Proposition~3.2]{MartinusiPD}, and on a domain where
the momentum derivatives do not vanish, the simultaneous vanishing of all
defects is equivalent to local additive separation of the Hamilton--Jacobi
equation in these variables \citep[Theorem~3.5]{MartinusiPD}, which is
Levi-Civita's theorem \citep{LeviCivita,Benenti2016}. The regularity condition
on the momentum derivatives is part of the theorem: where a momentum
derivative vanishes identically at zeroth order, as for the nodal pair of the
modified chart below, the defects alone do not decide separation
(Remark~\ref{rem:degenerate}). Three further results of \citet{MartinusiPD} are
used below.
\begin{enumerate}[label=(F\arabic*)]
 \item \emph{Exact expansion.} Since $\B$ is cubic in $H$, for every $V$ and
 every $\varepsilon$
 \begin{equation}
  b[H_0+\varepsilon V]=b[H_0]+\varepsilon\,b_1+\varepsilon^2b_2+\varepsilon^3b[V],
  \label{eq:cubic}
 \end{equation}
 with no remainder \citep[eq.~(4.2)]{MartinusiPD}. The coefficient
 $b_1=\Lop_{H_0}(V)$ defines the linearised separability operator, which is
 linear in $V$ and of differential order at most two.
 \item \emph{Covariance.} For a canonical diffeomorphism $\Phi$ and the
 product structure $\Phi_*\mathbf S$ transported by $\Phi$,
 $b^{\Phi_*\mathbf S}[H]=b^{\mathbf S}[H\circ\Phi]\circ\Phi^{-1}$
 \citep[Proposition~2.4]{MartinusiPD}. In a chart $(Q,P)$ of canonical functions
 of $(q,p)$, such as the Delaunay variables, the derivatives of $H$ are the
 nested brackets $H_{P_i}=\{Q_i,H\}$, $H_{Q_i}=-\{P_i,H\}$ and their iterates, so
 that the defects follow from the forward map alone, without the inversion of
 Kepler's equation.
 \item \emph{Melnikov--defect identity.} For Hamiltonians
 $H_0(q,p,I)+\varepsilon V(q,p,\vartheta,I)$ with a cyclic angle $\vartheta$ of
 $H_0$, rotor frequency $\nu_\vartheta=\partial H_0/\partial I$ and a homoclinic
 orbit $\gamma$ of the $(q,p)$ system, the Melnikov function $M(\vartheta_0)$
 \citep{HolmesMarsden1982,GuckenheimerHolmes} and the integrated defect
 \begin{equation}
  D_\gamma(\vartheta_0)=\int_{-\infty}^{\infty}\nu_\vartheta^{-2}\,\Lop_{H_0}(V)_{q\vartheta}
  \bigl(\gamma(t),\vartheta_0+\Phi(t),I_h\bigr)\,\dd t
  \label{eq:Dgamma}
 \end{equation}
 satisfy $M'=D_\gamma$, and if $V$ vanishes on the hyperbolic periodic orbit,
 $M=\Psi'$ and $D_\gamma=\Psi''$ with the Melnikov potential
 $\Psi(\vartheta_0)=\int V\,\dd t$ \citep[Theorem~4.7]{MartinusiPD}. The phase
 $\Phi(t)$ is the integral of $\nu_\vartheta$ along $\gamma$, and the hypotheses
 (H1)--(H3) of \citet{MartinusiPD} require a hyperbolic equilibrium of the
 $(q,p)$ system, a rotor frequency that does not vanish on the closure of
 $\gamma$ and a smooth perturbation.
\end{enumerate}

\subsection{Superposition of perturbations and kernel families}
\label{sec:sums}

Since the operator is cubic in the Hamiltonian, the defect of a sum of
perturbations, such as the harmonics of a planet or the potentials of an
intermediary, differs from the sum of the defects by interaction terms. For a
fixed pair of factors with canonical pairs $(q_i,p_i)$ and $(q_j,p_j)$, let
$T(F,G,K)$ denote the symmetric trilinear form obtained from \eqref{eq:LC} by
replacing the three factors of each term by the corresponding derivatives of
$F$, $G$ and $K$ and averaging over the six permutations of $(F,G,K)$, so that
$T(H,H,H)=b_{q_iq_j}[H]$. By trilinearity, the coefficients of \eqref{eq:cubic}
are $b_1=3T(H_0,H_0,V)$, $b_2=3T(H_0,V,V)$ and $b[V]=T(V,V,V)$, so that $b_1$ is
additive in the terms of $V=\sum_kV_k$, and the interaction defect of two
perturbations,
\begin{equation}
 \begin{aligned}
 I[V_1,V_2]&=b[H_0+V_1+V_2]-b[H_0+V_1]-b[H_0+V_2]+b[H_0]\\
 &=6T(H_0,V_1,V_2)+3T(V_1,V_1,V_2)+3T(V_1,V_2,V_2),
 \end{aligned}
 \label{eq:interaction}
\end{equation}
contains only terms with both $V_1$ and $V_2$, its lowest-order part being
bilinear in the two perturbations. The interaction depends on the product
structure: the zonal harmonics interact in the modified chart and do not
interact in the classical polar-nodal chart (Proposition~\ref{prop:axisym}).

Let $\Y_i=H_{p_i}\partial_{q_i}-H_{q_i}\partial_{p_i}$ be the component of the
Hamiltonian vector field of $H$ along the $i$th factor, and let $\mathbf H_{ij}$ be
the mixed block of the Hessian of $H$, with one derivative in the factor $i$
and one in the factor $j$. Then
\begin{equation}
	b_{q_iq_j}[H]=-\mathbf H_{ij}(\Y_i,\Y_j)
	\label{eq:mixed-Hessian}
\end{equation}
\citep[Proposition~3.2(a)]{MartinusiPD}, and this form gives the following
lemma, introduced here. Part (a) extends the Liouville Hamiltonians
\citep[after Theorem~3.5]{MartinusiPD}, classical in the theory of separation
\citep{Benenti2016}, to numerators and denominators that are arbitrary
functions of each canonical pair.

\begin{lemma}[Kernel families]
\label{lem:kernel}
On two canonical pairs $(q_1,p_1)$ and $(q_2,p_2)$, the following hold.
\begin{enumerate}[label=(\alph*)]
 \item For smooth functions $U_1,Z_1$ of $(q_1,p_1)$ and $U_2,Z_2$ of
 $(q_2,p_2)$, the quotient
 \begin{equation}
  H=\frac{U_1+U_2}{Z_1+Z_2},\qquad Z_1+Z_2\ne0,
  \label{eq:liouville}
 \end{equation}
 has $b_{q_1q_2}[H]=0$ identically.
 \item For a smooth function $K(q_2,p_2)$ and a smooth function $\Phi$ of three
 variables, $H=\Phi(q_1,p_1,K(q_2,p_2))$ has $b_{q_1q_2}[H]=0$ identically.
\end{enumerate}
No condition on the momentum derivatives of $H$ is required in either case.
\end{lemma}

\begin{proof}
\emph{(a)} Put $S=Z_1+Z_2$. For a derivative $x$ in the first factor and a
derivative $y$ in the second, $H_x=P_x/S$ with $P_x=(U_1)_x-H(Z_1)_x$, and
$H_y=Q_y/S$ with $Q_y=(U_2)_y-H(Z_2)_y$, so that
\[
 H_{xy}=-\frac{Q_y(Z_1)_x+P_x(Z_2)_y}{S^2}.
\]
The covector $P=S\,\dd_1H$ annihilates $\Y_1$, since $\Y_1H=0$, and similarly
$Q=S\,\dd_2H$ annihilates $\Y_2$, where $\dd_iH$ is the differential of $H$
restricted to the $i$th factor. Each term of $\mathbf H_{12}(\Y_1,\Y_2)$ therefore
contains $P(\Y_1)$ or $Q(\Y_2)$, and \eqref{eq:mixed-Hessian} gives $b=0$.

\emph{(b)} The second field is $\Y_2=\Phi_K\,(K_{p_2}\partial_{q_2}-K_{q_2}\partial_{p_2})$, so that
$\dd K(\Y_2)=0$. The mixed block of the Hessian of $H$ is the product of the
first-factor differential of $\Phi_K$ and $\dd K$, and its value on $(\Y_1,\Y_2)$
vanishes.
\end{proof}

On the level $H=E$, the family \eqref{eq:liouville} gives the separated form
$(U_1-EZ_1)+(U_2-EZ_2)=0$ of the Poincar\'e--Sundman Hamiltonian
$(Z_1+Z_2)(H-E)$, which covers the Kepler and Stark problems in parabolic and
Levi-Civita coordinates \citep{LeviCivita1920,LantoineRussell2011} and is used in
Section~\ref{sec:rcv}. The family (b) contains the polar-nodal Hamiltonians
$\frac12R^2+K(r,G)$ with an angular integral $G(\theta,\Theta)$ at fixed $N$, which
are the separable case of Proposition~\ref{prop:radial-angular}(d).

\subsection{Polar-nodal variables}

The polar-nodal variables $(r,\theta,\nu,R,\Theta,N)$, also called Hill or
Whittaker variables \citep{Hill1913,Deprit1981,Whittaker1988}, are the radial
distance, the argument of latitude, the longitude of the node, the radial
velocity, the angular momentum and its polar component, respectively, and they are
canonical where $r>0$ and $\Theta>|N|$. With $c=N/\Theta$ and $s=\sqrt{1-c^2}$ the
cosine and sine of the inclination, $p=\Theta^2/\mu$, $e$ the eccentricity,
$\eta=\sqrt{1-e^2}$, $f$ the true anomaly, $\kappa=e\cos f=p/r-1$ and
$\sigma=e\sin f=pR/\Theta$, the Kepler Hamiltonian is
\begin{equation}
 H_{00}=\frac12\Bigl(R^2+\frac{\Theta^2}{r^2}\Bigr)-\frac{\mu}{r}
 =-\frac{\mu^2}{2L^2},
 \label{eq:kepler}
\end{equation}
with $L$ the Delaunay action conjugate to the mean anomaly $\ell$; the
quantities $e$, $\eta$ and $L$ are functions on phase space through $H_{00}$ and
$\Theta$. On the prograde equatorial orbits, $\Theta=N$, the function $\Theta-N\ge0$
attains its minimum, so that $\dd\Theta=\dd N$ there, the momenta of the
factors $(\theta,\Theta)$ and $(\nu,N)$ are dependent, and the polar-nodal
product structure degenerates; the retrograde equatorial orbits, $\Theta=-N$,
behave in the same way.

\subsection{Modified polar-nodal variables}
\label{sec:modified}

The chart used here follows from the polar-nodal variables by the two steps
that lead from Delaunay to Poincar\'e variables \citep{MurrayDermott2000}. The
angles are combined into
\begin{equation}
 \psi=\theta+\nu,\qquad \chi=-\nu,
 \label{eq:psi}
\end{equation}
with the momenta $\Theta$ and $\Theta-N$, by the generating function
$(\theta+\nu)\Theta'-\nu X$, and the angle $\psi$, the argument of latitude
measured from the reference direction along the equator and then along the
orbit, reduces to the true longitude at prograde equatorial orbits. The pair
$(\chi,\Theta-N)$, whose action vanishes there, is then replaced by
\begin{equation}
 \begin{aligned}
 \xi&=\sqrt{2(\Theta-N)}\,\sin\chi=-\sqrt{2(\Theta-N)}\,\sin\nu,\\
 \zeta&=\sqrt{2(\Theta-N)}\,\cos\chi=\sqrt{2(\Theta-N)}\,\cos\nu,
 \end{aligned}
 \label{eq:xizeta}
\end{equation}
with $\rho^2=\xi^2+\zeta^2=2(\Theta-N)$. The following statement is introduced here.

\begin{proposition}[Modified polar-nodal chart]
\label{prop:chart}
The variables $(r,\psi,\xi,R,\Theta,\zeta)$ have the following properties.
\begin{enumerate}[label=(\alph*)]
 \item They are canonical, with the pairs $(r,R)$, $(\psi,\Theta)$ and $(\xi,\zeta)$.
 \item They form a chart on the set $r>0$, $\Theta>0$, $\rho^2<4\Theta$, which
 contains the prograde equatorial orbits $\xi=\zeta=0$. The orbits excluded
 from this set are the collisions $r=0$, the rectilinear orbits $\Theta=0$ and
 the retrograde equatorial orbits $\rho^2=4\Theta$.
 \item The inclination functions and the sine of the latitude $\varphi$ are
 \begin{equation}
  c=1-\frac{\rho^2}{2\Theta},\qquad
  s^2=\frac{\rho^2}{\Theta}\Bigl(1-\frac{\rho^2}{4\Theta}\Bigr),\qquad
  \sin\varphi=s\sin\theta=\Bigl(1-\frac{\rho^2}{4\Theta}\Bigr)^{1/2}
  \frac{\zeta\sin\psi+\xi\cos\psi}{\sqrt\Theta},
  \label{eq:sinphi}
 \end{equation}
 and $s^2\cos2\theta=-(4\Theta-\rho^2)\bigl[(\xi^2-\zeta^2)\cos2\psi+2\xi\zeta\sin2\psi\bigr]/(4\Theta^2)$,
 so that every zonal potential is a smooth function of the chart variables.
\end{enumerate}
\end{proposition}

\begin{proof}
\emph{(a)} The map \eqref{eq:psi} is canonical by the generating function given
above, and the map from an angle--action pair $(\chi,X)$ with $X>0$ to
$(\sqrt{2X}\sin\chi,\sqrt{2X}\cos\chi)$ is canonical because its bracket
$\{\xi,\zeta\}_{\chi X}=\cos^2\chi+\sin^2\chi=1$ \citep{MurrayDermott2000}, so that
the composition of the two maps is canonical.

\emph{(b)} The inverse map is $\Theta=\Theta$, $N=\Theta-\frac12\rho^2$,
$\nu=\operatorname{atan2}(-\xi,\zeta)$ and $\theta=\psi-\nu$, and the condition
$|N|<\Theta$ becomes $0<\rho^2<4\Theta$. The unit vector $\hat{\mathbf u}$ of the
position has $\hat u_x+\mathrm i\hat u_y=\mathrm e^{\mathrm i\nu}(\cos\theta+\mathrm ic\sin\theta)$ and
$\hat u_z=s\sin\theta$ \citep{Deprit1981}, and with
$\zeta-\mathrm i\xi=\rho\,\mathrm e^{\mathrm i\nu}$ and $\frac12(1\pm c)$ from
\eqref{eq:sinphi} the Cartesian position $\mathbf x=r\hat{\mathbf u}$ and velocity are
\begin{equation}
 \begin{gathered}
 x+\mathrm iy=r\Bigl[\Bigl(1-\frac{\rho^2}{4\Theta}\Bigr)\mathrm e^{\mathrm i\psi}
 +\frac{(\zeta-\mathrm i\xi)^2}{4\Theta}\,\mathrm e^{-\mathrm i\psi}\Bigr],\qquad
 z=r\sin\varphi,\\
 \dot{\mathbf x}=R\,\hat{\mathbf u}+\frac{\Theta}{r}\,\partial_\psi\hat{\mathbf u},
 \end{gathered}
 \label{eq:cartesian}
\end{equation}
with $\sin\varphi$ of \eqref{eq:sinphi}, since the transverse unit vector is the
derivative of $\hat{\mathbf u}$ in $\theta$ at fixed node, which is $\partial_\psi$ at fixed
$(\xi,\zeta)$. These expressions are smooth on the whole set of (b), including
$\rho=0$, in the same way as Poincar\'e variables extend Delaunay variables to
circular and equatorial orbits \citep{MurrayDermott2000}. At $\rho^2=4\Theta$ the
orbit is retrograde and equatorial, and the angle $\psi=\theta+\nu$ is not defined
there.

\emph{(c)} With $N=\Theta-\frac12\rho^2$, $c=N/\Theta$ and $s^2=(1-c)(1+c)$ give the
first two expressions. Since $\theta=\psi-\nu$, $\sqrt2(\Theta-N)^{1/2}\sin\theta=
\zeta\sin\psi+\xi\cos\psi$, which gives $\sin\varphi$, and the expression for
$s^2\cos2\theta$ follows from the same substitution and is a polynomial in
$(\xi,\zeta)$ with coefficients that are smooth for $\Theta>0$.
\end{proof}

The defects $b_{r\psi}$, $b_{r\xi}$ and $b_{\psi\xi}$ of the modified chart are
evaluated with \eqref{eq:LC} on the pairs $(r,R)$, $(\psi,\Theta)$ and $(\xi,\zeta)$.
Since the Kepler Hamiltonian depends only on $(r,R,\Theta)$, they have the
following structure for arbitrary perturbations, introduced here.

\begin{proposition}[Defects of perturbed Kepler problems]
\label{prop:structure}
Let $H=H_{00}+kV$ with a smooth function $V$ of all six variables of the
modified chart and a parameter $k$, and denote by $\{\cdot,\cdot\}_{\xi\zeta}$
the Poisson bracket in the pair $(\xi,\zeta)$. Then the following hold.
\begin{enumerate}[label=(\alph*)]
 \item The defects involving the Poincar\'e pair have no first-order term,
 \begin{align}
  b_{r\xi}[H]&=k^2\bigl(R\,\{V,\partial_rV\}_{\xi\zeta}
  -\partial_rH_{00}\,\{V,\partial_RV\}_{\xi\zeta}\bigr)+k^3\,b_{r\xi}[V],
  \label{eq:brxi}\\
  b_{\psi\xi}[H]&=k^2\,\frac{\Theta}{r^2}\,\{V,\partial_\psi V\}_{\xi\zeta}+k^3\,b_{\psi\xi}[V].
  \label{eq:bpsixi}
 \end{align}
 \item The first-order radial defect is
 \begin{equation}
  \Lop_{H_{00}}(V)_{r\psi}=-\frac{R\,\Theta}{r^4}\,\partial_r\bigl(r^2\partial_\psi V\bigr)
  +\frac{\Theta}{r^2}\,\partial_rH_{00}\,\partial_R\partial_\psi V .
  \label{eq:Lrpsi}
 \end{equation}
 \item If $V$ depends on $(\xi,\zeta)$ only through $\rho^2$, then
 $b_{r\xi}[H]=b_{\psi\xi}[H]=0$ identically, for every $k$.
\end{enumerate}
\end{proposition}

\begin{proof}
\emph{(a)} Each of the four terms of \eqref{eq:LC} for the pairs $(r,R)$ and
$(\xi,\zeta)$ contains one factor differentiated with respect to $\xi$ and another
with respect to $\zeta$, and $H_{00}$ does not depend on these variables, so both
factors come from $kV$ and there is no term of order $k$. The terms of order
$k^2$ take the third factor from $H_{00}$, whose derivatives are $\partial_RH_{00}=R$
and $\partial_rH_{00}$, and their sum is \eqref{eq:brxi}. For the pairs $(\psi,\Theta)$
and $(\xi,\zeta)$ the same argument applies, and since $H_{00}$ does not depend on
$\psi$, only the derivative $\partial_\Theta H_{00}=\Theta r^{-2}$ remains, which gives
\eqref{eq:bpsixi}. The order-$k^3$ terms are $b[V]$ by \eqref{eq:cubic}.

\emph{(b)} In \eqref{eq:LC} with the pairs $(r,R)$ and $(\psi,\Theta)$ the
derivatives $\partial_\psi H_{00}$, $\partial_r\partial_\psi H_{00}$, $\partial_R\partial_\psi H_{00}$ and
$\partial_R\partial_\Theta H_{00}$ vanish, and the terms linear in $V$ are
$-R\,\Theta r^{-2}\partial_r\partial_\psi V+\partial_rH_{00}\,\Theta r^{-2}\partial_R\partial_\psi V
-2R\,\Theta r^{-3}\partial_\psi V$, which is \eqref{eq:Lrpsi}.

\emph{(c)} If $F$ depends on $(\xi,\zeta)$ only through $\rho^2$, then
$\partial_\xi F=2\xi F'$ and $\partial_\zeta F=2\zeta F'$, with $F'$ the derivative with
respect to $\rho^2$. Applied to $H$ and to its first derivatives, this makes the
first and third terms of \eqref{eq:LC} cancel, and similarly the second and
fourth, for both pairs of factors.
\end{proof}

Statement (c) applies to every axisymmetric perturbation independent of the
argument of latitude, which is a function of $(r,R,\Theta,N)$ with
$N=\Theta-\frac12\rho^2$. The first-order defect \eqref{eq:Lrpsi} contains no
derivative with respect to $\Theta$, the only derivative of the radial and
angular pairs that changes between the two charts, so that it equals the
first-order radial--angular defect of the polar-nodal chart with $\theta$
replaced by $\psi$; for $V$ independent of $R$, it is the first-order term of
\citep[Proposition~4.2(a)]{MartinusiPD}.

\subsection{Numerical methods}
\label{sec:verification}

The numerical results were computed with NumPy \citep{NumPy2020} and SciPy
\citep{SciPy2020}, and all orbits were integrated with the eighth-order
Dormand--Prince method of SciPy with relative tolerances between $10^{-13}$
and $10^{-11}$. Appendix~\ref{app:numerics} describes the procedures and
collects the diagnostics, and the codes and the data of every table and figure
are archived \citep{CodeCMDA}. Statements that a defect vanishes identically
were decided by exact reduction of the numerators with SymPy \citep{SymPy2017},
and numerical samples serve only as a cross-check (Appendix~\ref{app:zero}).
During the preparation of this work the author used Claude (Anthropic) in
order to check derivations, write and run the numerical code, and edit the
language of the manuscript. After using this tool, the author reviewed and
edited the content as needed and takes full responsibility for the content of
the article.

\section{Exact defects of intermediaries}
\label{sec:intermediaries}

\subsection{General intermediary potentials}
\label{sec:general2020}

The classical intermediaries
\citep{Sterne1958,Garfinkel1958,Brouwer1959,Aksnes1965,CidLahulla1969,Deprit1981}
add to the Kepler Hamiltonian potentials with one of the radial powers
$r^0,\dots,r^{-3}$ and at most a term in twice the argument of latitude, with
coefficients that depend on the inclination. The general intermediary of
\citet{Martinusi2020} collects them in the form
$K=H_{00}+\mathcal V_0+\mathcal V_1+\mathcal V_2+\mathcal V_3$ with
\begin{equation}
 \begin{aligned}
 \mathcal V_0&=\frac{k_2\eta^3}{p^3}\bigl(A_{00}+T_0\bigr),&\qquad
 \mathcal V_1&=\frac{k_2\eta}{p^2r}\bigl(A_{10}+T_1\bigr),\\
 \mathcal V_2&=\frac{k_2}{pr^2}\bigl(A_{20}+T_2\bigr),&\qquad
 \mathcal V_3&=\frac{k_2}{r^3}\bigl(A_{30}+T_3\bigr),
 \end{aligned}
 \qquad T_k=A_{k1}\sin2\theta+A_{k2}\cos2\theta,
 \label{eq:general-full}
\end{equation}
where the coefficients $A_{km}$ are functions of $(\Theta,N)$ and
$k_2=\frac14\mu J_2R_e^2$, with $R_e$ the equatorial radius, so that the zonal
term is $H_{10}=k_2r^{-3}(1-3c^2-3s^2\cos2\theta)$. The prefactors make the
constant parts average to $k_2\eta^3p^{-3}A_{k0}$ over the mean anomaly. Along a
Kepler orbit $\dd\ell=n\,r^2\Theta^{-1}\dd f$, with $n$ the mean motion
\citep{Brouwer1959}, so that $T_2$ and $T_3$ average to zero, whereas $\theta=g+f$,
$\langle\sin2f\rangle_\ell=0$ and $\langle\cos2f\rangle_\ell=(1-\eta)(1+2\eta)/(1+\eta)$ give
\begin{equation}
 \begin{aligned}
 \langle\mathcal V_0\rangle_\ell&=\frac{k_2\eta^3}{p^3}\Bigl[A_{00}
 +\frac{(1-\eta)(1+2\eta)}{1+\eta}\bigl(A_{01}\sin2g+A_{02}\cos2g\bigr)\Bigr],\\
 \langle\mathcal V_1\rangle_\ell&=\frac{k_2\eta^3}{p^3}\Bigl[A_{10}
 +\frac{1-\eta}{1+\eta}\bigl(A_{11}\sin2g+A_{12}\cos2g\bigr)\Bigr].
 \end{aligned}
 \label{eq:V01-avg}
\end{equation}
Integrability is imposed in \citet{Martinusi2020} through the St\"ackel-type form
\begin{equation}
 K=\frac12\Bigl(R^2+\frac{\Theta^2}{r^2}\Bigr)-\frac{\mu}{r}+U_1(r,\Theta,N)
 +\frac{1}{r^2}\,U_2(\theta,\Theta,N),
 \label{eq:stackel2020}
\end{equation}
which admits the angular integral $\Theta^2+2U_2$ when $U_1$ does not depend on
$\Theta$ (Proposition~\ref{prop:radial-angular}(d) gives the exact condition) and
keeps the argument of latitude only in $\mathcal V_2$,
\begin{equation}
 A_{k1}=A_{k2}=0,\qquad k\in\{0,1,3\}.
 \label{eq:ansatz}
\end{equation}
The general intermediary is then
\begin{equation}
 \begin{aligned}
 \mathcal V_0&=\frac{k_2\eta^3}{p^3}A_{00},&\qquad
 \mathcal V_1&=\frac{k_2\eta}{p^2r}A_{10},\\
 \mathcal V_2&=\frac{k_2}{pr^2}\bigl(A_{20}+A_{21}\sin2\theta+A_{22}\cos2\theta\bigr),&\qquad
 \mathcal V_3&=\frac{k_2}{r^3}A_{30},
 \end{aligned}
 \label{eq:general}
\end{equation}
with the averages $\langle\mathcal V_k\rangle_\ell=k_2\eta^3p^{-3}A_{k0}$, $k=0,\dots,3$, and
Deprit's efficiency condition, which requires the intermediary and the zonal
term to have the same average over the mean anomaly \citep{Deprit1981},
becomes, with $\langle H_{10}\rangle_\ell=k_2\eta^3p^{-3}(1-3c^2)$
\citep{Kozai1959,Brouwer1959},
\begin{equation}
 A_{00}+A_{10}+A_{20}+A_{30}=1-3c^2 .
 \label{eq:efficiency}
\end{equation}
The phase-space transcriptions of the historical intermediaries into this
form, and the originals that differ from them, are listed in
Table~\ref{tab:intermediaries}.

\begin{remark}[Corrections to \citet{Martinusi2020}]
\label{rem:erratum}
The following misprints of \citet{Martinusi2020}, whose equation numbers are
used in this remark, are corrected here, and all formulas of this paper are
given in the corrected form, with the coefficients of
\citet{Sterne1958} and \citet{Garfinkel1958} taken from the original papers.
(i) In eq.~(15) the coefficients of Sterne, Brouwer, Cid and Lahulla and
Deprit are printed with the sign opposite to \eqref{eq:efficiency}, and the
coefficients $A_{20}$ and $A_{22}$ of Garfinkel are printed as $3(3c^2-1)$ and
$3s^2$, whereas \citet[eq.~(5)]{Garfinkel1958} gives $A_{20}=3(1-3c^2)$ and
$A_{22}=-3s^2$; in eq.~(19) the sum is therefore $A_{10}+A_{20}=0$, and Garfinkel's
transcription is not efficient, as stated there. (ii) The zonal term of eqs.~(9) and (21), to which the efficiency condition
eq.~(18) refers, is twice the potential of eq.~(4) for the same $k_2$. (iii)
The first-order generator, eq.~(26), is written for the normalisation of
eq.~(4), in which $1-3c^2$ is replaced by $\frac12(1-3c^2)$, and with the opposite
sign of the symplectic matrix, eq.~(24). Up to these conventions it coincides
with \eqref{eq:W1} below once its prefactor $k_2\eta^3p^{-3}$ is replaced by
$k_2p^{-1}\Theta^{-1}$, which corrects the printed relation $\dd\ell=r^2\Theta^{-1}\dd f$
to $\dd\ell=n\,r^2\Theta^{-1}\dd f$, and its coefficient $e\,s^2$ of $\cos2\theta$ is
replaced by $\sigma s^2$. (iv) Eq.~(9) prints the potential of Sterne with the
overall sign opposite to the zonal term of the same equation, and in the
potential of Garfinkel the bracket of the term in $r^{-2}$ should read
$3s^2-2-s^2\cos2\theta$; Table~\ref{tab:intermediaries} uses the sign of $H_{10}$. (v) The
potential of Aksnes in eq.~(9), $2k_2p^{-1}r^{-2}(1-3c^2-3s^2\cos2\theta)$, is twice
the entry of eq.~(15); only the latter satisfies the efficiency condition
eq.~(18), and it is used here.
\end{remark}

The operator gives the ansatz \eqref{eq:ansatz} as the exact condition for the
vanishing of the linearised defects, without assuming the form
\eqref{eq:stackel2020}, and the same condition governs additive separation of
the first-order Hamilton--Jacobi equation in the polar-nodal chart, by the
following proposition, introduced here. Parts (a)--(c) hold also in the
polar-nodal chart with $\psi$ replaced by $\theta$.

\begin{proposition}[First-order defect of the general form]
\label{prop:general-form}
Let $K=H_{00}+\sum_k\mathcal V_k$ with \eqref{eq:general-full} and smooth coefficients
$A_{km}(\Theta,N)$, on a domain of elliptic orbits, $H_{00}<0$, with $\Theta>|N|$.
Then the following hold.
\begin{enumerate}[label=(\alph*)]
 \item For every function $a(E,\Theta,N)$, every integer $k$ and every smooth
 $T(\theta,\Theta,N)$, the potential $V=a(H_{00},\Theta,N)\,r^{-k}\,T$ has the first-order
 radial defect
 \begin{equation}
  \Lop_{H_{00}}(V)_{r\psi}=(k-2)\,a(H_{00},\Theta,N)\,\frac{R\,\Theta}{r^{k+3}}\,\partial_\theta T .
  \label{eq:Lop-power}
 \end{equation}
 \item The first-order defects of $K$ are
 \begin{equation}
  \Lop_{H_{00}}(K-H_{00})_{r\psi}=k_2R\,\Theta\Bigl(-\frac{2\eta^3}{p^3r^3}\,\partial_\theta T_0
  -\frac{\eta}{p^2r^4}\,\partial_\theta T_1+\frac{1}{r^6}\,\partial_\theta T_3\Bigr),
  \label{eq:Lop-general-full}
 \end{equation}
 and $\Lop_{H_{00}}(K-H_{00})_{r\xi}=\Lop_{H_{00}}(K-H_{00})_{\psi\xi}=0$. The term $T_2$ of
 $\mathcal V_2$ does not contribute.
 \item All first-order defects of $K$ vanish identically if and only if
 \eqref{eq:ansatz} holds.
 \item In the polar-nodal chart, fix $N$, let $(E_0,\Theta_0)$ range over an open set
 with $E_0<0$ and $\Theta_0>|N|$, let $I_r$ be an open interval of radii on which
 $R_0(r)^2=2E_0+2\mu r^{-1}-\Theta_0^2r^{-2}>0$, and let $I_\theta$ be an open interval or circle
 of values of $\theta$. For a solution
 $W=\int R_0\,\dd r+\Theta_0\theta+k_2W_1+O(k_2^2)$ of $K(r,\theta,\partial_rW,\partial_\theta W,N)=E_0+k_2E_1$,
 the first-order equation
 \begin{equation}
  R_0\,\partial_rW_1+\frac{\Theta_0}{r^2}\,\partial_\theta W_1
  +k_2^{-1}(K-H_{00})\big|_{R=R_0(r),\,\Theta=\Theta_0}=E_1
  \label{eq:HJ1}
 \end{equation}
 has an additive solution $W_1=W_r(r)+W_\theta(\theta)$ on $I_r\times I_\theta$, for some constant
 $E_1$ and every such $(E_0,\Theta_0)$, if and only if \eqref{eq:ansatz} holds.
\end{enumerate}
\end{proposition}

\begin{proof}
\emph{(a)} In \eqref{eq:Lrpsi} the derivative with respect to $r$ acts also on
$a$ through $\partial_rH_{00}$, and $\partial_R\partial_\psi V=a_E\,R\,r^{-k}\,\partial_\theta T$, since
$\partial_RH_{00}=R$ and $\partial_\psi=\partial_\theta$ at fixed $(\xi,\zeta)$. Hence
\[
 \Lop_{H_{00}}(V)_{r\psi}=-\frac{R\,\Theta}{r^4}\Bigl[a\,\partial_r\bigl(r^{2-k}\bigr)
 +r^{2-k}a_E\,\partial_rH_{00}\Bigr]\partial_\theta T+\frac{\Theta}{r^2}\,\partial_rH_{00}\,a_E\,R\,r^{-k}\,\partial_\theta T,
\]
in which the two terms with $a_E$ cancel, and $\partial_r(r^{2-k})=(2-k)r^{1-k}$ gives
\eqref{eq:Lop-power}.

\emph{(b)} With $\eta=\Theta\sqrt{-2H_{00}}/\mu$ and $p=\Theta^2/\mu$, the coefficients of
$\mathcal V_0,\dots,\mathcal V_3$ are of the form of (a) with $k=0,1,2,3$, and the constant
parts $A_{k0}$ do not depend on $\theta$. Formula \eqref{eq:Lop-power} with
$(k-2)=-2,-1,0,1$ gives \eqref{eq:Lop-general-full}. The defects involving the
Poincar\'e pair have no first-order term by Proposition~\ref{prop:structure}(a).
In the polar-nodal chart, $\nu$ is cyclic for $K$, every term of
\eqref{eq:LC} with the pair $(\nu,N)$ contains a derivative with respect to
$\nu$, and the corresponding defects vanish to all orders.

\emph{(c)} ($\Leftarrow$) is immediate from \eqref{eq:Lop-general-full}.
($\Rightarrow$) Near a point with $R\ne0$, the functions
$(r,E,\psi,\Theta,\xi,\zeta)$ with $E=H_{00}$ are local coordinates, since
$\partial_RH_{00}=R\ne0$, and $\eta>0$ and $p$ depend only on $(E,\Theta)$. After division by
$k_2R\Theta$, the right-hand side of \eqref{eq:Lop-general-full} is a combination of
$r^{-3}$, $r^{-4}$ and $r^{-6}$ with coefficients independent of $r$ in these
coordinates, so each coefficient vanishes, which gives $\partial_\theta T_0=\partial_\theta T_1=\partial_\theta T_3=0$, and the linear independence
of $\cos2\theta$ and $\sin2\theta$ gives \eqref{eq:ansatz}.

\emph{(d)} On the zeroth-order solution $H_{00}=E_0$ and $\Theta=\Theta_0$, so that $\eta$ and
$p$ are constants, and multiplication of \eqref{eq:HJ1} by $r^2$ gives
\[
 r^2R_0\,W_r'-r^2E_1+\sum_{k=0}^3a_k\,r^{2-k}\bigl(A_{k0}+T_k(\theta)\bigr)+\Theta_0W_\theta'=0,
\]
with the nonzero constants $a_0=\eta^3p^{-3}$, $a_1=\eta p^{-2}$, $a_2=p^{-1}$ and $a_3=1$.
An additive solution exists if and only if the sum over $k$ is a function of $r$
plus a function of $\theta$, since $W_r'$ and $W_\theta'$ can then be read off, that is,
if and only if $\partial_r\partial_\theta$ of the sum, $\sum_k(2-k)\,a_k\,r^{1-k}\,\partial_\theta T_k$,
vanishes on $I_r\times I_\theta$. The functions $r$, $1$ and $r^{-2}$ are linearly independent
on $I_r$, so $\partial_\theta T_0=\partial_\theta T_1=\partial_\theta T_3=0$, and as in (c) this gives
\eqref{eq:ansatz} at $(\Theta_0,N)$; conversely, under \eqref{eq:ansatz} the
term $a_2T_2$ is a function of $\theta$ alone.
\end{proof}

\begin{remark}[Linearised defects in the modified chart]
\label{rem:degenerate}
In the modified chart $H_{00}$ does not depend on $(\xi,\zeta)$, so that
$\partial_\zeta H_{00}\equiv0$ and the regularity hypothesis of the Levi-Civita
theorem fails for the nodal pair at zeroth order; by
Proposition~\ref{prop:structure}(a), the linearised nodal defects of every
perturbation vanish. Vanishing linearised defects therefore do not imply
first-order separation in this chart. The member of \eqref{eq:general} with
$A_{22}=2\Theta^2(\Theta-N)$ ($\mu=1$) is, by Proposition~\ref{prop:chart}(c),
$\mathcal V_2=k_2V_2$ with
\[
 V_2=\frac{(\zeta^2-\xi^2)\cos2\psi-2\xi\zeta\sin2\psi}{r^2}=\frac{2(\Theta-N)\cos2\theta}{r^2},
\]
and its linearised defects vanish by (c). In an expansion
$W=W_0+k_2W_1$ in which $W_0$ and $W_1$ are sums of functions of $r$, $\psi$ and $\xi$, the
zeroth-order equation fixes $\partial_rW_0=R_0(r)$ and $\partial_\psi W_0=\Theta_0$ and leaves
$\zeta_0(\xi)=\partial_\xi W_0$ free, and the nodal part of $W_1$ does not enter the
first-order equation, since $\partial_\zeta H_{00}=0$. After multiplication by $r^2$,
the first-order equation requires $(\zeta_0^2-\xi^2)\cos2\psi-2\xi\zeta_0\sin2\psi$ to be
independent of $\xi$, so that $\zeta_0^2-\xi^2$ and $\xi\zeta_0$ are constant on an interval
of $\xi$. Then $(\xi^2+\zeta_0^2)^2=(\zeta_0^2-\xi^2)^2+4\xi^2\zeta_0^2$ and hence $\xi^2$ are constant
there, which is impossible, and an angle-independent term added for
\eqref{eq:efficiency} does not change this argument. In the polar-nodal chart
at fixed $N$ the same member separates by (d), with the angular integral
$\Theta^2+4k_2(\Theta-N)\cos2\theta$.
\end{remark}

The dependence of $\mathcal V_0$ and $\mathcal V_1$ on the radial velocity, which
enters through $\eta$, does not contribute to the first-order defect, since the
terms with $a_E$ cancel. The factor $k-2$ in \eqref{eq:Lop-power} vanishes for
$k=2$ alone, so that $r^{-2}$ is the only radial weight for which an arbitrary
angular dependence produces no first-order defect, the operator counterpart of
the classical observation that a potential $U_2(\theta,\Theta,N)\,r^{-2}$ only
modifies the centrifugal term of \eqref{eq:kepler} \citep{Deprit1981,Lara2021}.
For the zonal harmonics, $a$ is constant and $k=n+1$, and \eqref{eq:Lop-power}
gives the factor $n-1$ of \eqref{eq:zonal-defect-n} below.

\begin{table}[ht]
\centering
\caption{Phase-space transcriptions of the historical intermediaries into the
form \eqref{eq:general}, and the original potentials of Sterne and Garfinkel.
Column 2: nonzero coefficient functions, with the sign of $H_{10}$
(Remark~\ref{rem:erratum}); the originals are the potentials of position
\eqref{eq:originals}, with the physical latitude and the reference constants
$c_0$, $p_0$ and $\eta_0$. Column 3: efficiency condition \eqref{eq:efficiency},
the equality of the averages over the mean anomaly. Columns 4--6:
second-order coefficients of the defects of the modified chart, whose
first-order coefficients vanish (Proposition~\ref{prop:general-form} and
Proposition~\ref{prop:axisym}). An entry $0$ means identically zero by the
exact test of Appendix~\ref{app:zero}. Column 7: exact separation in the
polar-nodal chart at fixed $N$ (Proposition~\ref{prop:radial-angular}(d)).}
\label{tab:intermediaries}
\footnotesize\setlength{\tabcolsep}{3.5pt}
\begin{tabular}{@{}l>{\raggedright\arraybackslash}p{3.7cm}cccccc@{}}
\toprule
Intermediary & coefficients & eff. & $b_{r\psi}$ & $b_{r\xi}$ & $b_{\psi\xi}$ & polar-nodal\\
\midrule
\citet{Sterne1958} & $A_{22}=-3s^2$, $A_{30}=1-3c^2$ & yes & $\ne0$ & $\ne0$ & $\ne0$ & no\\
\citet{Garfinkel1958} & $A_{10}=3(3c^2-1)$, $A_{20}=-A_{10}$, $A_{22}=-3s^2$ & no & $\ne0$ & $\ne0$ & $\ne0$ & no\\
\citet{Brouwer1959} & $A_{00}=1-3c^2$ & yes & $0$ & $0$ & $0$ & yes\\
\citet{Aksnes1965} & $A_{20}=1-3c^2$, $A_{22}=-3s^2$ & yes & $0$ & $0$ & $\ne0$ & yes\\
\citet{CidLahulla1969} & $A_{30}=1-3c^2$ & yes & $0$ & $0$ & $0$ & yes\\
\citet{Deprit1981} & $A_{20}=1-3c^2$ & yes & $0$ & $0$ & $0$ & yes\\
\midrule
Sterne, original & eq.~\eqref{eq:originals}, $c_0$, $p_0$ & -- & $0$ & $0$ & $\ne0$ & yes\\
Garfinkel, original & eq.~\eqref{eq:originals}, $c_0$, $p_0$, $\eta_0$ & -- & $0$ & $0$ & $\ne0$ & yes\\
\bottomrule
\end{tabular}
\end{table}

\subsection{Intermediaries with radial and angular potentials}
\label{sec:radial-angular}

The intermediaries of the main problem modify the Kepler Hamiltonian by a
potential that depends on the radial distance and by a correction to the
centrifugal term that depends on the argument of latitude
\citep{Deprit1981}. The following proposition, introduced here,
gives the exact defects of this class in the modified chart and, in part (d),
in the polar-nodal chart.

\begin{proposition}[Radial and angular potentials]
\label{prop:radial-angular}
Let
\begin{equation}
 H=\frac12R^2-\frac{\mu}{r}+\frac{\Theta^2+2U_2}{2r^2}+U_1,
 \label{eq:radial-angular}
\end{equation}
with smooth functions $U_1(r,\Theta,\xi,\zeta)$ and $U_2(\psi,\Theta,\xi,\zeta)$, on a
domain $\{r\in I_r,\ R\in I_R,\ \psi\in I_\psi,\ \Theta\in I_\Theta,\ (\xi,\zeta)\in D\}$ with open
intervals $I_r$, $I_R$ and $I_\Theta$, an open interval or circle $I_\psi$ and an open connected
set $D$, and put $V=U_1+r^{-2}U_2$. Then the
following hold.
\begin{enumerate}[label=(\alph*)]
 \item For all values of $U_1$ and $U_2$, with $\partial_\Theta$ taken at fixed
 $(\xi,\zeta)$,
 \begin{align}
  b_{r\psi}[H]&=\frac{R}{r^4}\,\frac{\partial U_2}{\partial\psi}\,
  \frac{\partial}{\partial r}\Bigl(r^2\frac{\partial U_1}{\partial\Theta}\Bigr),
  \label{eq:radial-angular-defect}\\
  b_{r\xi}[H]&=R\Bigl(\frac{1}{r^4}\Bigl\{U_2,\frac{\partial(r^2U_1)}{\partial r}\Bigr\}_{\xi\zeta}
  +\Bigl\{U_1,\frac{\partial U_1}{\partial r}\Bigr\}_{\xi\zeta}\Bigr),
  \label{eq:radial-xi}\\
  b_{\psi\xi}[H]&=\frac{\Theta}{r^2}\,\{V,\partial_\psi V\}_{\xi\zeta}+b_{\psi\xi}[V].
  \label{eq:angular-xi}
 \end{align}
 \item On the open set where $\partial_\psi U_2\ne0$, the radial defect vanishes if and
 only if $\partial_r(r^2\partial_\Theta U_1)=0$ there, that is, if and only if
 $U_1=\phi_1(r,\xi,\zeta)+r^{-2}\phi_2(\Theta,\xi,\zeta)$ locally, with smooth functions
 $\phi_1$ and $\phi_2$. If $U_1$ and $U_2$ are real-analytic, $b_{r\psi}$ vanishes on the
 whole domain if and only if $U_2$ does not depend on $\psi$, or $U_1$ has this form
 on the whole domain.
 \item If $U_1$ and $U_2$ depend on $(\xi,\zeta)$ only through $\rho^2$, that is,
 on the node only through $N$, then $b_{r\xi}=b_{\psi\xi}=0$, and if, in addition,
 $U_2$ does not depend on $\psi$, all three defects vanish identically.
 \item In the polar-nodal chart, let $U_1=U_1(r,\Theta,N)$ and
 $U_2=U_2(\theta,\Theta,N)$ on a domain
 $\{r\in I_r,\ R\in I_R,\ \theta\in I_\theta,\ \Theta\in I_\Theta,\ N\in I_N\}$ with an open interval
 $I_N$, and take $\partial_\Theta$ at fixed $N$. Then the defects that involve the pair
 $(\nu,N)$ vanish identically, and
 \begin{equation}
  b_{r\theta}[H]=\frac{R}{r^4}\,\frac{\partial U_2}{\partial\theta}\,
  \frac{\partial}{\partial r}\Bigl(r^2\frac{\partial U_1}{\partial\Theta}\Bigr)\Big|_N .
  \label{eq:radial-angular-pn}
 \end{equation}
 On the open set where $\partial_\theta U_2\ne0$, $b_{r\theta}$ vanishes if and only if
 $U_1=\phi_1(r,N)+r^{-2}\phi_2(\Theta,N)$ locally, and $\Theta^2+2U_2+2\phi_2$ is then an
 integral of $H$; for real-analytic $U_1$ and $U_2$ the global statement of (b)
 holds with $(\xi,\zeta)$ replaced by $N$.
\end{enumerate}
\end{proposition}

\begin{proof}
\emph{(a)} Write $K=H-\frac12R^2=-\mu r^{-1}+\frac12Gr^{-2}+U_1$ with
$G=\Theta^2+2U_2$. Since $K$ does not depend on $R$, $H_R=R$ and
$H_{R\psi}=H_{R\Theta}=0$, so that the second and fourth terms of \eqref{eq:LC} for
the pairs $(r,R)$ and $(\psi,\Theta)$ vanish and
\[
 b_{r\psi}[H]=R\,\bigl(K_\psi\,\partial_rK_\Theta-K_\Theta\,\partial_rK_\psi\bigr)=R\,\{K,\partial_rK\}_{\psi\Theta},
\]
the form given in \citet[Proposition~4.2]{MartinusiPD}. By bilinearity of the bracket, using
$\{G,G\}_{\psi\Theta}=0$ and $\{U_1,\partial_rU_1\}_{\psi\Theta}=0$ because $U_1$ does not depend
on $\psi$,
\[
 \{K,\partial_rK\}_{\psi\Theta}
 =\frac{1}{2r^2}\{G,\partial_rU_1\}_{\psi\Theta}+\frac{1}{r^3}\{G,U_1\}_{\psi\Theta}
 =\frac{\partial_\psi U_2}{r^4}\,\partial_r\bigl(r^2\partial_\Theta U_1\bigr),
\]
since $\{G,F\}_{\psi\Theta}=2\,\partial_\psi U_2\,\partial_\Theta F$ for a function $F$ that does not
depend on $\psi$. This gives \eqref{eq:radial-angular-defect}. The same
reduction for the pairs $(r,R)$ and $(\xi,\zeta)$ gives
$b_{r\xi}[H]=R\,\{K,\partial_rK\}_{\xi\zeta}$, and expanding with $\{\Theta^2,\cdot\}_{\xi\zeta}=0$ and
$\{U_2,U_2\}_{\xi\zeta}=0$ leads to \eqref{eq:radial-xi}. Finally,
\eqref{eq:angular-xi} is \eqref{eq:bpsixi} with $k=1$, because $H=H_{00}+V$.

\emph{(b)} The factor $R\,r^{-4}$ of \eqref{eq:radial-angular-defect} vanishes
only at $R=0$, so on the open set where $\partial_\psi U_2\ne0$ the defect vanishes if
and only if $\partial_r(r^2\partial_\Theta U_1)=0$ at the corresponding values of
$(r,\Theta,\xi,\zeta)$, by continuity in $R$. Locally, $r^2\partial_\Theta U_1$ then
depends only on $(\Theta,\xi,\zeta)$, and integration in $\Theta$ gives the stated
form; conversely, this form makes the second factor vanish. For real-analytic
$U_1$ and $U_2$, suppose that $\partial_\psi U_2$ does not vanish identically. It is
then nonzero on a nonempty open set, on which the second factor vanishes, and
this factor is real-analytic on the connected set $I_r\times I_\Theta\times D$, so that it
vanishes identically by the identity theorem for real-analytic functions
\citep{KrantzParks2002}; integration over the interval $I_\Theta$ gives the global
form. Without analyticity the global statement fails: for $U_2=f(\Theta)\sin\psi$
and $U_1=r\,h(\Theta)$, with smooth functions $f$ and $h$ whose supports are
disjoint, the defect vanishes identically, although $U_2$ depends on $\psi$ and
$U_1$ does not have the stated form where $h'\ne0$.

\emph{(c)} The first statement is Proposition~\ref{prop:structure}(c) applied
to $V$, and the second follows from \eqref{eq:radial-angular-defect}.

\emph{(d)} In the polar-nodal chart $\nu$ is cyclic for $H$, so every term of
\eqref{eq:LC} with the pair $(\nu,N)$ contains a derivative with respect to
$\nu$ and vanishes. The computation of (a) for the pairs $(r,R)$ and
$(\theta,\Theta)$ uses only that $K$ does not depend on $R$, that $U_1$ does not
depend on the angle and that $\{G,F\}_{\theta\Theta}=2\,\partial_\theta U_2\,\partial_\Theta F$ for a
function $F$ independent of $\theta$, with the bracket taken at fixed $(\nu,N)$,
and it gives \eqref{eq:radial-angular-pn}. The proof of (b) applies with
$(\xi,\zeta)$ replaced by $N$. If $U_1=\phi_1(r,N)+r^{-2}\phi_2(\Theta,N)$, then
$H=\frac12R^2-\mu r^{-1}+\phi_1(r,N)+\frac12\tilde G\,r^{-2}$ with
$\tilde G=\Theta^2+2U_2+2\phi_2$, which depends only on $(\theta,\Theta,N)$, so that
$\{\tilde G,H\}=0$.
\end{proof}

Since $N=\Theta-\frac12\rho^2$, the derivative $\partial_\Theta$ at fixed $(\xi,\zeta)$ in
(b) equals $\partial_\Theta+\partial_N$ in polar-nodal variables, and the conditions of
(b) and (d) differ for potentials that depend on $N$: for $U_2=\sin2\theta$ and
$U_1=r\,N$ the polar-nodal defect \eqref{eq:radial-angular-pn} vanishes,
whereas in the modified chart $\partial_r(r^2\partial_\Theta U_1)=3r^2$. Part (d)
sharpens the St\"ackel-type form \eqref{eq:stackel2020}: with a term in the
argument of latitude, the radial potential may depend on $\Theta$ at fixed $N$
only through a term proportional to $r^{-2}$. The intermediaries of Brouwer,
Cid and Lahulla and Deprit, whose potentials depend only on the inclination
and on the radial variables, have identically vanishing defects in both
charts and separate exactly in the polar-nodal chart. Those with a term in the
argument of latitude can separate in the polar-nodal chart, with the integral
$\Theta^2+2U_2+2\phi_2$, whereas in the modified chart the angle
$\theta=\psi-\nu$ couples $(\psi,\Theta)$ with the Poincar\'e pair and
\eqref{eq:angular-xi} is in general of the order of $U_2^2$. Near the prograde
equatorial orbits, where $\partial_\zeta K$ is of the order of $k_2\rho$, the
vanishing of the modified-chart defects does not by itself give separation in
that chart (Remark~\ref{rem:degenerate}).

\subsection{Higher-order defects and the classical intermediaries}
\label{sec:general-defects}

In the modified chart the terms in $2\theta$ of \eqref{eq:general} are the smooth
functions of Proposition~\ref{prop:chart}(c) multiplied by $s^{-2}$, and for the
historical intermediaries $A_{21}$ and $A_{22}$ contain the factor $s^2$. Since $\eta$
and $p$ are functions of $H_{00}$ and $\Theta$, the potentials $\mathcal V_0$ and
$\mathcal V_1$ depend on the radial velocity, and the intermediary is not of the
form \eqref{eq:radial-angular} when $A_{00}$ or $A_{10}$ is nonzero. Writing
\begin{equation}
 K=H_{00}+k_2\Bigl(F_0(H_{00},\Theta,\rho^2)+\frac{F_1(H_{00},\Theta,\rho^2)}{r}
 +\frac{g(\psi,\Theta,\xi,\zeta)}{r^2}+\frac{a_3(\Theta,\rho^2)}{r^3}\Bigr),
 \label{eq:general-F}
\end{equation}
with $F_0=\eta^3p^{-3}A_{00}$, $F_1=\eta p^{-2}A_{10}$,
$g=p^{-1}(A_{20}+A_{21}\sin2\theta+A_{22}\cos2\theta)$ and $a_3=A_{30}$, the following
proposition, introduced here, gives all defects beyond first order.

\begin{proposition}[Defects of the general intermediary]
\label{prop:general}
For the Hamiltonian \eqref{eq:general-F} with smooth $F_0$, $F_1$, $g$ and $a_3$
the following hold.
\begin{enumerate}[label=(\alph*)]
 \item The three linearised defects vanish for all coefficients $A_{km}$.
 \item The radial defect is $b_{r\psi}[K]=k_2^2\,\beta_2+k_2^3\,b_{r\psi}[V]$, with
 $V=k_2^{-1}(K-H_{00})$ and
 \begin{equation}
  \beta_2=\frac{R}{r^6}\,\frac{\partial g}{\partial\psi}
  \Bigl(2r^3\frac{\partial F_0}{\partial\Theta}+r^2\frac{\partial F_1}{\partial\Theta}
  -\Theta\frac{\partial F_1}{\partial E}-\frac{\partial a_3}{\partial\Theta}\Bigr),
  \label{eq:beta2}
 \end{equation}
 where $E$ denotes the first argument of $F_0$ and $F_1$, the derivatives with
 respect to $\Theta$ are taken at fixed $E$ and $\rho^2$, and all functions are
 evaluated at $E=H_{00}$. The defects $b_{r\xi}$ and $b_{\psi\xi}$ are given by
 \eqref{eq:brxi} and \eqref{eq:bpsixi} with $kV=K-H_{00}$.
 \item If $A_{21}=A_{22}=0$, all defects vanish identically. On the open set where
 $\partial_\psi g\ne0$, the second-order radial defect $\beta_2$ vanishes if and only if
 $\partial_\Theta F_0$, $\partial_\Theta F_1$ and $\Theta\,\partial_EF_1+\partial_\Theta a_3$ vanish at the
 corresponding values of $(E,\Theta,\rho^2)$. If $F_0$, $F_1$, $g$ and $a_3$ are
 real-analytic on a connected domain and $\partial_\psi g\not\equiv0$, then $\beta_2$
 vanishes identically if and only if these three functions vanish identically.
\end{enumerate}
\end{proposition}

\begin{proof}
\emph{(a)} The general intermediary satisfies \eqref{eq:ansatz}, and
Proposition~\ref{prop:general-form}(c) applies.

\emph{(b)} By \eqref{eq:cubic}, $b_{r\psi}[K]$ is a cubic polynomial in $k_2$ with
constant term $b_{r\psi}[H_{00}]=0$ and cubic coefficient $b_{r\psi}[V]$, and its
linear term vanishes by (a). The second-order coefficient follows from
\eqref{eq:LC} by the chain rule through $H_{00}$, using $\partial_RH_{00}=R$,
$\partial_rH_{00}=\mu r^{-2}-\Theta^2r^{-3}$ and $\partial_\Theta H_{00}=\Theta r^{-2}$; it collects
the products of one derivative of $g$ with one derivative of $F_0$, $F_1$ or
$a_3$, which gives \eqref{eq:beta2}.

\emph{(c)} If $A_{21}=A_{22}=0$, then $K$ does not depend on $\psi$ and depends on
$(\xi,\zeta)$ only through $\rho^2$, so that every term of \eqref{eq:LC} for
$b_{r\psi}$ contains a vanishing derivative with respect to $\psi$, and
Proposition~\ref{prop:structure}(c) gives the other two defects. For the
second statement, ($\Leftarrow$) is immediate from \eqref{eq:beta2}.
($\Rightarrow$) Near a point with $R\ne0$ and $\partial_\psi g\ne0$, the functions
$(r,E,\psi,\Theta,\xi,\zeta)$ with $E=H_{00}$ are local coordinates, since
$\partial_RH_{00}=R\ne0$, and in these coordinates the bracket in \eqref{eq:beta2} is a
polynomial in $r$ of degree three whose coefficients depend on $(E,\Theta,\rho^2)$.
Its vanishing for all $r$ in an interval forces the coefficients of $r^3$, $r^2$
and $r^0$ to vanish, and the points with $R=0$ follow by continuity. In the
real-analytic case these coefficients vanish on a nonempty open set and
therefore identically, by the identity theorem \citep{KrantzParks2002}. The
smooth example in the proof of Proposition~\ref{prop:radial-angular}(b), here
with $g=f(\Theta)\sin\psi$, $F_0=h(\Theta)$ and $F_1=a_3=0$, is a counterexample to the
global statement without analyticity.
\end{proof}

For real-analytic coefficients of \eqref{eq:general} on a connected domain, the
global statement of (c) applies. With $\mu=1$,
$F_0=(-2E)^{3/2}\Theta^{-3}A_{00}$ and $F_1=(-2E)^{1/2}\Theta^{-3}A_{10}$, and since
$\rho^2=2(\Theta-N)$, their derivatives with respect to $\Theta$ at fixed $E$ and
$\rho^2$ vanish identically if and only if $A_{00}=\Theta^3h_0(\Theta-N)$ and $A_{10}=\Theta^3h_1(\Theta-N)$,
with functions $h_0$ and $h_1$ of one variable. The third condition of (c) then
reads $-\Theta(-2E)^{-1/2}h_1+\partial_\Theta A_{30}=0$ for all $E$ in an interval, which
forces $h_1=0$ and $A_{30}=h_3(\Theta-N)$. The members with a term in the argument
of latitude and $\beta_2\equiv0$ therefore have $A_{10}=0$ and $A_{00}$, $A_{30}$ of these
forms, and for coefficients that are polynomials in $c$, as in
Table~\ref{tab:intermediaries}, these forms leave $A_{00}=0$ and a constant
$A_{30}$. The bracket of \eqref{eq:beta2} equals $6c(1-c)\Theta^{-1}$ for the
transcription of Sterne, contains $A_{10}$ for that of Garfinkel and vanishes for
that of Aksnes, whose coupling with the Poincar\'e pair is
$b_{\psi\xi}=144\,k_2^2\,c\sin^2\varphi\,\Theta^{-4}r^{-6}+O(k_2^3)$ ($\mu=1$).

The transcriptions of \citet{Brouwer1959}, \citet{CidLahulla1969} and
\citet{Deprit1981} have identically vanishing defects, that of
\citet{Aksnes1965} separates in the polar-nodal chart and acquires a
second-order coupling with the Poincar\'e pair, and those of \citet{Sterne1958}
and \citet{Garfinkel1958}, which combine a term in $\cos2\theta$ with an
inclination-dependent $A_{30}$ or $A_{10}$, acquire second-order defects in both
charts. These verdicts refer to the transcriptions, in which $c$, $s$, $p$ and
$\eta$ are functions on phase space. The originals are potentials of position:
\citet[eq.~(4)]{Sterne1958} and \citet[eqs.~(5)--(6)]{Garfinkel1958} write them
in spherical coordinates with the physical latitude $\varphi$, and only the
elements that fix their coefficients are reference constants, ``mere
constants'' in the words of \citet[p.~30]{Sterne1958}, whereas
\citet{Aksnes1965} uses a dynamical coefficient, a distinction drawn by
\citet[Sect.~II]{AksnesJPL1970}. With $k_2=\frac14J_2$ and $\mu=R_e=1$, they read
\begin{equation}
 \begin{aligned}
 V_{\mathrm S}&=k_2\Bigl[\frac{1-3c_0^2}{r^3}+\frac{6\sin^2\varphi-3s_0^2}{p_0r^2}\Bigr],\\
 V_{\mathrm G}&=k_2\Bigl[\frac{3(3c_0^2-1)\,\eta_0}{p_0^2\,r}+\frac{6(\sin^2\varphi-c_0^2)}{p_0r^2}\Bigr],
 \end{aligned}
 \label{eq:originals}
\end{equation}
with the constants $c_0$, $s_0$, $p_0$ and $\eta_0$ of a reference orbit, and they
reduce to the transcriptions when these constants are replaced by the
functions $c$, $s$, $p$ and $\eta$ and $\sin^2\varphi$ by
$s^2\sin^2\theta=\frac12s^2(1-\cos2\theta)$. Both are
axisymmetric potentials $F(r,\sin\varphi)$ with the latitude term
$6k_2p_0^{-1}r^{-2}\sin^2\varphi$. Since $r^2\partial_\psi F$ does not depend on $r$,
Proposition~\ref{prop:axisym} below gives $b_{r\psi}=b_{r\xi}=0$ identically
and, by \eqref{eq:axisym-xi},
\begin{equation}
 b_{\psi\xi}=\frac{144\,k_2^2\,c\,\sin^2\varphi}{p_0^2\,r^6}+O(k_2^3),
 \label{eq:originals-xi}
\end{equation}
the coupling of Aksnes's transcription with $\Theta^4$ replaced by $p_0^2$. In the
polar-nodal chart at fixed $N$, $\sin^2\varphi=(1-N^2\Theta^{-2})\sin^2\theta$ makes the
latitude term of the form $r^{-2}U_2(\theta,\Theta,N)$ with a radial part that
depends on $r$ alone, and Proposition~\ref{prop:radial-angular}(d) gives exact
separation, the separation of \eqref{eq:originals} in spherical coordinates.
The coupling with the Poincar\'e pair is therefore produced by the dependence
on the physical latitude, which the zonal term, the transcription of Aksnes
and the originals of Sterne and Garfinkel share, whereas the intermediaries of
Brouwer, Cid and Lahulla and Deprit, which depend on the inclination alone,
have no coupling.

\section{Zonal perturbations and the first-order Lie generator}
\label{sec:generator}

The defects of Section~\ref{sec:intermediaries} are evaluated on the
intermediaries; this section evaluates the same radial defect on the zonal
problem itself, first as a function on phase space, then along propagated
orbits, and finally as the response of the product structure to the
first-order Lie generator.

\subsection{Axisymmetric potentials}
\label{sec:axisym}

Every potential of an axisymmetric planet, in particular the zonal sum
\[
 \sum_{n\ge2}\mu J_nR_e^n\,r^{-(n+1)}P_n(\sin\varphi)
\]
with the Legendre polynomials
$P_n$ \citep{Kozai1959,Lara2021}, is a function $F(r,\varsigma)$ of the radial
distance and of the sine of the latitude $\varsigma=\sin\varphi$, which is the single
smooth function \eqref{eq:sinphi} of $(\psi,\Theta,\xi,\zeta)$, and this structure
determines all defects exactly. Under $H_{00}+F$, $\dot\Theta=-\partial_\psi F=r\,f_T$, with
$f_T$ the transverse component of the perturbing acceleration $-\nabla F$ in the
orbital plane, which is the torque equation of the Gauss form of the
perturbation equations \citep{MurrayDermott2000,MontenbruckGill2000}, and the
latitudinal component is $f_\varphi=-r^{-1}\partial_\varphi F=-r^{-1}\cos\varphi\,\partial_\varsigma F$. The
following proposition is introduced here.

\begin{proposition}[Axisymmetric potentials]
\label{prop:axisym}
Let $H=H_{00}+F(r,\varsigma)$ with a smooth function $F$, $\varsigma=\sin\varphi$ given by
\eqref{eq:sinphi}, and denote by $F_\varsigma$ the partial derivative with respect to
$\varsigma$. Then the following hold exactly in the modified polar-nodal chart,
with no smallness assumption on $F$.
\begin{enumerate}[label=(\alph*)]
 \item The radial defect is linear in $F$,
 \begin{equation}
  b_{r\psi}[H]=\Lop_{H_{00}}(F)_{r\psi}=-\frac{R\,\Theta}{r^4}\,\partial_r\bigl(r^2\partial_\psi F\bigr)
  =\frac{R\,\Theta}{r^4}\,\partial_r\bigl(r^3f_T\bigr),
  \label{eq:axisym-r}
 \end{equation}
 where $f_T=-r^{-1}\partial_\psi F$ is regarded as a function of $(r,\psi,\Theta,\xi,\zeta)$.
 \item The defect $b_{r\xi}[H]$ vanishes identically.
 \item The coupling with the Poincar\'e pair is
 \begin{equation}
  b_{\psi\xi}[H]=\frac{c}{r^2}\,F_\varsigma^2+F_\varsigma^3\,b_{\psi\xi}[\varsigma],\qquad
  b_{\psi\xi}[\varsigma]=-\frac{\sqrt{1-c}\,\bigl(c^2-c\cos^2\theta+\cos^2\theta\bigr)\sin\theta}
  {\Theta^2\sqrt{1+c}},
  \label{eq:axisym-xi}
 \end{equation}
 with $c=1-\rho^2/(2\Theta)$ and $\theta=\psi-\nu$; away from the poles,
 $F_\varsigma=-r f_\varphi/\cos\varphi$.
 \item For a sum $F=\sum_kF_k$ of such potentials, $b_{r\psi}$ and $b_{r\xi}$ are additive,
 and the interaction defect \eqref{eq:interaction} of two terms is
 \begin{equation}
  I_{\psi\xi}[F_1,F_2]=\frac{2c}{r^2}\,\partial_\varsigma F_1\,\partial_\varsigma F_2
  +3\,\partial_\varsigma F_1\,\partial_\varsigma F_2\,\bigl(\partial_\varsigma F_1+\partial_\varsigma F_2\bigr)\,b_{\psi\xi}[\varsigma].
  \label{eq:axisym-interaction}
 \end{equation}
 \item In the polar-nodal chart all defects of $H$ are linear in $F$, and
 $b_{r\nu}=b_{\theta\nu}=0$.
\end{enumerate}
\end{proposition}

\begin{proof}
\emph{(a)} Put $K=H-\frac12R^2=-\mu r^{-1}+\frac12\Theta^2r^{-2}+F$. Since $K$ does not
depend on $R$, $H_R=R$ and $H_{R\psi}=H_{R\Theta}=0$, and \eqref{eq:LC} for the pairs
$(r,R)$ and $(\psi,\Theta)$ reduces to $b_{r\psi}[H]=R\,\{K,\partial_rK\}_{\psi\Theta}$, as in the
proof of Proposition~\ref{prop:radial-angular}(a). With
$K_\psi=F_\varsigma\varsigma_\psi$, $K_\Theta=\Theta r^{-2}+F_\varsigma\varsigma_\Theta$, $\partial_rK_\psi=F_{r\varsigma}\varsigma_\psi$ and
$\partial_rK_\Theta=-2\Theta r^{-3}+F_{r\varsigma}\varsigma_\Theta$, the terms quadratic in $F$ cancel,
\[
 \{K,\partial_rK\}_{\psi\Theta}=F_\varsigma\varsigma_\psi\bigl(-2\Theta r^{-3}+F_{r\varsigma}\varsigma_\Theta\bigr)
 -\bigl(\Theta r^{-2}+F_\varsigma\varsigma_\Theta\bigr)F_{r\varsigma}\varsigma_\psi=-\frac{\Theta\,\varsigma_\psi}{r^4}\,\partial_r\bigl(r^2F_\varsigma\bigr),
\]
which is \eqref{eq:axisym-r}, since $\partial_\psi F=F_\varsigma\varsigma_\psi$ and $\varsigma$ does not depend on $r$.

\emph{(b)} The same reduction gives $b_{r\xi}[H]=R\{K,\partial_rK\}_{\xi\zeta}$, and
$\{K,\partial_rK\}_{\xi\zeta}=F_\varsigma F_{r\varsigma}(\varsigma_\xi \varsigma_\zeta-\varsigma_\zeta \varsigma_\xi)=0$.

\emph{(c)} By Proposition~\ref{prop:structure}(a) with $k=1$,
$b_{\psi\xi}[H]=\Theta r^{-2}\{F,\partial_\psi F\}_{\xi\zeta}+b_{\psi\xi}[F]$. The first term is
$\Theta r^{-2}F_\varsigma^2\{\varsigma,\varsigma_\psi\}_{\xi\zeta}$, because $\{\varsigma,\varsigma\}_{\xi\zeta}=0$. In $b_{\psi\xi}[F]$, every
second derivative is $F_{\varsigma\varsigma}\varsigma_x\varsigma_y+F_\varsigma\varsigma_{xy}$, and the four terms with $F_{\varsigma\varsigma}$
cancel in pairs, which leaves $F_\varsigma^3b_{\psi\xi}[\varsigma]$. The bracket $\{\varsigma,\varsigma_\psi\}_{\xi\zeta}=c/\Theta$
and the closed form of $b_{\psi\xi}[\varsigma]$ follow from \eqref{eq:sinphi} by direct
computation, and the relation
between $F_\varsigma$ and $f_\varphi$ is $\partial_\varphi=\cos\varphi\,\partial_\varsigma$.

\emph{(d)} Additivity follows from (a) and (b), and \eqref{eq:axisym-interaction}
follows from (c) by expanding $(\partial_\varsigma F_1+\partial_\varsigma F_2)^2$ and $(\partial_\varsigma F_1+\partial_\varsigma F_2)^3$.

\emph{(e)} The node $\nu$ is cyclic, which gives the last two defects as in the
proof of Proposition~\ref{prop:general-form}(b), and the computation of (a)
with $\theta$ in place of $\psi$ and $N$ fixed gives a defect linear in $F$.
\end{proof}

For the zonal harmonics, $\partial_\psi\sin\varphi=s\cos\theta$ at fixed $(\xi,\zeta)$,
with $s$ the sine of the inclination. In satellite theory the harmonics are
added to the Kepler Hamiltonian in order of increasing degree
\citep{Brouwer1959,Kozai1959}, and the zonal problem truncated at degree
$n_{\max}\ge2$ has the Hamiltonian
\begin{equation}
	H_{\mathrm Z}=H_{00}+\sum_{n=2}^{n_{\max}}\frac{\mu J_nR_e^n}{r^{n+1}}\,P_n(\sin\varphi).
	\label{eq:zonal-HZ}
\end{equation}
Since the radial defect is linear in $F$ by Proposition~\ref{prop:axisym}(a),
each harmonic contributes a separate term, and \eqref{eq:axisym-r} gives
\begin{align}
	b_{r\psi}[H_{\mathrm Z}]
	&=\mu\,\frac{R\,\Theta}{r^4}\,s\cos\theta\sum_{n=2}^{n_{\max}}(n-1)\,J_n
	\Bigl(\frac{R_e}{r}\Bigr)^{n}P_n'(\sin\varphi)
	\label{eq:zonal-defect-n}\\
	&=-\frac{R\,\Theta}{r^2}\sum_{n=2}^{n_{\max}}(n-1)\,f_{T,n},
	\label{eq:torque-form}
\end{align}
with $f_{T,n}$ the transverse acceleration of the harmonic of degree $n$. The
contribution of each harmonic does not depend on the truncation, and the
radial defect is the product of the radial velocity, the angular momentum and
a transverse force in which the harmonic of degree $n$ has the weight $n-1$.
For the main problem, $n_{\max}=2$, and with $k_2=\frac14\mu J_2R_e^2$
\begin{equation}
	\Lop_{H_{00}}(H_{10})_{r\psi}=\frac{6k_2\,R\,\Theta\,s^2\sin2\theta}{r^6},
	\label{eq:zonal-defect}
\end{equation}
which coincides with the polar-nodal value.

\begin{figure}[ht]
\centering
\includegraphics[width=\textwidth]{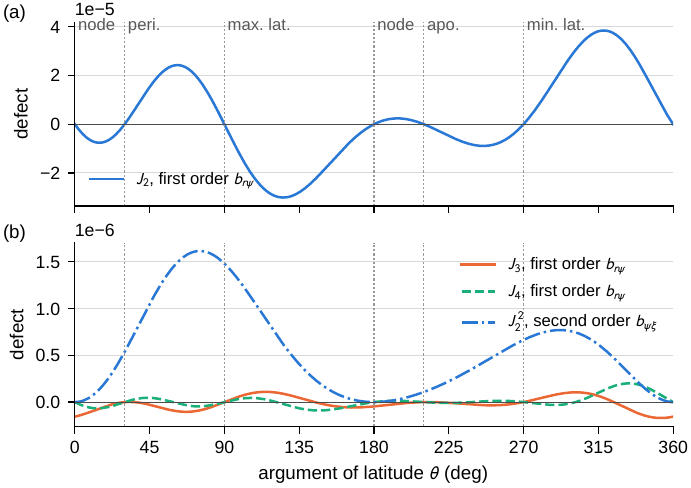}
\caption{Defects of the zonal terms of the Earth along a Kepler orbit with
$a=1.2\,R_e$, $e=0.1$, $i=50^\circ$, argument of pericentre $30^\circ$ and node
$0.4$~rad, in units $\mu=R_e=1$. (a) Radial defect \eqref{eq:zonal-defect} of
$J_2$. (b) Contributions of $J_3$ and $J_4$ to the radial defect \eqref{eq:zonal-defect-n}, and coupling $b_{\psi\xi}$ of $J_2$ from \eqref{eq:axisym-xi}. Dotted vertical lines
mark the nodes, the points of extreme latitude and the apsides}
\label{fig:defects}
\end{figure}

Although the operator is cubic, the radial defect of every axisymmetric
potential is exactly linear, and the harmonics interact only in the coupling
with the Poincar\'e pair, through the total latitudinal derivative
$\partial_\varsigma F=\sum_k\partial_\varsigma F_k$; in the polar-nodal chart they do not interact at
all. The coupling vanishes to second order where the latitudinal force
vanishes, which for $J_2$ alone is the equatorial plane and is moved off the
equator by the odd harmonics (Sections~\ref{sec:propagated} and
\ref{sec:vinti}). For the intermediaries with a term in the argument of
latitude, the coefficient $\beta_2$ of Proposition~\ref{prop:general}(b) is the
corresponding interaction between $g$ and $F_0$, $F_1$, $a_3$, in the sense of
\eqref{eq:interaction}. Along the Kepler orbit of
Figure~\ref{fig:defects}, the first-order defect of $J_2$ vanishes at the
apsides, at the nodes and at the points of extreme latitude, and the
second-order coupling of $J_2$ exceeds the first-order defects of $J_3$ and
$J_4$, which are of the same nominal order $J_2^2$, so that a consistent
second-order treatment has to include all three (Section~\ref{sec:second}).

\subsection{Radial defect along numerically integrated orbits}
\label{sec:propagated}

\begin{figure}[htbp]
\centering
\includegraphics[width=\textwidth]{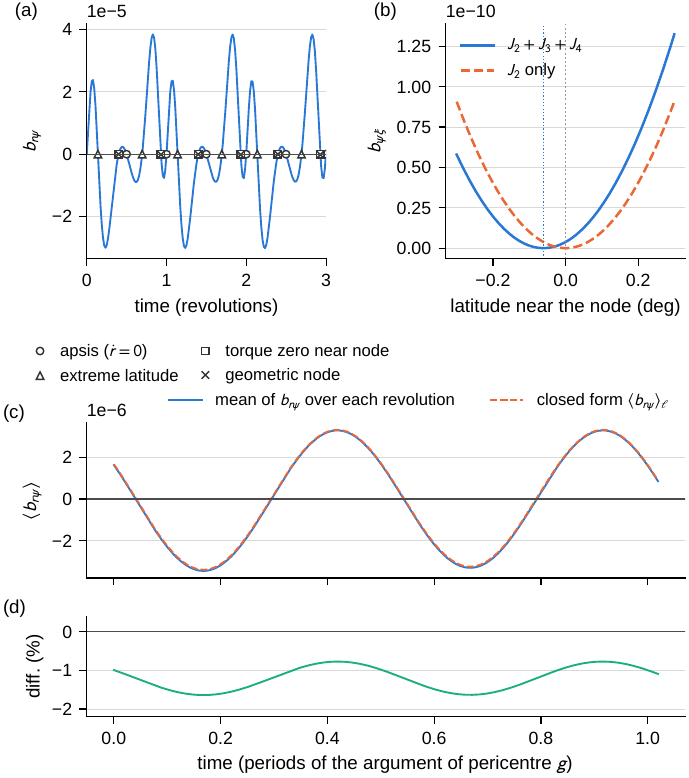}
\caption{Defects along an orbit of the Earth zonal problem with $J_2$, $J_3$
and $J_4$, $a=1.2\,R_e$, $e=0.1$, $i=50^\circ$, initial argument of pericentre
$30^\circ$, $\mu=R_e=1$. (a) Radial defect \eqref{eq:torque-form} over the first
three revolutions, with its zeros located on the dense solution: apsides,
points of extreme latitude and zeros of the weighted torque near the nodes;
the geometric nodes are marked at the value of the defect there. (b) Coupling
$b_{\psi\xi}$ of \eqref{eq:axisym-xi} near the first ascending node, for the
three harmonics and for $J_2$ alone; the dotted lines mark the zeros of the
latitudinal force. (c) Mean of $b_{r\psi}$
over each revolution, between consecutive pericentre passages, and the
closed-form average over the mean anomaly \eqref{eq:brpsi-avg} with the $J_3$ and
$J_4$ terms, at the elements averaged over the same revolution, over one period
of the argument of pericentre. (d) Difference between the two curves of (c), in percent of the amplitude of
the closed-form average; its second-order part is given by
Proposition~\ref{prop:second-mean}}
\label{fig:propagated}
\end{figure}

Since the radial defect of the zonal problem is exactly linear, its value
along an orbit of the full problem is the scalar \eqref{eq:torque-form},
computed from the position and velocity alone. Figure~\ref{fig:propagated}
shows $b_{r\psi}$ and $b_{\psi\xi}$, evaluated with \eqref{eq:torque-form} and
\eqref{eq:axisym-xi}, along an orbit integrated under the $J_2$, $J_3$ and $J_4$
terms of the Earth \citep{Pavlis2012} over one period of the argument of
pericentre; Appendix~\ref{app:orbit} describes the procedure.

The average of the radial defect over the mean anomaly at fixed Delaunay
elements follows by integration over the true anomaly with
$\dd\ell=n\,r^2\Theta^{-1}\dd f$. For the $J_2$ term, with $\mu=R_e=1$\footnote{Here and in the rest of the
	paper, units with $\mu=R_e=1$ are used, in which lengths are measured in
	planetary radii and times in units of $(R_e^3\mu^{-1})^{1/2}$; in the closed-form
	expressions $\mu$ and $R_e$ are recovered by dimensional analysis.},
\begin{equation}
 \langle b_{r\psi}\rangle_\ell=\frac{3J_2}{4}\,
 \frac{(G^2-H^2)(G^2-L^2)(G^2-3L^2)}{G^{11}L^7}\cos2g
 =\frac{3J_2}{4}\,\frac{s^2e^2(2+e^2)}{G^9L^3}\cos2g,
 \label{eq:brpsi-avg}
\end{equation}
and the $J_3$ and $J_4$ terms add harmonics in $g$ and $3g$, and in $2g$ and $4g$,
respectively, with no constant part, and the coefficients of $\sin g$ and of
$\cos2g$ in these terms contain the inclination factors $G^2-5H^2$ and
$G^2-7H^2$ of the long-period terms \eqref{eq:J3avg} and \eqref{eq:J4avg}. At
first order in the zonal coefficients, the mean of the radial defect over the
fast angle is therefore a long-period function of the argument of pericentre
with the harmonics of the long-period terms of Section~\ref{sec:second}, and it
vanishes for circular and for equatorial orbits. Along the propagated orbit,
the mean of $b_{r\psi}$ over each revolution follows the sum of the closed-form
averages at the elements averaged over the same revolution up to a
difference of less than $2\%$ of its amplitude, which consists of a constant
and a harmonic in $2g$ (Figure~\ref{fig:propagated}(c) and (d) and
Table~\ref{tab:orbit-diag}). This difference is the second-order part of the
mean defect, which \eqref{eq:brpsi-avg} does not contain, and
Section~\ref{sec:second-mean} gives it in closed form.

The fast oscillation of the defect carries the event structure: it vanishes
at the apsides and at the zeros of the weighted torque $\sum_n(n-1)f_{T,n}$ of
\eqref{eq:torque-form}, which for $J_2$ alone are the nodes and the points of
extreme latitude. With an odd harmonic the zero near a node leaves the
equatorial plane: near the equator the weighted torque is proportional to
$3J_2\varsigma r^{-2}-3J_3r^{-3}+O(\varsigma J_4)$, which vanishes at the height
$z=r\varsigma\approx R_eJ_3J_2^{-1}$, $14.9$~km south of the equator for the Earth
($14.8$~km along the orbit of Figure~\ref{fig:propagated}, where the term in $J_4$
contributes, Table~\ref{tab:orbit-diag}), and the geometric node becomes the zero of the degree-two term
\eqref{eq:zonal-defect} evaluated on the full motion. The coupling $b_{\psi\xi}$
vanishes to second order where the latitudinal force vanishes
(Figure~\ref{fig:propagated}(b)), and the terms linear in $\varsigma$ of
\[
F_\varsigma=\frac{3J_2\,\varsigma}{r^3}+\frac{3J_3\,(5\varsigma^2-1)}{2r^4}
+\frac{5J_4\,(7\varsigma^3-3\varsigma)}{2r^5}\qquad(\mu=R_e=1)
\]
place this zero at $\varsigma\approx\frac12J_3J_2^{-1}r^{-1}$, at the height
$z\approx\frac12R_eJ_3J_2^{-1}$, about $7.4$~km south of the equatorial plane, which is
the centre of the shifted oblate structure of Proposition~\ref{prop:shifted}
(Table~\ref{tab:orbit-diag}). The operator thus distinguishes three nearby
planes: the equatorial plane of the geometric node, the force equator at
$\frac12R_eJ_3J_2^{-1}$ and the zero of the weighted torque at $R_eJ_3J_2^{-1}$.

\subsection{First-order generator}

In the Lie-transform approach of \citet{Deprit1969}, the first-order generator
$W_1$ satisfies the homological equation
\begin{equation}
 \{W_1,H_{00}\}=H_{10}-H_{01},\qquad H_{01}=\mathcal V_0+\mathcal V_1+\mathcal V_2+\mathcal V_3,
 \label{eq:homological}
\end{equation}
which in Delaunay variables reads $n\,\partial W_1/\partial\ell=H_{10}-H_{01}$, with
$n=\mu^2L^{-3}$. A solution periodic in $\ell$ exists if and only if the average
of $H_{10}-H_{01}$ over $\ell$ vanishes \citep{Deprit1969,Lara2021}, which is
Deprit's efficiency condition \eqref{eq:efficiency}.

\begin{proposition}[First-order generator]
\label{prop:W1}
Under \eqref{eq:efficiency}, with $\mu=1$ and $E$ the eccentric anomaly, the
function
\begin{multline}
 W_1=\frac{k_2}{p\,\Theta}\Bigl[(A_{00}+A_{10})(f-\ell)-A_{10}\,e\sin E
 +(1-3c^2-A_{30})\,\sigma
 -2s^2\Bigl(\kappa+\frac34\Bigr)\sin2\theta\\
 +s^2\sigma\cos2\theta+\frac12\bigl(A_{21}\cos2\theta-A_{22}\sin2\theta\bigr)\Bigr]
 \label{eq:W1}
\end{multline}
solves \eqref{eq:homological} and is periodic in the mean anomaly.
\end{proposition}

\begin{proof}
Along a Kepler orbit with fixed elements, $\dd\ell=n\,r^2\Theta^{-1}\dd f$, and
with $\mu=1$, $n=\eta^3p^{-3/2}$ and $\Theta=p^{1/2}$, the solution of
$n\,\partial_\ell W_1=Q$ is $W_1=\Theta^{-1}\int Q\,r^2\,\dd f$ up to a function of
the elements. The integrals of the terms of $Q=H_{10}-H_{01}$ follow from
$r=p(1+e\cos f)^{-1}$, $\theta=g+f$, $\int r\,\dd f=pE\eta^{-1}$ and
$\int r^2\,\dd f=\Theta\ell n^{-1}$ \citep{Brouwer1959,Lara2021}. The secular
terms combine into $(A_{00}+A_{10})(f-\ell)-A_{10}(E-\ell)$ under
\eqref{eq:efficiency}, and $E-\ell=e\sin E$ by Kepler's equation. The
trigonometric terms give the remaining entries of \eqref{eq:W1}, with
$\frac12e[\sin(2\theta-f)+\frac13\sin(2\theta+f)]=
\frac23\kappa\sin2\theta-\frac13\sigma\cos2\theta$.
\end{proof}

Up to the normalisation and the misprints of Remark~\ref{rem:erratum},
\eqref{eq:W1} is the generator of \citet[eq.~(26)]{Martinusi2020}.

\subsection{Second-order mean of the radial defect}
\label{sec:second-mean}

The difference between the two curves of Figure~\ref{fig:propagated}(c) is
produced by the $J_2$ term alone: the integration of the same initial state
with $J_2$ only reproduces its constant and its harmonic in $2g$, and halving
$J_2$ halves both parts relative to the amplitude of \eqref{eq:brpsi-avg}
(Table~\ref{tab:second-check}). The difference is therefore of second order in
$J_2$, and the generator of Proposition~\ref{prop:W1} gives it in closed form. In this subsection
$\mu=R_e=1$, the Hamiltonian is $H_{00}+J_2V$, with $V=r^{-3}P_2(\sin\varphi)$ the
term of degree two in \eqref{eq:zonal-HZ}, and, by
Proposition~\ref{prop:axisym}(a) and \eqref{eq:zonal-defect}, the radial defect
is exactly $b_{r\psi}=J_2b_1$, with
\begin{equation}
 b_1=\Lop_{H_{00}}(V)_{r\psi}=\frac{3R\,\Theta\,s^2\sin2\theta}{2r^6}.
 \label{eq:b1-J2}
\end{equation}
In the Delaunay variables $(\ell,g,h,L,G,H)$, with $G=\Theta$ and $H=N$, the
generator \eqref{eq:W1} with $A_{00}=1-3c^2$ and all other coefficients zero,
which corresponds to the normalisation $H_{01}=\langle H_{10}\rangle_\ell$ of
\citet{Brouwer1959}, is $W_1=J_2w$, with
\begin{equation}
 w=\frac{1}{4G^3}\Bigl[(1-3c^2)(f-\ell+e\sin f)
 -\frac32\,s^2\Bigl(\sin2\theta+e\sin(f+2g)+\frac e3\sin(3f+2g)\Bigr)\Bigr].
 \label{eq:W1-J2}
\end{equation}
To first order in $J_2$, the osculating elements $y$ follow from the mean
elements $y'$ by the flow of $w$ over the time $J_2$, so that
$F(y)=F(y')+J_2\{F,w\}(y')+O(J_2^2)$ for every smooth function $F$, and the
transformed Hamiltonian is $H_{00}+J_2\langle V\rangle_\ell+O(J_2^2)$. Over a
revolution, the mean momenta $L'$, $G'$ and $H'$ therefore change by $O(J_2^2)$,
and the mean anomaly $\ell'$ and the mean argument of pericentre $g'$ advance at
rates that are constant up to $O(J_2^2)$, the latter at
\begin{equation}
 \dot g=\frac34\,J_2\,n\,\frac{5c^2-1}{G^4}+O(J_2^2)
 \label{eq:gdot}
\end{equation}
\citep{Brouwer1959,Kozai1959}. A revolution is the interval between two
consecutive passages through the osculating pericentre; over it, $\bar b$ is the
time mean of $b_{r\psi}$, $\bar L$, $\bar G$ and $\bar H$ are the time means of the
osculating momenta, and $\bar g$ is the polar angle of the time mean of
$(\cos g,\sin g)$. Finally, $\tilde b_1=b_1-\langle b_1\rangle_\ell$ is the
short-period part of $b_1$, and $\mathcal Z$ is the primitive of
$\partial_g\tilde b_1$ in $\ell$ with zero mean over $\ell$, which is periodic
because $\partial_g\tilde b_1$ has zero mean.

\begin{proposition}[Second-order mean of the radial defect]
\label{prop:second-mean}
For the $J_2$ problem, consider non-equatorial orbits, $s>0$, on which the
Delaunay chart is defined, with eccentricity $0<e<1$ fixed as $J_2\to0$, and
revolutions delimited by transverse passages through the osculating
pericentre, that is, simple zeros of $R$ with $\dot R>0$. Let
$\langle b_{r\psi}\rangle_\ell$ be the first-order average \eqref{eq:brpsi-avg}. With all
functions on the right-hand sides evaluated at $(\bar g,\bar L,\bar G,\bar H)$, the
following hold.
\begin{enumerate}[label=(\alph*)]
 \item The mean over a revolution is given by
 \begin{equation}
  \bar b-\langle b_{r\psi}\rangle_\ell=J_2^2\Bigl[\bigl\langle\{\tilde b_1,w\}\bigr\rangle_\ell
  +\frac34\,\frac{5c^2-1}{G^4}\,\mathcal Z\big|_{\ell=0}\Bigr]+O(J_2^3),
  \label{eq:second-mean}
 \end{equation}
 and the first term in the brackets does not change when a function of
 $(g,L,G,H)$ is added to $w$.
 \item In closed form,
 \begin{equation}
  \bar b-\langle b_{r\psi}\rangle_\ell=\frac{3J_2^2s^2}{G^{13}L^3}
  \bigl(\Gamma_0+\Gamma_2\cos2g+\Gamma_4\cos4g\bigr)+O(J_2^3),
  \label{eq:second-mean-cf}
 \end{equation}
 where $\Gamma_2=\Gamma_2^{\mathrm b}+\Gamma_2^{\mathrm d}$, the part
 $\Gamma_2^{\mathrm d}$ is the contribution of the second term of
 \eqref{eq:second-mean}, and
 \begin{equation}
 \begin{aligned}
  \Gamma_0={}&-\frac{1}{64}\bigl[c^2(5\eta^6+44\eta^5-162\eta^4-156\eta^3+525\eta^2-264)\\
  &\qquad{}-3\eta^6-28\eta^5+108\eta^4+108\eta^3-375\eta^2+198\bigr],\\
  \Gamma_2^{\mathrm b}={}&\frac{1-\eta}{8960(1+\eta)}\bigl[c^2(1995\eta^6-7290\eta^5-61235\eta^4-76540\eta^3\\
  &\qquad{}+73845\eta^2+181110\eta+82995)-595\eta^6+2634\eta^5+20283\eta^4\\
  &\qquad{}+24156\eta^3-28461\eta^2-65958\eta-30459\bigr],\\
  \Gamma_2^{\mathrm d}={}&\frac{5c^2-1}{280(1+\eta)}\bigl[3\eta^6-67\eta^5-109\eta^4+101\eta^3+136\eta^2-32\eta-32\\
  &\qquad{}-2e(1+\eta)(9\eta^4-60\eta^2+16)\bigr],\\
  \Gamma_4={}&-\frac{s^2(1-\eta)^2}{128(1+\eta)}\,(4\eta^5-12\eta^4-75\eta^3-9\eta^2+99\eta+33).
 \end{aligned}
 \label{eq:Gammas}
 \end{equation}
 \item Relative to the amplitude of \eqref{eq:brpsi-avg}, each $\Gamma_k$ enters
 \eqref{eq:second-mean-cf} with the weight
 \begin{equation}
  \frac{4J_2}{G^4e^2(2+e^2)}\,\Gamma_k,
  \label{eq:second-mean-rel}
 \end{equation}
 and, as $e\to0$ at fixed $G$ and $c$,
 \begin{equation}
 \begin{gathered}
  \Gamma_0=-\frac{s^2}{8}+O(e^2),\qquad
  \Gamma_2^{\mathrm d}=(5c^2-1)\Bigl(\frac e4+\frac{3e^2}{16}\Bigr)+O(e^3),\\
  \Gamma_2^{\mathrm b}=\frac{(87c^2-35)\,e^2}{16}+O(e^4),\qquad
  \Gamma_4=O(e^4),
 \end{gathered}
  \label{eq:Gammas-small-e}
 \end{equation}
 so that, relative to that amplitude, the constant is
 $-J_2s^2(4e^2G^4)^{-1}(1+O(e^2))$ and the coefficient of $\cos2g$ is
 \begin{equation}
  \frac{(5c^2-1)J_2}{2eG^4}+\frac{(51c^2-19)J_2}{4G^4}+O\Bigl(\frac{eJ_2}{G^4}\Bigr),
  \label{eq:second-mean-small-e}
 \end{equation}
 whose bracket part, from $\Gamma_2^{\mathrm b}$, is $(87c^2-35)J_2(8G^4)^{-1}+O(e^2J_2G^{-4})$.
\end{enumerate}
\end{proposition}

\begin{proof}
\emph{(a)} Along the orbit,
$b_{r\psi}=J_2b_1(y)=J_2b_1(y')+J_2^2\{b_1,w\}(y')+O(J_2^3)$, and the time mean of
the second term over a revolution is $J_2^2\langle\{b_1,w\}\rangle_\ell+O(J_2^3)$.
In the first term, $b_1=\langle b_1\rangle_\ell+\tilde b_1$. The average
$\langle b_1\rangle_\ell$ is a function of $(g',L',G',H')$, and, since $g'$ is
linear in time up to $O(J_2^2)$, its time mean is its value at the time means
of these elements up to $O(J_2^2)$. For the short-period part, let $\mathcal P$
be the primitive of $\tilde b_1$ in $\ell$ with zero mean, so that
$\partial_g\mathcal P=\mathcal Z$ and, along the mean motion,
$\dd\mathcal P/\dd t=\dot\ell'\,\tilde b_1+\dot g\,\mathcal Z$. The osculating
pericentre $\ell=0$ lies at $\ell'=-J_2\,\partial_Lw+O(J_2^2)$, so that over a
revolution of duration $T$ the mean anomaly $\ell'$ advances by $2\pi+O(J_2^2)$,
the argument $g'$ by $\dot gT$, and $\mathcal P$, which is periodic in $\ell$, by
$\dot gT\,\mathcal Z|_{\ell=0}+O(J_2^2)$. The time mean of $\mathcal Z$ is
$\langle\mathcal Z\rangle_\ell+O(J_2)=O(J_2)$, and the time mean of
$\tilde b_1(y')$ is therefore
\[
 \frac{\dot g}{\dot\ell'}\,\mathcal Z\big|_{\ell=0}+O(J_2^2),
\]
which, with $\dot\ell'=n+O(J_2)$ and \eqref{eq:gdot}, is the second term of
\eqref{eq:second-mean} divided by $J_2$. The time means of the osculating
momenta are $X'+J_2\langle\{X,w\}\rangle_\ell+O(J_2^2)$, $X\in\{L,G,H\}$, and
$\bar g=g'+J_2\langle\{g,w\}\rangle_\ell+O(J_2^2)$, with $g'$ taken at the middle of
the revolution. Since $\{\cdot,w\}$ is a derivation, the chain rule gives
\[
 \langle b_1\rangle_\ell(\bar g,\bar L,\bar G,\bar H)=\langle b_1\rangle_\ell(y')
 +J_2\bigl\langle\{\langle b_1\rangle_\ell,w\}\bigr\rangle_\ell+O(J_2^2).
\]
Subtraction of $J_2$ times this expression from the sum of the three means
gives \eqref{eq:second-mean}, since
$\{b_1,w\}-\{\langle b_1\rangle_\ell,w\}=\{\tilde b_1,w\}$, and the evaluation of the
right-hand side at the averaged elements in place of $y'$ changes it by
$O(J_2^3)$. If a function $\mathcal U$ of $(g,L,G,H)$ is added to $w$, the first
term changes by
\[
 \bigl\langle\partial_\ell\tilde b_1\,\partial_L\mathcal U+\partial_g\tilde b_1\,\partial_G\mathcal U
 -\partial_G\tilde b_1\,\partial_g\mathcal U\bigr\rangle_\ell,
\]
which vanishes because $\partial_\ell\tilde b_1$ has zero mean by periodicity and
the derivatives in $g$ and $G$ at fixed $\ell$ commute with the average, under
which $\tilde b_1$ has zero mean.

\emph{(b)} The derivatives in \eqref{eq:second-mean} are taken at fixed $\ell$,
with $c=H/G$ and with $e$ and $f$ functions of $(\ell,L,G)$ through
\[
 \frac{\partial e}{\partial L}=\frac{\eta^3}{eG},\qquad
 \frac{\partial e}{\partial G}=-\frac{\eta^2}{eG},\qquad
 \frac{\partial f}{\partial e}=\frac{(2+e\cos f)\sin f}{\eta^2},\qquad
 \frac{\partial f}{\partial\ell}=\frac{(1+e\cos f)^2}{\eta^3}.
\]
Since $b_1$ contains the factor $r^{-6}=G^{-12}(1+e\cos f)^6$, the terms of
$\{b_1,w\}$ are trigonometric polynomials in $f$ or products $(f-\ell)\,Y$, which
come from the first term of \eqref{eq:W1-J2} and in which $Y$ and
$Y\,\dd\ell/\dd f$ are trigonometric polynomials; in
$\{\langle b_1\rangle_\ell,w\}$, the angle $f-\ell$ multiplies functions independent
of $\ell$, and its mean over $\ell$ vanishes. The means over $\ell$ follow from
\[
 \langle\cos kf\rangle_\ell=\Bigl(-\frac{e}{1+\eta}\Bigr)^k(1+k\eta),\qquad
 \langle\sin kf\rangle_\ell=0,\qquad k\ge0,
\]
which the residue theorem gives with $z=\mathrm e^{\mathrm if}$, since
$\dd\ell/\dd f=\eta^3(1+e\cos f)^{-2}$ has a double pole inside the unit circle
at $z=-e/(1+\eta)$, and from
\[
 \bigl\langle(f-\ell)\,Y\bigr\rangle_\ell=\langle\mathcal T_Y\rangle_\ell,\qquad
 \mathcal T_Y=\sum_{k\ne0}\frac{\hat Y_k}{\mathrm ik}\,\mathrm e^{\mathrm ikf},
\]
where $\hat Y_k$ are the Fourier coefficients in $f$ of $Y\,\dd\ell/\dd f$. The
last identity follows by integration by parts over $0\le\ell\le2\pi$ with the
primitive $\hat Y_0f+\mathcal T_Y$ of $Y$ in $\ell$, since $f-\ell$ vanishes at
both ends, the mean of $f$ over $\ell$ is $\pi$ and the mean of $\mathcal T_Y$
over $f$ vanishes. For $Y=\partial_gb_1$, the primitive of $\partial_g\tilde b_1$
in $\ell$ is $\hat Y_0(f-\ell)+\mathcal T_Y$, so that
$\mathcal Z|_{\ell=0}=\mathcal T_Y|_{f=0}-\langle\mathcal T_Y\rangle_\ell$. With these
rules every mean is a rational function of $c$, $G$ and $e/(1+\eta)$, of which
$e$ and $\eta$ are rational functions, and the computation, carried out with
SymPy \citep{SymPy2017} and included in \citet{CodeCMDA}, gives
\eqref{eq:Gammas}.

\emph{(c)} The weight \eqref{eq:second-mean-rel} is the ratio of
\eqref{eq:second-mean-cf} to the amplitude of \eqref{eq:brpsi-avg}. At $\eta=1$,
the bracket of $\Gamma_0$ equals $8s^2$, and the expansions of $\Gamma_2^{\mathrm d}$
and $\Gamma_2^{\mathrm b}$ follow from \eqref{eq:Gammas} with $\eta=1-\frac12e^2+O(e^4)$, the
factor $1-\eta=\frac12e^2+O(e^4)$ and the value $194880c^2-78400$ of the bracket of
$\Gamma_2^{\mathrm b}$ at $\eta=1$; the factor $(1-\eta)^2$ gives the order of $\Gamma_4$.
Substitution in \eqref{eq:second-mean-rel}, with $4e^{-2}(2+e^2)^{-1}=2e^{-2}-1+O(e^2)$,
gives the constant and \eqref{eq:second-mean-small-e}. The constant part is
multiplicative only for $s>0$, which is assumed.
\end{proof}

A quadrature of \eqref{eq:second-mean} over $\ell$ reproduces the closed form
\eqref{eq:second-mean-cf}, and at the elements averaged over each revolution
the closed form agrees with the constant and the coefficient of $\cos2g$ of
the integrations with $J_2$, with $J_2/2$ and with $e=0.05$
(Table~\ref{tab:second-check}). The largest misfit, relative to the largest
value of $\bar b-\langle b_{r\psi}\rangle_\ell$, halves with $J_2$, so that it
scales with $J_2^3$, as the remainder in \eqref{eq:second-mean-cf} requires; for
the orbit of Figure~\ref{fig:propagated} it is consistent with the terms of
order $J_2^3$, among them those in $J_2J_3$ and $J_2J_4$, which
\eqref{eq:second-mean-cf} does not contain.

By Proposition~\ref{prop:second-mean}(c), the constant part of the difference
does not vanish with the eccentricity, whereas the first-order average
\eqref{eq:brpsi-avg} carries the factor $e^2(2+e^2)$. The limit has a direct
origin: by \eqref{eq:W1-J2},
\[
 \lim_{e\to0}\{R,w\}=-\frac{s^2\sin2\theta}{2G^5},
\]
so that the radial velocity of a near-circular orbit has a short-period part of
first order in $J_2$ in the harmonic $2\theta$, and the mean of $J_2b_1$ with
this radial velocity, $r=G^2$ and $\Theta=G$ is $-\frac38J_2^2s^4G^{-16}$, the
limit of \eqref{eq:second-mean-cf}. Relative to the amplitude, the constant part
therefore grows as $e^{-2}$. For small $e$ the coefficient of $\cos2g$ is
dominated by $\Gamma_2^{\mathrm d}$, which is produced by the advance of the
pericentre during the revolution, depends on the choice of the osculating
pericentre as the endpoint of the revolution, and vanishes identically at the
critical inclination with the factor $5c^2-1$ of \eqref{eq:gdot}. There the
bracket part remains, and \eqref{eq:second-mean-small-e} gives
$-\frac{11}{5}J_2G^{-4}+O(e^2J_2G^{-4})$ relative to the amplitude.

The derivation in Delaunay variables assumes $J_2\ll e^2G^4$. At the scale
$e^2\sim J_2G^{-4}$ the transformation generated by $w$ is no longer close to the
identity in $(L,G)$, the first-order averaged Hamiltonian, reduced to one
degree of freedom, has for $c^2>\frac13$ a Keplerian action smaller than $G$
\citep[Sect.~2.1.4]{Martinusi2023}, and, by \eqref{eq:second-mean-small-e}, the
second-order term of \eqref{eq:second-mean-cf} becomes as large as the
first-order average. The bound $O(J_2^3)$ is therefore established at fixed $e$;
the integration with $e=0.05$ is consistent with a uniform bound for
$J_2\ll e^2G^4$ (Appendix~\ref{app:orbit}), and smaller eccentricities call for
variables regular at $e=0$, such as the polar-nodal or the resonant variables
\citep{Martinusi2023}. For $e$ of the order of $J_2s^2G^{-4}$, the short-period
part $J_2\{R,w\}$ of the radial velocity produces additional zeros of $R$, and the
passages through the osculating pericentre no longer delimit revolutions.

\subsection{Coordinate response and first-order defect}

A near-identity canonical transformation changes the product structure on
which the defects are evaluated, and the following theorem, introduced here,
identifies the part of a first-order defect that a family of product
structures can remove. Let
$\mathbf S_\lambda$ be a smooth family of product structures with parameter
$\lambda\in\mathbb R^m$, let $U$ be a fixed domain with a measure $\dd m$ and
positive weights $w_{ij}$, and let
\begin{equation}
 \begin{gathered}
 \langle\mathbf f,\mathbf h\rangle_w=\int_U\sum_{i<j}w_{ij}f_{ij}h_{ij}\,\dd m,\\
 \mathbf b(\lambda,\varepsilon)=\bigl(b^{\mathbf S_\lambda}_{ij}[H_0+\varepsilon V]\bigr)_{i<j},\qquad
 \mathcal E(\lambda,\varepsilon)=\|\mathbf b(\lambda,\varepsilon)\|_w^2.
 \end{gathered}
\end{equation}

\begin{theorem}[Coordinate response]
\label{thm:response}
Suppose that $(\lambda,\varepsilon)\mapsto\mathbf b(\lambda,\varepsilon)$ is of class $C^3$ near
$(\lambda_0,0)$ as a map into the weighted space $L^2_w$ of arrays with
$\|\mathbf f\|_w<\infty$, which holds, for instance, when the components and their
derivatives up to order three are continuous and bounded by a function whose
square is integrable with respect to $w_{ij}\,\dd m$; suppose further that
$\mathbf b(\lambda_0,0)=0$, and let
\begin{equation}
 \begin{gathered}
 \mathbf a_\alpha=\partial_{\lambda_\alpha}\mathbf b(\lambda_0,0),\qquad
 \mathbf r=\partial_\varepsilon\mathbf b(\lambda_0,0)=\Lop^{\mathbf S_{\lambda_0}}_{H_0}(V),\\
 G_{\alpha\beta}=\langle\mathbf a_\alpha,\mathbf a_\beta\rangle_w,\qquad
 h_\alpha=\langle\mathbf a_\alpha,\mathbf r\rangle_w .
 \end{gathered}
\end{equation}
If the matrix $G$ is positive definite, the following hold.
\begin{enumerate}[label=(\alph*)]
 \item For all sufficiently small $\varepsilon$, $\mathcal E(\cdot,\varepsilon)$ has a
 unique critical point near $\lambda_0$, which is a strict local minimum at
 \begin{equation}
  \lambda_*(\varepsilon)=\lambda_0-\varepsilon\,G^{-1}h+O(\varepsilon^2).
  \label{eq:lambda-star}
 \end{equation}
 \item With the projection residual
 $\mathbf r_\perp=\mathbf r-\sum_\alpha(G^{-1}h)_\alpha\mathbf a_\alpha$, which is
 orthogonal to every $\mathbf a_\alpha$,
 \begin{equation}
  \mathcal E(\lambda_*(\varepsilon),\varepsilon)
  =\varepsilon^2\|\mathbf r_\perp\|_w^2+O(\varepsilon^3)
  =\varepsilon^2\bigl(\|\mathbf r\|_w^2-h^{\mathsf T}G^{-1}h\bigr)+O(\varepsilon^3).
 \end{equation}
\end{enumerate}
\end{theorem}

\begin{proof}
\emph{(a)} Since $\mathcal E=\langle\mathbf b,\mathbf b\rangle_w$ and $\mathbf b$ is $C^3$ into
$L^2_w$, $\mathcal E$ is of class $C^3$. The gradient $\mathbf g=\partial_\lambda\mathcal E$ satisfies
$\mathbf g(\lambda_0,0)=0$, $\partial_\lambda\mathbf g(\lambda_0,0)=2G$ and
$\partial_\varepsilon\mathbf g(\lambda_0,0)=2h$, because $\mathbf b(\lambda_0,0)=0$. The
implicit function theorem gives a $C^2$ branch $\lambda_*(\varepsilon)$ of critical
points with $G\lambda_*'(0)+h=0$, which is \eqref{eq:lambda-star}, and
continuity of the Hessian keeps it positive definite on a fixed
neighbourhood for small $\varepsilon$, so that the critical point is unique there
and is a strict minimum.

\emph{(b)} Taylor expansion along the branch gives
$\mathbf b(\lambda_*(\varepsilon),\varepsilon)=\varepsilon\mathbf r_\perp+O(\varepsilon^2)$, and
$\langle\mathbf a_\alpha,\mathbf r_\perp\rangle_w=h_\alpha-(GG^{-1}h)_\alpha=0$, so that
$\mathbf r_\perp$ is the orthogonal residual and Pythagoras' theorem gives the
second form.
\end{proof}

\begin{proposition}[Response to a canonical generator]
\label{prop:canonical-generator}
Suppose that $b^{\mathbf S_0}[H_0]=0$ and that, in the direction $\lambda_\alpha$, the
family $\mathbf S_\lambda$ is transported from $\mathbf S_0$ by the Hamiltonian flow
$\Phi^\alpha_t$ of a function $K_\alpha$, with $\frac{\dd}{\dd t}(F\circ\Phi^\alpha_t)=\{F,K_\alpha\}\circ\Phi^\alpha_t$. Then
$\mathbf a_\alpha=\Lop^{\mathbf S_0}_{H_0}(\{H_0,K_\alpha\})$.
\end{proposition}

\begin{proof}
By the covariance (F2),
$b^{(\Phi_t^\alpha)_*\mathbf S_0}[H_0]=b^{\mathbf S_0}[H_0\circ\Phi^\alpha_t]\circ(\Phi^\alpha_t)^{-1}$.
At $t=0$ the derivative of the outer composition acts on
$b^{\mathbf S_0}[H_0]=0$ and vanishes, and the derivative of
$b^{\mathbf S_0}[H_0\circ\Phi^\alpha_t]$ is the linearised operator applied to
$\frac{\dd}{\dd t}H_0\circ\Phi^\alpha_t|_{t=0}=\{H_0,K_\alpha\}$.
\end{proof}

The two statements apply to the intermediaries as follows.

\begin{proposition}[First-order defect and Lie generator]
\label{prop:generator}
For every general intermediary \eqref{eq:general} satisfying
\eqref{eq:efficiency}, with $W_1$ of Proposition~\ref{prop:W1}, the following
hold in the modified chart on the non-equatorial orbits, $\rho>0$, and on the
whole chart when $A_{21}$ and $A_{22}$ are products of $s^2$ with smooth functions
of $(\Theta,N)$, as for the historical intermediaries.
\begin{enumerate}[label=(\alph*)]
 \item $\Lop_{H_{00}}(H_{10}-H_{01})=\Lop_{H_{00}}(H_{10})$, whose only nonzero
 component is \eqref{eq:zonal-defect}, independently of the coefficients
 $A_{km}$.
 \item $\Lop_{H_{00}}(H_{10})=-\Lop_{H_{00}}(\{H_{00},W_1\})$, so the first-order
 defect of the zonal problem is the negative of the operator response, in the
 sense of Proposition~\ref{prop:canonical-generator}, to the transport of the
 modified polar-nodal structure by the flow of $W_1$.
 \item For the one-parameter family transported by the flow of $W_1$, the
 response array is $\mathbf a=-\mathbf r$, so that, for every weight with
 $\|\mathbf r\|_w>0$, the matrix $G=\|\mathbf r\|_w^2$ of Theorem~\ref{thm:response} is
 positive and $\mathbf r_\perp=0$, and the linearised defects of the zonal problem
 vanish on the transported structure.
\end{enumerate}
\end{proposition}

\begin{proof}
\emph{(a)} This is Proposition~\ref{prop:general}(a) together with linearity.

\emph{(b)} This follows from \eqref{eq:homological} and linearity, since
$\{W_1,H_{00}\}=-\{H_{00},W_1\}$.

\emph{(c)} By Proposition~\ref{prop:canonical-generator} with $K_\alpha=W_1$ the response array
is $\Lop_{H_{00}}(\{H_{00},W_1\})=-\mathbf r$, which lies in the span of the
response arrays, and the projection residual vanishes.
\end{proof}

Since the linearised defects of every member of the family vanish and the
first-order defect of the zonal problem is entirely removable by the Lie
transform, the first-order defect cannot rank intermediaries; in the modified
chart, statement (c) concerns the defects, whose vanishing does not imply
separation of the Hamilton--Jacobi equation there (Remark~\ref{rem:degenerate}). They are distinguished at
second order, through the defects of Proposition~\ref{prop:general}(b), the
second-order terms of the transformation (Section~\ref{sec:second}) and the
choice of chart (Section~\ref{sec:vinti}).

\section{Second-order efficiency and the critical inclination}
\label{sec:second}

By Proposition~\ref{prop:generator}, the first-order generator removes the
first-order defect of the zonal problem; at second order the same generator
enters the averaged Hamiltonian, and this section follows it there. The
results are stated in the gauge of the generator \eqref{eq:W1}, and the
effect of a change of gauge is given after Corollary~\ref{cor:critical}.

For the Earth, $J_3J_2^{-2}=-2.2$ and $J_4J_2^{-2}=-1.4$ \citep{Pavlis2012}, so
that $J_3$ and $J_4$ enter a second-order theory. The usual ordering
\citep{Brouwer1959,Kozai1959,CoffeyDepritMiller1986,Lara2021} introduces a
bookkeeping parameter $\varepsilon$ with
\begin{equation}
 H=H_{00}+\varepsilon H_1+\varepsilon^2H_2,\qquad H_1=H_{10},\qquad
 H_2=\frac{\mu J_3R_e^3}{r^4}P_3(\sin\varphi)+\frac{\mu J_4R_e^4}{r^5}P_4(\sin\varphi),
 \label{eq:ordering}
\end{equation}
so that $J_3$ and $J_4$ enter at second order. With the generator
$W=\varepsilon W_1+\varepsilon^2W_2$ and the Lie series $H\mapsto H+\{H,W\}+\frac12\{\{H,W\},W\}+\dots$
\citep{Deprit1969}, the transformed Hamiltonian is $H_{00}+\varepsilon K_1+\varepsilon^2K_2+O(\varepsilon^3)$
with $K_1=H_{01}$ under \eqref{eq:homological} and
\begin{equation}
 K_2=Q_2+\{H_{00},W_2\},\qquad Q_2=H_2+\tfrac12\{H_1+K_1,W_1\}.
 \label{eq:K2}
\end{equation}
A generator $W_2$ that is periodic in the mean anomaly exists if and only if
$\langle K_2\rangle_\ell=\langle Q_2\rangle_\ell$, which is the second-order form of the
efficiency condition. In Delaunay variables $(\ell,g,h,L,G,H)$, with
$\mu=R_e=1$, the averages of the zonal terms over the mean anomaly are
\begin{gather}
 \Bigl\langle\frac{P_3(\sin\varphi)}{r^4}\Bigr\rangle_\ell
 =\frac{3\sqrt{(L^2-G^2)(G^2-H^2)}\,(G^2-5H^2)}{8G^8L^4}\,\sin g,
 \label{eq:J3avg}\\
 \Bigl\langle\frac{P_4(\sin\varphi)}{r^5}\Bigr\rangle_\ell
 \begin{aligned}[t]
 &=-\frac{3(3G^2-5L^2)(3G^4-30G^2H^2+35H^4)}{128\,G^{11}L^5}\\
 &\quad+\frac{15(G^2-H^2)(G^2-L^2)(G^2-7H^2)}{64\,G^{11}L^5}\cos2g,
 \end{aligned}
 \label{eq:J4avg}
\end{gather}
in agreement with the classical first-order theory of the odd and even
harmonics \citep{Kozai1959,Brouwer1959}. The following proposition is
introduced here.

\begin{proposition}[Second-order efficiency]
\label{prop:second}
Let $K_1$ be an efficient general intermediary and $W_1$ the generator of
Proposition~\ref{prop:W1}. Then the following hold.
\begin{enumerate}[label=(\alph*)]
 \item For every function of the form $F(r,H_{00},\Theta,N)+r^{-2}U_2(\theta,\Theta,N)$,
 and in particular for every member of the family \eqref{eq:general} and every
 Hamiltonian \eqref{eq:radial-angular} with $U_1=U_1(r,\Theta,N)$, the average
 over the mean anomaly does not depend on the argument of pericentre $g$.
 \item The second-order homological equation \eqref{eq:K2} has a solution
 $W_2$ periodic in $\ell$ with $K_2$ in the class of (a) if and only if
 $\langle Q_2\rangle_\ell$ does not depend on $g$. The term
 $\langle H_2\rangle_\ell$ contains the long-period terms $\sin g$ of $J_3$ and
 $\cos2g$ of $J_4$ given by \eqref{eq:J3avg} and \eqref{eq:J4avg}.
 \item If $K_1$ and $K_1'$ are two efficient general intermediaries with
 $A_{21}=A_{22}=0$, and $W_1$, $W_1'$ the corresponding generators of
 Proposition~\ref{prop:W1}, then $\langle Q_2'\rangle_\ell-\langle Q_2\rangle_\ell$ is a
 function of $(L,G,H)$ only, so the long-period part of $\langle Q_2\rangle_\ell$
 does not depend on the intermediary.
\end{enumerate}
\end{proposition}

\begin{proof}
\emph{(a)} The function $F$ evaluated along a Kepler orbit depends on $\ell$
and on the elements $(L,G,H)$ only. For the second term,
$\dd\ell=n\,r^2\Theta^{-1}\dd f$ gives $\langle r^{-2}U_2\rangle_\ell=
(n/\Theta)\,(2\pi)^{-1}\int_0^{2\pi}U_2(g+f,\Theta,N)\,\dd f$, which is the mean of $U_2$
over the argument of latitude and does not depend on $g$.

\emph{(b)} In Delaunay variables $\{H_{00},W_2\}=-n\,\partial_\ell W_2$, so a periodic
$W_2$ exists if and only if $\langle K_2\rangle_\ell=\langle Q_2\rangle_\ell$, and by
(a) the left-hand side does not depend on $g$. The second statement is
\eqref{eq:J3avg}--\eqref{eq:J4avg}.

\emph{(c)} Put $\delta K=K_1'-K_1$ and $\delta W=W_1'-W_1$. By \eqref{eq:homological},
$\{\delta W,H_{00}\}=-\delta K$, and by \eqref{eq:W1} with $A_{21}=A_{22}=0$ the
function $\delta W$ is periodic in $\ell$ and does not depend on $g$; $\delta K$,
$K_1$ and $K_1'$ do not depend on $g$ either. From \eqref{eq:K2},
\[
 Q_2'-Q_2=\tfrac12\{H_1+K_1,\delta W\}+\tfrac12\{\delta K,W_1\}+\tfrac12\{\delta K,\delta W\}.
\]
With $\delta K=\{H_{00},\delta W\}$, the Jacobi identity gives
$\{\delta K,W_1\}=\{H_{00},\{\delta W,W_1\}\}-\{H_1-K_1,\delta W\}$, where
\eqref{eq:homological} was used in the form $\{W_1,H_{00}\}=H_1-K_1$. The
average over $\ell$ of $\{H_{00},X\}=-n\,\partial_\ell X$ vanishes for every $X$ that
is periodic in $\ell$, so that
\[
 \langle Q_2'-Q_2\rangle_\ell=\bigl\langle\{K_1,\delta W\}+\tfrac12\{\delta K,\delta W\}\bigr\rangle_\ell .
\]
Both brackets involve only functions that do not depend on $g$ and $h$, so
their average is a function of $(L,G,H)$.
\end{proof}

With the generator \eqref{eq:W1}, Deprit's efficiency condition, the scalar
relation \eqref{eq:efficiency} at first order, can therefore not be satisfied at
second order by any intermediary of the class (a) as soon as $J_3\ne0$ or the
total coefficient of $\cos2g$ in $\langle Q_2\rangle_\ell$ is nonzero. The secular part,
the average of $\langle Q_2\rangle_\ell$ over $g$, can be matched by terms without the
argument of latitude, which keep the defects identically zero by
Proposition~\ref{prop:general}(c). It depends on the mean motion through
factors other than $\eta^3p^{-3}$, as \eqref{eq:J4avg} shows, so that within the
four radial weights of \eqref{eq:general} the second-order coefficients have to
depend on the Kepler energy, as $\mathcal V_0$ and $\mathcal V_1$ already do; the
wider class of (a) also admits other radial powers. The long-period part
remains in the transformed Hamiltonian, and it drives the frozen orbits
through the $J_3$ term \citep{CoffeyDepritDeprit1994} and the libration at the
critical inclination through the $\cos2g$ terms
\citep{CoffeyDepritMiller1986,Jupp1969}.

\begin{figure}[ht]
\centering
\includegraphics[width=\textwidth]{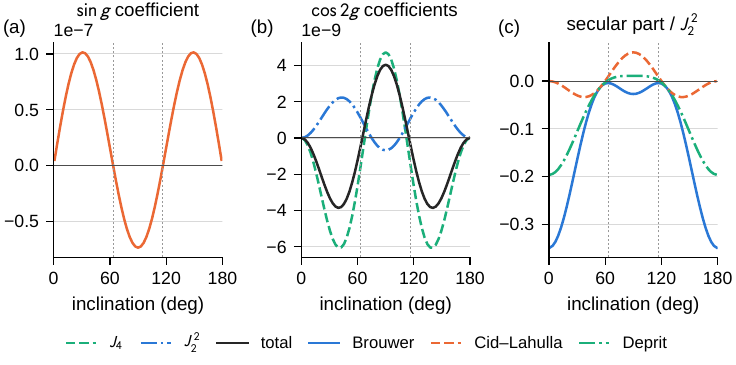}
\caption{Second-order averages over the mean anomaly for the Earth at
$a=1.3\,R_e$, $e=0.2$ (pericentre $1.04\,R_e$), versus inclination; dotted lines mark the critical
inclinations. (a) Coefficient of $\sin g$ from $J_3$, eq.~\eqref{eq:J3avg}.
(b) Coefficients of $\cos2g$ from $J_4$, eq.~\eqref{eq:J4avg}, from the $J_2^2$
bracket, eq.~\eqref{eq:J22lp}, and their sum. (c) Secular part of the $J_2^2$
bracket, in units of $J_2^2$, for three efficient intermediaries. Panels (a) and (b)
plot the closed forms, and panel (c) plots the quadratures of
Appendix~\ref{app:second}}
\label{fig:second}
\end{figure}

By Proposition~\ref{prop:second}(c), the coefficient of $\cos2g$ in the $J_2^2$
bracket of $\langle Q_2\rangle_\ell$ is the same for all efficient intermediaries without terms in the
argument of latitude, whereas the secular parts differ
(Figure~\ref{fig:second}). For the normalisation of \citet{Brouwer1959} it was
derived analytically by \citet[eq.~(12)]{Lara2021redux} in the Brouwer
short-period gauge; with his factorial convention the second-order term carries the factor
$\frac12J_2^2$, and with $H_{00}=-\frac12L^{-2}$, $p=G^2$ and $\eta=GL^{-1}$ his term reads
\begin{equation}
 \frac{3J_2^2}{64}\,\frac{(G^2-H^2)(G^2-L^2)(G^2-15H^2)}{G^{11}L^5}\cos2g
 \qquad(\mu=R_e=1).
 \label{eq:J22lp}
\end{equation}
Proposition~\ref{prop:second}(c) transfers it to the family, and a quadrature
over the mean anomaly for the intermediaries of Brouwer, Cid and Lahulla and
Deprit reproduces it on a grid of states to rounding error
(Appendix~\ref{app:second}). The coefficient depends on the $\ell$-independent
part of $W_1$, which is fixed by \eqref{eq:W1}: replacing $W_1$ by $W_1+\phi(g,L,G,H)$
changes $\langle Q_2\rangle_\ell$ by $\{\langle H_{10}\rangle_\ell,\phi\}=-\partial_G\langle H_{10}\rangle_\ell\,\partial_g\phi$,
because the averages of $H_1$ and $K_1$ coincide under \eqref{eq:efficiency},
and $\partial_G\langle H_{10}\rangle_\ell$, the secular rate of the argument of
pericentre, is proportional to $1-5c^2$ \citep{Brouwer1959,Kozai1959} and
vanishes at the critical inclination. This gives the following consequence.

\begin{corollary}[Critical inclination]
\label{cor:critical}
At the critical inclination $c^2=\frac15$, the coefficient of $\cos2g$ in
$\langle Q_2\rangle_\ell$ does not depend on the intermediary of
Proposition~\ref{prop:second}(c), and it does not change when a regular function
$\phi(g,L,G,H)$, independent of $\ell$ and of the node $h$, is added to $W_1$.
\end{corollary}

\begin{proof}
The statement follows from Proposition~\ref{prop:second}(c) and from the
vanishing of $\partial_G\langle H_{10}\rangle_\ell$ at $c^2=\frac15$.
\end{proof}

At $c^2=\frac15$, $G^2-15H^2=-2G^2$ and $G^2-7H^2=-\frac25G^2$, so \eqref{eq:J22lp}
equals the $J_4$ coefficient of \eqref{eq:J4avg} with $J_4$ replaced by $J_2^2$, and
the total coefficient of $\cos2g$ is
\begin{equation}
 \frac{15\,(J_4+J_2^2)}{64}\,\frac{(G^2-H^2)(G^2-L^2)(G^2-7H^2)}{G^{11}L^5}
 \qquad(\mu=R_e=1),
 \label{eq:critical}
\end{equation}
which the direct quadrature of the $J_2^2$ bracket confirms to rounding error
(Appendix~\ref{app:second}). For $e>0$ the resonant coefficient at the critical
inclination vanishes if and only if $J_4=-J_2^2$; at $e=0$ the factor $G^2-L^2$
vanishes, and the argument of pericentre is undefined. The condition $J_4=-J_2^2$ is Vinti's relation
(Section~\ref{sec:vinti}) and the condition identified by \citet{Brouwer1959} for
the cancellation of the terms with the critical divisor \citep{Lara2026}.
\citet[p.~106]{Izsak1963} observed that no small divisors appear in the
second-order solution of Vinti's problem, and in the $J_2$--$J_4$ problem the
critical divisor of the long-period generator carries the factor $J_4+J_2^2$
\citep[eq.~(32)]{Lara2026}. For the Earth, $J_4+J_2^2=-4.5\times10^{-7}$, which is
$28\%$ of $J_4$, so that the $J_2^2$ term reduces the resonant coefficient by a
factor of $3.6$ (Figure~\ref{fig:second}(b)).

The restriction to the gauge \eqref{eq:W1} matters for the efficiency
statement. Away from the critical inclination, a function $\phi(g,L,G,H)$ with
$\partial_g\phi$ equal to the zero-mean long-period part of $\langle Q_2\rangle_\ell$ divided
by $\partial_G\langle H_{10}\rangle_\ell$ removes that part locally, as in the elimination of
the perigee \citep{AlfriendCoffey1984,Lara2021}, at the price of the divisor
$1-5c^2$. The $J_3$ term of \eqref{eq:J3avg} shares the factor $G^2-5H^2$ with
the apsidal rate, and the addition
\begin{equation}
 \phi_3=\frac{J_3\,e\,s}{2J_2G}\cos g\qquad(\mu=R_e=1)
 \label{eq:J3gauge}
\end{equation}
to $W_1$ removes it smoothly across the critical inclination, as in
\citet[eq.~(45)]{Lara2026}. The terms in $\cos2g$ have no such factor, and at the
critical inclination their coefficient \eqref{eq:critical} is unchanged by every
regular addition to $W_1$ that depends on $(g,L,G,H)$, by
Corollary~\ref{cor:critical}. For the efficient intermediaries of
Proposition~\ref{prop:second}(c), the term proportional to $J_4+J_2^2$ is
therefore the obstruction to second-order efficiency under regular axially
symmetric changes of the generator.

\section{Oblate spheroidal coordinates and Vinti's focal radius}
\label{sec:vinti}

The defect depends on the product structure on which it is evaluated, and
Theorem~\ref{thm:response} turns this dependence into a selection rule. In the
meridional plane the oblate spheroidal structures form a two-parameter family,
with a focal radius and a centre on the axis, and this section applies the
rule to it and tests the selected structures along propagated orbits.

For axisymmetric potentials $\Phi(R,z)$ in cylindrical coordinates,
\citet{MartinusiPD} introduced the St\"ackel residuals
\begin{equation}
 g=\Phi_{Rz},\qquad
 A=3z\Phi_R-3R\Phi_z+Rz(\Phi_{RR}-\Phi_{zz})+(z^2-R^2)g,
 \label{eq:residual}
\end{equation}
and showed that the defect of the prolate structure with focal distance
$\Delta$ is proportional to $A-g\Delta^2$ \citep[Proposition~6.1]{MartinusiPD}, in
line with the classical conditions for St\"ackel potentials
\citep{Stackel,DeZeeuw1985}. Here $R$ denotes the cylindrical radius and $g$ the
residual of \eqref{eq:residual}, in this section only. The residuals $A$ and $g$ and the residual
\begin{equation}
 P=-3\Phi_R-R\,(\Phi_{RR}-\Phi_{zz})-2z\,g,
 \label{eq:residual-P}
\end{equation}
which measures the effect of a displacement of the centre along the axis,
appear in \citet[eqs.~(40)--(41)]{An2013} as $\mathfrak s=A$, $\mathfrak h=-P$ and
$\mathfrak k=g$, and An's condition for separation in spheroidal coordinates
whose centre is translated along the axis \citep[Sect.~5]{An2013} reads
$A+ag+\delta P=0$ in the notation below. The oblate spheroidal coordinates
centred at $z=\delta$ are $R=c\cosh u\sin v$, $z-\delta=c\sinh u\cos v$, with
focal radius $c>0$ and $D_c=c^2(\sinh^2u+\cos^2v)$, and the following
statement, introduced here, expresses the defect of these structures through
the same residuals.

\begin{proposition}[Oblate factorisation]
\label{prop:oblate}
Let $H=\frac12(p_R^2+p_z^2)+\frac12N^2R^{-2}+\Phi(R,z)$ be the meridional
Hamiltonian at fixed polar angular momentum $N$. Then the following hold.
\begin{enumerate}[label=(\alph*)]
 \item For the oblate structure centred at the origin,
 \begin{equation}
  \frac{\partial^2(D_c\Phi)}{\partial u\,\partial v}=D_c\,(A+c^2g),\qquad
  b_{uv}[H]=-\frac{p_up_v}{D_c^2}\,(A+c^2g).
  \label{eq:oblate}
 \end{equation}
 \item For the oblate structure centred at $z=\delta$,
 \begin{equation}
  b_{uv}[H]=-\frac{p_up_v}{D_c^2}\,\bigl(A+a\,g+\delta P\bigr),\qquad a=c^2+\delta^2 .
  \label{eq:oblate-shifted}
 \end{equation}
\end{enumerate}
\end{proposition}

\begin{proof}
\emph{(a)} The coordinate fields are
$\partial_u=c\sinh u\sin v\,\partial_R+c\cosh u\cos v\,\partial_z$ and
$\partial_v=c\cosh u\cos v\,\partial_R-c\sinh u\sin v\,\partial_z$, and the chain rule
gives the first identity. The kinetic energy is
$\frac12D_c^{-1}(p_u^2+p_v^2)$, and $\partial_{uv}D_c=0$. The centrifugal term
contributes $D_c\cdot\frac12N^2R^{-2}=\frac12N^2(\sin^{-2}v-\cosh^{-2}u)$ to $D_cH$,
which has no mixed derivative. The second identity then follows as in the
proof of \citet[Proposition~6.1]{MartinusiPD}.

\emph{(b)} The kinetic and centrifugal terms do not change under the
translation $z\mapsto z-\delta$, so (a) applies with $A$ replaced by the
expression $A_\delta$ obtained from \eqref{eq:residual} by replacing $z$ with
$z-\delta$ in the coefficients. Expanding,
$A_\delta=A+\delta P+\delta^2g$, and $A_\delta+c^2g$ is the residual of
\eqref{eq:oblate-shifted}.
\end{proof}

The residual $A+ag+\delta P$ is affine in $(a,\delta)$, and for $\delta=0$ and
$a=-\Delta^2$ it is the prolate residual of \citet[Proposition~6.1]{MartinusiPD}.
The local focal estimate $c^2_{\mathrm{loc}}=-A/g$ is the value of $c^2$ at
which the defect of the centred structure vanishes at a point. For the
potential of \citet{Vinti1961}, the Euler two-centre potential with centres at
$z=\pm\mathrm ic$, the product $D_c\Phi=-\mu c\sinh u$ depends on $u$ alone, so
that \eqref{eq:oblate} gives $A+c^2g=0$ and the estimate equals $c^2$ wherever
$g\ne0$. For the zonal harmonics the residuals follow from a ladder structure,
introduced here, in which $\varphi$ denotes the latitude, $R=r\cos\varphi$ and
$z=r\sin\varphi$.

\begin{lemma}[Residual ladder]
\label{lem:ladder}
Let $\phi_n=r^{-(n+1)}P_n(\sin\varphi)$, $n\ge0$, be the exterior zonal harmonics,
and denote by $A[\Phi]$, $g[\Phi]$ and $P[\Phi]$ the residuals of a potential $\Phi$.
Then the following hold.
\begin{enumerate}[label=(\alph*)]
 \item For every smooth $\Phi$,
 \begin{equation}
  A[\Phi]=-\frac1r\,\frac{\partial^2(r^2\Phi)}{\partial r\,\partial\varphi}.
  \label{eq:A-polar}
 \end{equation}
 \item $A[\phi_0]=A[\phi_1]=P[\phi_0]=0$, and for every $n\ge0$
 \begin{equation}
  A[\phi_{n+2}]=g[\phi_n],\qquad P[\phi_{n+1}]=-2\,g[\phi_n].
  \label{eq:ladder}
 \end{equation}
\end{enumerate}
\end{lemma}

\begin{proof}
\emph{(a)} The Euler operator $r\partial_r=R\partial_R+z\partial_z$ and the
rotation $T=R\partial_z-z\partial_R=\partial_\varphi$ commute, and $Tr=0$, so that
$r^{-1}\partial_r\partial_\varphi(r^2\Phi)=(2+r\partial_r)T\Phi$. Expanding
$(2+R\partial_R+z\partial_z)(R\Phi_z-z\Phi_R)$ gives $-A$.

\emph{(b)} With $x=\sin\varphi$, the identities
$(1-x^2)P_n'=(n+1)(xP_n-P_{n+1})$ and $P_{n+2}'=xP_{n+1}'+(n+2)P_{n+1}$
\citep[Chap.~22]{AbramowitzStegun1964} give $\partial_z\phi_n=-(n+1)\phi_{n+1}$ and
\[
 \begin{aligned}
 g[\phi_n]&=-(n+1)\,\partial_R\phi_{n+1}
 =(n+1)\,r^{-(n+3)}\cos\varphi\,\bigl[(n+2)P_{n+1}+xP_{n+1}'\bigr]\\
 &=(n+1)\,r^{-(n+3)}\cos\varphi\,P_{n+2}' ,
 \end{aligned}
\]
while \eqref{eq:A-polar} gives
$A[\phi_{n+2}]=(n+1)\,r^{-(n+3)}\cos\varphi\,P_{n+2}'$. Since $r^2\phi_0=r$ and
$r^2\phi_1=\sin\varphi$, \eqref{eq:A-polar} gives $A[\phi_0]=A[\phi_1]=0$. For a
harmonic $\Phi$, $\Phi_{zz}=-\Phi_{RR}-R^{-1}\Phi_R$, so that
$P[\Phi]=-4\Phi_R-2r\partial_r\Phi_R$; for $\Phi=\phi_{n+1}$ the function $\Phi_R$ is
homogeneous of degree $-(n+3)$, hence $P[\phi_{n+1}]=2(n+1)\,\partial_R\phi_{n+1}=-2g[\phi_n]$,
and $n=-1$ gives $P[\phi_0]=0$.
\end{proof}

For a planet with the zonal potential
\begin{equation}
 \Phi=-\frac{\mu}{r}\Bigl[1-J_2\Bigl(\frac{R_e}{r}\Bigr)^2P_2(\sin\varphi)
 -J_4\Bigl(\frac{R_e}{r}\Bigr)^4P_4(\sin\varphi)\Bigr],
\end{equation}
the residuals are linear in $\Phi$, and with $\mu=1$ Lemma~\ref{lem:ladder}
and a direct computation of $g[\phi_2]$ give
\begin{equation}
 \begin{gathered}
 A_K=0,\qquad
 A_2=-R_e^2\,g_K,\qquad
 g_2=\beta\,g_K,\qquad
 A_4=R_e^2\,\beta\,g_K,\\
 \beta=\frac52\Bigl(\frac{R_e}{r}\Bigr)^2\bigl(3-7\sin^2\varphi\bigr),
 \end{gathered}
 \label{eq:zonal-residuals}
\end{equation}
where the subscripts $K$, $2$ and $4$ refer to the Kepler term and to the
coefficients of $J_2$ and $J_4$. With the Vinti parameter
\begin{equation}
 \kappa_V=\frac{J_4+J_2^2}{J_2},
\end{equation}
which measures the departure of the planet from Vinti's relation
\citep{Vinti1961}, the following proposition is introduced here.

\begin{proposition}[Vinti's focal radius as a coordinate response]
\label{prop:vinti-response}
Let $K$ be a compact set of positive area in the meridional plane outside the
planet, $r\ge r_1>0$, let $w>0$ be integrable on $K$, so that the weighted norms
below are finite, and let $\lambda=c^2$. Then the following hold.
\begin{enumerate}[label=(\alph*)]
 \item The weighted square of the oblate residual,
 $\mathcal E(\lambda)=\int w\,(A+\lambda g)^2$, is a quadratic function of
 $\lambda\in\mathbb R$ with the unique minimiser
 $\lambda_*=-\langle g,A\rangle_w/\langle g,g\rangle_w$ whenever $\langle g,g\rangle_w>0$, which
 defines an oblate structure, $c^2=\lambda_*$, when $\lambda_*>0$.
 \item For $J_4=O(J_2^2)$,
 \begin{equation}
  \lambda_*=J_2R_e^2\bigl[1-\kappa_V\,\langle\beta\rangle_w+O(J_2^2)\bigr],\qquad
  \langle\beta\rangle_w=\frac{\int w\,g_K^2\,\beta}{\int w\,g_K^2},
  \label{eq:lambda-vinti}
 \end{equation}
 so that, for every weight and sufficiently small $J_2>0$, $\lambda_*>0$ and the
 first-order coordinate response is Vinti's choice $c^2=J_2R_e^2$.
 \item The residual at the minimiser is
 $A+\lambda_*g=J_2R_e^2\,\kappa_V\,g_K\,(\beta-\langle\beta\rangle_w)+O(J_2^3)$, and its
 leading term vanishes if and only if $J_4=-J_2^2$.
 \item Let
 \begin{equation}
  Q_\lambda=\frac{p_up_v}{D_c^2}
  =\frac{Rz\,(p_R^2-p_z^2)+(z^2-R^2+\lambda)\,p_Rp_z}{(r^2-\lambda)^2+4\lambda z^2},
  \qquad r^2=R^2+z^2,
  \label{eq:Qlambda}
 \end{equation}
 so that, by \eqref{eq:oblate}, $b_{uv}[H]=-Q_\lambda(A+\lambda g)$ at every point of
 phase space. Let $w\ge0$ be a measurable weight on $K\times\mathbb R^2$, with $K$ as
 above and the momenta in the second factor, with $\int w\,(1+|\mathbf p|^4)<\infty$ and
 $\int wQ_0^2g_K^2>0$. For $J_4=O(J_2^2)$, the defect objective
 $\mathcal E_B(\lambda)=\int w\,Q_\lambda^2(A+\lambda g)^2$ has a critical point
 \begin{equation}
  \lambda_B=J_2R_e^2\bigl[1-\kappa_V\,\langle\beta\rangle_{wQ_0^2}+O(J_2^2)\bigr],
  \label{eq:lambda-B}
 \end{equation}
 where $\langle\beta\rangle_{wQ_0^2}$ is the average of \eqref{eq:lambda-vinti} with
 the weight $wQ_0^2$.
\end{enumerate}
\end{proposition}

\begin{proof}
\emph{(a)} The residual is affine in $\lambda$, so $\mathcal E$ is quadratic with
leading coefficient $\langle g,g\rangle_w$, and its stationary point is the stated
minimiser. For the zonal field, $g=g_K(1+O(J_2))$ and $g_K=-3\mu Rz\,r^{-5}$ vanish
only on the axes, so that $\langle g,g\rangle_w>0$ for small $J_2$.

\emph{(b)} By \eqref{eq:zonal-residuals},
$A=-J_2R_e^2g_K+J_4R_e^2\beta g_K$ and $g=g_K(1+J_2\beta+O(J_2^2))$, hence
\[
 \frac{A}{g}=-J_2R_e^2\bigl[1-(J_4J_2^{-1}+J_2)\beta+O(J_2^2)\bigr]
 =-J_2R_e^2\bigl[1-\kappa_V\beta+O(J_2^2)\bigr].
\]
Inserting $A=g\cdot(A/g)$ into the minimiser of (a) gives
$\lambda_*=-\langle g^2,A/g\rangle_w/\langle g,g\rangle_w$, and the expansion
$g^2=g_K^2(1+O(J_2))$ yields \eqref{eq:lambda-vinti}.

\emph{(c)} At $\lambda_*$, $A+\lambda_*g=g\,(\lambda_*+A/g)$, and (b) gives
$\lambda_*+A/g=J_2R_e^2\kappa_V(\beta-\langle\beta\rangle_w)+O(J_2^3)$. The function
$\beta$ is not constant on any open set, so the leading term vanishes if and
only if $\kappa_V=0$, that is, $J_4=-J_2^2$.

\emph{(d)} The factor \eqref{eq:Qlambda} follows from
$p_u=\partial_uR\,p_R+\partial_uz\,p_z$, $p_v=\partial_vR\,p_R+\partial_vz\,p_z$ and the
coordinate fields given in the proof of Proposition~\ref{prop:oblate}. For
$|\lambda|\le\frac12r_1^2$ the denominator of \eqref{eq:Qlambda} is at least $\frac14r_1^4$
on $K$, so that $Q_\lambda$ and its derivatives in $\lambda$ are smooth and bounded
by a constant times $|\mathbf p|^2$; the integrands of $\mathcal E_B$ and of its
derivatives in $\lambda$ are then bounded by a constant times $w|\mathbf p|^4$, and
$\mathcal E_B$ is smooth near $\lambda=0$, with derivatives taken under the integral
\citep[Theorem~2.27]{Folland1999}. Put $J_4=\kappa'J_2^2$ with $\kappa'$
fixed and $\lambda=J_2R_e^2(1+x)$. By \eqref{eq:zonal-residuals},
$A+\lambda g=J_2R_e^2B$ exactly, with
\[
 B=g_K\bigl[x+J_2(1+\kappa'+x)\beta\bigr]+\kappa'(1+x)J_2^2\,g_4 ,
\]
where $g_4$ is the coefficient of $J_4$ in $g$, and the condition
$\partial_\lambda\mathcal E_B=0$, divided by $2J_2R_e^2$, becomes
\[
 F(x,J_2)=\int wQ_\lambda^2\,g\,B
 +\frac{J_2R_e^2}{2}\int w\,\partial_\lambda(Q_\lambda^2)\,B^2=0 .
\]
Here $F(x,0)=x\int wQ_0^2g_K^2$, so that $\partial_xF(0,0)>0$, and the implicit
function theorem gives a solution $x(J_2)$ with $x(0)=0$. Since $B$ vanishes at
$(x,J_2)=(0,0)$ and $\partial_{J_2}B=(1+\kappa')\beta g_K$ there, the second term
and the derivatives of $Q_\lambda$ and $g$ do not contribute to $\partial_{J_2}F(0,0)$,
which gives $x'(0)=-(1+\kappa')\langle\beta\rangle_{wQ_0^2}$. Since $\kappa_V=J_2(1+\kappa')$, this is
\eqref{eq:lambda-B}.
\end{proof}

\begin{figure}[ht]
\centering
\includegraphics[width=0.75\textwidth]{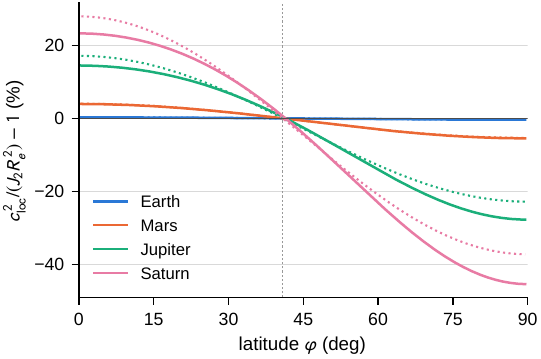}
\caption{Relative departure of the local focal estimate from Vinti's value
$J_2R_e^2$ at $r=1.05\,R_e$ versus latitude, from the exact residuals (solid)
and from the first-order formula $-\kappa_V\beta$ (dotted). The dotted vertical
line marks the latitude $\arcsin\sqrt{3/7}=40.9^\circ$, where $\beta=0$; the Earth
curve stays within $0.4\%$ of zero}
\label{fig:vinti}
\end{figure}

Since $Q_\lambda$ depends on $\lambda$, the oblate defect $-Q_\lambda(A+\lambda g)$ is not
affine in $\lambda$, although the residual is. Parts (a)--(c) concern the
residual objective, whose minimiser is explicit, and part (d) the defect
objective of Theorem~\ref{thm:response}, applied at $\lambda_0=0$ with
$\mathbf a=-Q_0g_K$; the two minimisers differ by the replacement of $w$ by
$wQ_0^2$ and by terms of relative order $J_2^2$ (Appendix~\ref{app:vinti}).
Since $A_2$ is proportional to $g_K$ pointwise, the first-order response does
not depend on the weight for either objective, and Vinti's choice
$c^2=J_2R_e^2$ is the first-order coordinate response of the oblate family. By
(c), a constant $\lambda$ makes the residual of order $J_2^3$ on the whole domain
only if $J_4+J_2^2=O(J_2^3)$, so that Vinti's relation $J_4=-J_2^2$
\citep{Vinti1961} is the condition under which the oblate family removes the
defect of the zonal problem to second order; by \eqref{eq:critical}, the same
combination controls the resonant term at the critical inclination. To
leading order the local estimate departs from Vinti's value by
$\frac52|\kappa_V|R_e^2r^{-2}|3-7\sin^2\varphi|$, which vanishes at $\sin^2\varphi=\frac37$
and is largest at the poles (Figure~\ref{fig:vinti} and
Table~\ref{tab:planets}).

\begin{table}[ht]
\centering
\caption{Oblate defect of four planets, for the even zonal field with $J_2$ and
$J_4$. Columns 5 and 6: maximum over latitude of the relative departure
$|1-c^2_{\mathrm{loc}}J_2^{-1}R_e^{-2}|$ at $r=1.05\,R_e$, from the exact residuals
and from the first-order formula $10|\kappa_V|R_e^2r^{-2}$. Columns 7 and 8:
minimiser $\lambda_*$ of the residual objective of
Proposition~\ref{prop:vinti-response}(a) on a uniform grid of $120$ radii in
$1.05\le r/R_e\le3$ and $161$ latitudes in $5^\circ$--$85^\circ$, and the relative
residual $\|A+\lambda_*g\|/\|A\|$ on the same grid; the formula
\eqref{eq:lambda-vinti} agrees with $\lambda_*$ to $3\times10^{-6}$ for the Earth and
$5\times10^{-3}$ for Saturn. For Jupiter and Saturn the higher measured moments
change the departures near the poles materially. Zonal coefficients from
\citet{Pavlis2012} (Earth), \citet{Konopliv2016} (Mars), \citet{Iess2018}
(Jupiter) and \citet{Iess2019} (Saturn).}
\label{tab:planets}
\footnotesize\setlength{\tabcolsep}{3.5pt}
\begin{tabular}{@{}lrrrrrrr@{}}
\toprule
Planet & $J_2$ & $J_4J_2^{-2}$ & $\kappa_V$ & exact & first order &
$\lambda_*J_2^{-1}R_e^{-2}$ & residual\\
\midrule
Earth & $1.0826\times10^{-3}$ & $-1.38$ & $-4.1\times10^{-4}$ & $0.38\%$ & $0.37\%$ & $0.99966$ & $1.3\times10^{-3}$\\
Mars & $1.957\times10^{-3}$ & $-4.02$ & $-5.9\times10^{-3}$ & $5.5\%$ & $5.4\%$ & $0.99521$ & $1.8\times10^{-2}$\\
Jupiter & $1.4697\times10^{-2}$ & $-2.72$ & $-2.52\times10^{-2}$ & $27.8\%$ & $22.9\%$ & $0.98199$ & $8.0\times10^{-2}$\\
Saturn & $1.6291\times10^{-2}$ & $-3.52$ & $-4.11\times10^{-2}$ & $45.5\%$ & $37.3\%$ & $0.97115$ & $0.13$\\
\bottomrule
\end{tabular}
\end{table}

With the third harmonic the residual acquires the term $J_3\,g[\phi_1]$, which
is not proportional to $g_K$, and the centred structures can no longer remove
it. The shifted structures of
Proposition~\ref{prop:oblate}(b) and Lemma~\ref{lem:ladder} give the following
statement, introduced here, in which $J_0=-1$, $J_1=0$ and $\mu=R_e=1$, so that
the zonal potential is $\Phi=\sum_{n\ge0}J_n\phi_n$.

\begin{proposition}[Third zonal harmonic and shifted oblate structures]
\label{prop:shifted}
Let $\Phi=\sum_{n\ge0}J_n\phi_n$, with $J_0=-1$ and $J_1=0$, be a finite sum or a series that converges,
together with its derivatives, on a shell $r_1<r<r_2$. Then the following hold.
\begin{enumerate}[label=(\alph*)]
 \item For all $(a,\delta)$,
 \begin{equation}
  A+a\,g+\delta P=\sum_{m\ge0}\bigl(J_{m+2}-2\delta J_{m+1}+aJ_m\bigr)\,g[\phi_m].
  \label{eq:shifted-residual}
 \end{equation}
 \item The oblate structure with parameters $(a,\delta)$, $a>\delta^2$, separates the
 meridional Hamiltonian, $b_{uv}[H]\equiv0$, if and only if
 \begin{equation}
  J_{m+2}=2\delta J_{m+1}-aJ_m\qquad\text{for all }m\ge0 .
  \label{eq:recurrence}
 \end{equation}
 In that case
 \begin{equation}
  a=J_2,\qquad \delta=\frac{J_3}{2J_2},\qquad
  c^2=J_2\Bigl(1-\frac{J_3^2}{4J_2^3}\Bigr),\qquad J_4=\frac{J_3^2}{J_2}-J_2^2,
  \label{eq:vinti1966}
 \end{equation}
 and $\Phi$ is the potential of two centres at $z=\delta\pm\mathrm ic$ with the
 complex conjugate masses $\frac12(1\pm\mathrm i\delta c^{-1})$.
 \item Let $K$ be a compact exterior set of positive area, symmetric under
 $z\mapsto-z$, and $w>0$ integrable on $K$ and even in $z$. For $J_3,J_4=O(J_2^2)$
 and sufficiently small $J_2>0$ the weighted square
 $\mathcal E(a,\delta)=\int w\,(A+ag+\delta P)^2$ of the truncated potential with
 $J_2$, $J_3$ and $J_4$ has a unique minimiser, with
 \begin{equation}
  a_*=J_2\bigl[1-\kappa_V\,\langle\beta\rangle_w+O(J_2^2)\bigr],\qquad
  \delta_*=\frac{J_3}{2J_2}+O(J_2^2),
  \label{eq:shifted-response}
 \end{equation}
 with $\langle\beta\rangle_w$ of \eqref{eq:lambda-vinti}, and $a_*-\delta_*^2=J_2+O(J_2^2)>0$, so
 that the minimiser defines an oblate structure.
\end{enumerate}
\end{proposition}

\begin{proof}
\emph{(a)} The series may be differentiated term by term on the shell. By
Lemma~\ref{lem:ladder}, $A[\Phi]=\sum_{n\ge2}J_nA[\phi_n]=\sum_{m\ge0}J_{m+2}\,g[\phi_m]$
and $P[\Phi]=\sum_{n\ge1}J_nP[\phi_n]=-2\sum_{m\ge0}J_{m+1}\,g[\phi_m]$, and
$g[\Phi]=\sum_mJ_m\,g[\phi_m]$.

\emph{(b)} By Proposition~\ref{prop:oblate}(b), $b_{uv}[H]\equiv0$ if and only if
the residual vanishes. The functions $g[\phi_m]=-(m+1)\,\partial_R\phi_{m+1}$ are
not identically zero and are homogeneous of degree $-(m+3)$, so that
$g[\phi_m]=r^{-(m+3)}G_m(\varphi)$ with a trigonometric polynomial $G_m\not\equiv0$,
which has finitely many zeros. A latitude at which no $G_m$ vanishes exists,
since the zeros form a countable set, and the restriction of
\eqref{eq:shifted-residual} to the ray at that latitude is a power series in $r^{-1}$ with the coefficients $(J_{m+2}-2\delta J_{m+1}+aJ_m)G_m$, which
vanishes on an interval only if all its coefficients vanish; this gives
\eqref{eq:recurrence}. The cases $m=0,1,2$ give \eqref{eq:vinti1966}. The points
$z_\pm=\delta\pm\mathrm ic$ are the roots of $\zeta^2-2\delta\zeta+a$, so every sequence
$J_n=-(m_+z_+^n+m_-z_-^n)$ satisfies \eqref{eq:recurrence}, and $J_0=-1$, $J_1=0$
fix $m_\pm$. The expansion
$|\mathbf x-\zeta\mathbf e_z|^{-1}=\sum_n\zeta^n\phi_n$ for $r>|\zeta|$
\citep[Chap.~22]{AbramowitzStegun1964} identifies $\sum_nJ_n\phi_n$ with the
potential of the two centres.

\emph{(c)} The harmonic $\phi_n$ has the parity $(-1)^n$ in $z$, so $g_n=g[\phi_n]$ has
the parity $(-1)^{n+1}$, and odd and even functions are orthogonal for $w$. Put
$a=J_2+\tilde x$ and $\delta=\delta_0+\tilde dJ_2^{-1}$, with $\delta_0=\frac12J_3J_2^{-1}$. By
\eqref{eq:shifted-residual}, with $J_0=-1$, $J_1=0$ and $J_n=0$ for $n>4$, the residual
is affine in $(\tilde x,\tilde d)$,
\[
 A+ag+\delta P=\rho_0+\tilde x\,g+\tilde d\,\frac{P}{J_2},
\]
with $g=-g_0+J_2g_2+J_3g_3+J_4g_4$, $J_2^{-1}P=-2g_1-2J_3J_2^{-1}g_2-2J_4J_2^{-1}g_3$ and
\[
 \rho_0=\Bigl(J_4+J_2^2-\frac{J_3^2}{J_2}\Bigr)g_2+J_3\Bigl(J_2-\frac{J_4}{J_2}\Bigr)g_3+J_2J_4\,g_4 ,
\]
whose odd part is $O(J_2^2)$ and whose even part is $O(J_2^3)$. The odd parts of $g$
and $J_2^{-1}P$ are $-g_0+O(J_2)$ and $O(J_2)$, and their even parts are $O(J_2^2)$ and
$-2g_1+O(J_2)$. By parity the scaled Gram matrix of $(g,J_2^{-1}P)$ is
$\mathrm{diag}(\|g_0\|_w^2,4\|g_1\|_w^2)+O(J_2)$, positive definite for small $J_2$, since
$g_0=-g_K\not\equiv0$ and $g_1\not\equiv0$, so the minimiser is unique. The right-hand sides
of the normal equations are $\langle g,\rho_0\rangle_w=O(J_2^2)$ and
$\langle J_2^{-1}P,\rho_0\rangle_w=O(J_2^3)$, which gives $\tilde x=O(J_2^2)$, then
$\tilde d=O(J_2^3)$, that is, $\delta_*=\delta_0+O(J_2^2)$, and
$\tilde x=-\langle g,\rho_0\rangle_w\langle g,g\rangle_w^{-1}+O(J_2^4)$. With $g_2=\beta g_K=-\beta g_0$
by \eqref{eq:zonal-residuals} and $J_4+J_2^2=J_2\kappa_V$, this is
$\tilde x=-J_2\kappa_V\langle\beta\rangle_w+O(J_2^3)$, which is \eqref{eq:shifted-response};
since $\delta_*^2=O(J_2^2)$, $a_*-\delta_*^2=J_2+O(J_2^2)$.
\end{proof}

The translated spheroidal potential is classical. \citet[Sect.~8,
eq.~(50)]{Vinti1966inv} gives its entire zonal sequence,
$J_n=\sigma^n\sin((n-1)\gamma)/\sin\gamma$, which satisfies \eqref{eq:recurrence} with
$a=\sigma^2$ and $\delta=\sigma\cos\gamma$ by the addition identity of the sine, and his
eqs.~(53)--(55) give the displacement and the fourth harmonic of
\eqref{eq:vinti1966}; the reference potential of \citet{Vinti1966} is built on
it, is fitted exactly through $J_3$ and reproduces the fraction
$(J_3^2J_2^{-1}-J_2^2)J_4^{-1}$ of the fourth harmonic. The contributions of
Proposition~\ref{prop:shifted} are the direct proof of necessity and
sufficiency by the residual ladder and the weighted fit (c) for fields that
do not satisfy the recurrence. The recurrence also shows why tuning the fourth
harmonic alone can remove the second-order residual, as in
Proposition~\ref{prop:vinti-response}(c), whereas exact oblate separation
requires the entire sequence. For the centred structure of
Proposition~\ref{prop:vinti-response} the residual contains the term
$J_3\,g[\phi_1]$, which is even in $z$ and cannot be removed; the displacement
$\delta_*$ removes it to first order, and the leading residual at the minimiser
is that of Proposition~\ref{prop:vinti-response}(c), since the exact condition
$J_4=J_3^2J_2^{-1}-J_2^2$ of (b) differs from Vinti's relation $J_4=-J_2^2$ by
$J_3^2J_2^{-1}=O(J_2^3)$. To first order the plane $z=\delta_*$ is the zero set of
the latitudinal force near the equator found in Section~\ref{sec:propagated},
so that the shifted structure and the coupling $b_{\psi\xi}$ single out the same
plane (Tables~\ref{tab:shifted} and \ref{tab:orbit-diag}).

\begin{table}[ht]
\centering
\caption{Shifted oblate structures (Proposition~\ref{prop:shifted}). Columns
2--4: centre $\delta=\frac12R_eJ_3J_2^{-1}$ and focal radius $c$ of the separable
structure \eqref{eq:vinti1966}, and Vinti's centred value $R_e\sqrt{J_2}$.
Column 5: fourth harmonic of the separable potential relative to the planet's
$J_4$. Columns 6--8: minimiser of the two-parameter residual objective on a
uniform grid of $120$ radii in $1.05\le r/R_e\le3$ and $2\times161$ latitudes in
$\pm(5^\circ$--$85^\circ)$, against $1-\kappa_V\langle\beta\rangle_w$ of
\eqref{eq:shifted-response}. Columns 9 and 10: relative residual, the norm of
the residual divided by the norm $\|A\|$ of the uncorrected residual, at
$(a_*,\delta_*)$ and for the centred structure ($\delta=0$) with its own re-optimised
focal radius, on the same grid. Zonal coefficients from \citet{Pavlis2012}
(Earth) and \citet{Konopliv2016} (Mars, field MRO120D, $R_e=3396$~km).}
\label{tab:shifted}
\footnotesize\setlength{\tabcolsep}{2.8pt}
\begin{tabular}{@{}lrrrrrrrrr@{}}
\toprule
Planet & $\delta$ (km) & $c$ (km) & $R_e\sqrt{J_2}$ (km) & $J_4$ ratio &
$a_*J_2^{-1}R_e^{-2}$ & $1-\kappa_V\langle\beta\rangle_w$ & $\delta_*/\delta$ & residual & $\delta=0$\\
\midrule
Earth & $-7.46$ & $209.73$ & $209.86$ & $0.720$ & $0.999657$ & $0.999659$ & $1.0011$ & $1.3\times10^{-3}$ & $4.9\times10^{-3}$\\
Mars & $27.32$ & $147.71$ & $150.22$ & $0.216$ & $0.994999$ & $0.995120$ & $0.9934$ & $1.9\times10^{-2}$ & $3.7\times10^{-2}$\\
\bottomrule
\end{tabular}
\end{table}

The residual norms of Tables~\ref{tab:planets} and \ref{tab:shifted} are
spatial, and a dynamical output is needed to compare the structures along
orbits. For the oblate structure with centre $\delta$, focal radius $c$ and
$a=c^2+\delta^2$, with $R=c\cosh u\sin v$ and $z-\delta=c\sinh u\cos v$, the function
\begin{equation}
 I_3=p_v^2+\frac{N^2}{\sin^2v}+2\delta\cos v-2E\,c^2\cos^2v\qquad(\mu=R_e=1),
 \label{eq:I3}
\end{equation}
with $p_v$ the momentum conjugate to $v$ and $E$ the energy (along an orbit of a
zonal field, the conserved value of its Hamiltonian), is an exact integral
of the displaced potential of Proposition~\ref{prop:shifted}(b), for which
$D_c\Phi=-(c\sinh u-\delta\cos v)$ separates, as follows from the separation of
$D_c(H-E)$ \citep{Vinti1961}. For a zonal field $\Phi$ that does not satisfy the
recurrence, let $\Delta\Phi$ be its difference from this displaced potential.
On the energy surface $H=E$, the $v$ part of the separated function $D_c(H-E)$
is $\frac12I_3$, and the bracket of $I_3$ with $H$ gives
\begin{equation}
 \frac{\dd I_3}{\dd t}=-\frac{2p_v}{D_c}\,\frac{\partial(D_c\,\Delta\Phi)}{\partial v},
 \label{eq:I3dot}
\end{equation}
whereas the defect of Proposition~\ref{prop:oblate} contains the mixed
derivative $\partial_u\partial_v(D_c\Delta\Phi)$. A structure that minimises the
spatial residual does not in general minimise the drift of $I_3$, and
Table~\ref{tab:drift} compares four structures on the same initial states and
the same field: the centred structure with Vinti's value $a=J_2$ (C$_0$), the
centred structure whose focal radius minimises the residual of the field with
$J_3$ on the grid of Table~\ref{tab:shifted} (C$_*$), the displaced least-squares
pair of Proposition~\ref{prop:shifted}(c) (S$_*$) and the displaced structure
with the recurrence values \eqref{eq:vinti1966} (S$_1$). The output is the
approximate conservation of $I_3$ on the stated grid of orbits, which is not an
error of orbit propagation. Over $300$ periods the median of the per-orbit ratio
of the excursion of C$_0$ to that of the displaced structure is $44$ (S$_*$) and
$77$ (S$_1$) for the Earth and $23$ and $39$ for Mars. Between $30$ and $300$ periods the excursion of C$_0$
grows by the factors $6.1$ and $4.6$, through the long-period term of $J_3$,
which requires eccentricity, and that of S$_1$ by $1.70$ and $1.46$. The
re-optimised centred structure C$_*$ gives no gain, and the recurrence values
S$_1$ do better than the least-squares pair S$_*$ on all $18$ Earth orbits and on
$16$ of the $18$ Mars orbits, as \eqref{eq:I3dot} allows. For S$_1$ the values
$a=J_2$ and $\delta=\frac12J_3J_2^{-1}$ reproduce $J_2$ and $J_3$, and the difference
$\Delta\Phi$ in \eqref{eq:I3dot} is $\Delta J_4\,\phi_4$, with
$\Delta J_4=J_4-(J_3^2J_2^{-1}-J_2^2)$ equal to $0.28J_4$ for the Earth and $0.78J_4$
for Mars (Table~\ref{tab:shifted}), minus the terms of degree five and higher of
the displaced potential, whose coefficients follow from
\eqref{eq:recurrence} and \eqref{eq:vinti1966}, starting with
$J_5=J_3^3J_2^{-2}-2J_2J_3$ and $J_6=J_2^3-3J_3^2+J_3^4J_2^{-3}$. To separate the two contributions,
the orbits of Table~\ref{tab:drift} were propagated again in the field with $J_4$
replaced by its recurrence value $J_3^2J_2^{-1}-J_2^2$, so that $\Delta J_4=0$ and
$\Delta\Phi$ reduces to the terms of degree five and higher. Over $300$ periods
the excursion of S$_1$ then decreases by the median per-orbit factors $12.7$ for
the Earth and $15.1$ for Mars, and by at least $3.7$ (Earth) and $6.7$ (Mars) on
each orbit, while the median per-orbit ratio for C$_0$, whose excursion comes from
$J_3$, is $1.0$ for both planets (Appendix~\ref{app:vinti}). The term
$\Delta J_4\,\phi_4$ therefore causes most of the excursion of S$_1$ in
Table~\ref{tab:drift}.

\begin{table}[ht]
\centering
\caption{Excursion of the approximate third integral \eqref{eq:I3} along
orbits of the zonal field with $J_2$, $J_3$ and $J_4$, for $18$ orbits per planet
with $a\in\{1.15,1.5,2.5\}R_e$, $e\in\{0.02,0.1\}$, $i\in\{30^\circ,50^\circ,70^\circ\}$,
started at the osculating pericentre with argument of pericentre $30^\circ$ and
node on the $x$ axis. Columns 5 and 6: median over the orbits of
$\max_t|I_3(t)-I_3(0)|\,G_0^{-2}$ after $30$ and $300$ initial Kepler periods, with
$G_0$ the initial angular momentum. Columns 7 and 8: median over the orbits of
the ratio of the C$_0$ value to the value of the structure, at $30$ and $300$
periods. The per-orbit values are in the archive \citep{CodeCMDA}.}
\label{tab:drift}
\footnotesize\setlength{\tabcolsep}{3.5pt}
\begin{tabular}{@{}llrrrrrr@{}}
\toprule
& & & & \multicolumn{2}{c}{median excursion} & \multicolumn{2}{c}{ratio}\\
\cmidrule(lr){5-6}\cmidrule(lr){7-8}
Planet & Structure & $aJ_2^{-1}$ & $\delta$ (km) & 30 periods & 300 periods & 30 & 300\\
\midrule
Earth & C$_0$ & $1$ & $0$ & $3.05\times10^{-6}$ & $1.86\times10^{-5}$ & $1.00$ & $1.00$\\
 & C$_*$ & $0.999661$ & $0$ & $3.10\times10^{-6}$ & $1.87\times10^{-5}$ & $1.00$ & $1.00$\\
 & S$_*$ & $0.999657$ & $-7.47$ & $2.81\times10^{-7}$ & $3.51\times10^{-7}$ & $13.5$ & $44.2$\\
 & S$_1$ & $1$ & $-7.46$ & $1.55\times10^{-7}$ & $2.64\times10^{-7}$ & $29.8$ & $76.6$\\
Mars & C$_0$ & $1$ & $0$ & $3.59\times10^{-5}$ & $1.66\times10^{-4}$ & $1.00$ & $1.00$\\
 & C$_*$ & $0.995206$ & $0$ & $3.84\times10^{-5}$ & $1.67\times10^{-4}$ & $0.96$ & $0.99$\\
 & S$_*$ & $0.994999$ & $27.14$ & $7.36\times10^{-6}$ & $8.44\times10^{-6}$ & $5.49$ & $22.7$\\
 & S$_1$ & $1$ & $27.32$ & $4.32\times10^{-6}$ & $6.32\times10^{-6}$ & $12.2$ & $39.3$\\
\bottomrule
\end{tabular}
\end{table}

\section{Secular separatrices and homoclinic splitting}
\label{sec:separatrices}

\subsection{Coboundary perturbations}

The integrated defect of (F3) is the last form of the operator used here: on
a separatrix it is the derivative of the Melnikov function. The
intermediaries of Section~\ref{sec:intermediaries} have no homoclinic orbits
of physical interest. The Kepler problem has no hyperbolic periodic orbit, and
in units $\mu=R_e=1$ the angular systems $\Theta^2+2U_2(\theta,\Theta,N)$ of the
inverse-square normal forms whose coefficients, together with their
derivatives in $\Theta$, are bounded independently of $k_2$, as for the
historical intermediaries, have saddles only at small angular momentum,
$\Theta^2$ of the order of $k_2$ for coefficients independent of $\Theta$ and of the
order of $k_2^{1/2}$ for the efficient members of \eqref{eq:general}, whose
coefficient of $r^{-2}$ contains $p^{-1}=\Theta^{-2}$; these are nearly radial orbits
with pericentres deep inside the planet, and they supply no exterior satellite
separatrix. Coefficients whose derivatives in $\Theta$ are of order $k_2^{-1}$ are
not covered by this estimate: $A_{20}=1-3c^2$ and $A_{22}=\Theta^2(b+u\,(\Theta-2))$, with
$b>0$ and $u=\frac14-2k_2^{-1}$, satisfy \eqref{eq:efficiency} and give a saddle at
$\theta=0$, $\Theta=2$ for $N=0$. Separatrices appear in the averaged secular
problems, such as the libration of the argument of pericentre near the
critical inclination \citep{CoffeyDepritMiller1986,Jupp1969} and the
Kozai--Lidov cycles of hierarchical triples
\citep{Kozai1962,Lidov1962,Naoz2016}, where the Melnikov method
\citep{HolmesMarsden1982,GuckenheimerHolmes} applies. By the following
statement, introduced here, the part of a perturbation removed by a regular
generator does not contribute to the splitting.

\begin{proposition}[Coboundary perturbations]
\label{prop:coboundary}
Under the hypotheses (F3), let $V=\{W,H_0\}$ for a function $W$ that is
smooth near the closure of $\gamma\times S^1$ and whose restriction to the
hyperbolic periodic orbit $\{x_h\}\times S^1$ does not depend on $\vartheta$. Then
$M\equiv0$ and $D_\gamma\equiv0$, that is,
\begin{equation}
 \int_{-\infty}^{\infty}\nu_\vartheta^{-2}\,\Lop_{H_0}\bigl(\{W,H_0\}\bigr)_{q\vartheta}
 \bigl(\gamma(t),\vartheta_0+\Phi(t),I_h\bigr)\,\dd t=0
 \qquad\text{for every }\vartheta_0 .
\end{equation}
Consequently, if $V=\{W,H_0\}+V_\perp$ with such a $W$, the Melnikov function
and the integrated defect of $V$ equal those of $V_\perp$.
\end{proposition}

\begin{proof}
Along the unperturbed flow $\dd W/\dd t=\{W,H_0\}=V$. On the periodic orbit
$V=\nu_\vartheta\,\partial_\vartheta W=0$, since $W$ does not depend on $\vartheta$ there, and
(F3) gives $M=\Psi'$ and $D_\gamma=\Psi''$ with
\[
 \begin{aligned}
 \Psi(\vartheta_0)&=\int_{-\infty}^{\infty}\frac{\dd}{\dd t}
 W\bigl(\gamma(t),\vartheta_0+\Phi(t),I_h\bigr)\,\dd t\\
 &=\lim_{T\to\infty}\Bigl[W\bigl(\gamma(T),\vartheta_0+\Phi(T)\bigr)
 -W\bigl(\gamma(-T),\vartheta_0+\Phi(-T)\bigr)\Bigr].
 \end{aligned}
\]
Both points tend to the periodic orbit, on which $W$ takes a single value,
so $\Psi\equiv0$. The last statement follows from the linearity of $M$ and
$D_\gamma$ in $V$.
\end{proof}

The proposition is the homoclinic counterpart of the efficiency condition:
the part of a perturbation generated by a regular function affects neither
the averaged dynamics nor the first-order splitting, and the integrated
defect measures the remainder.

\subsection{Validity of the first-order splitting}

The first-order Melnikov function gives the splitting when it dominates the
remainder of the expansion in $\varepsilon$. Let $\nu_h$ be the rotor frequency on the
hyperbolic orbit and $\lambda$ its exponent. Two hypotheses give an exponentially
small Melnikov integral: the rotor frequency stays equal to $\nu_h$ along the
homoclinic orbit, and the coefficients of the harmonics of $V$ in the rotor
angle, evaluated along that orbit, extend analytically to a strip
$|\mathrm{Im}\,t|<\tau$ of complex time, tend to zero as $|\mathrm{Re}\,t|\to\infty$ in the
strip, and have integrals of their absolute values along the lines
$\mathrm{Im}\,t=\mathrm{const}$ bounded uniformly in $\nu_h$; a pointwise bound in the
strip does not suffice.
Under both, the harmonic of order $k$ of the Melnikov integral is
$O(\mathrm e^{-k|\nu_h|\tau'})$ for every $\tau'<\tau$, where $\tau$ is of the order of
$\lambda^{-1}$ ($\tau=\pi/(2\lambda)$ for a homoclinic orbit in $\mathrm{sech}\,\lambda t$), and once
this is smaller than $\varepsilon$ the estimate $O(\varepsilon^2)$ of the remainder no longer
certifies the first-order term \citep{Neishtadt1984,HMS1988}; the rigorous
theory establishes regimes in which the Melnikov term gives the correct
asymptotics as well as regimes in which it fails \citep{BaldomaEtAl2012}. The
uniform bound is part of the hypothesis: for
$H_0=\frac12p^2-\frac12q^2+\frac14q^4+\frac12I^2$ and $V=q\cos(\vartheta+Ip/\sqrt2)$, the rotor
frequency equals $\Lambda$ along the homoclinic orbit with $I=\Lambda$, whereas the
phase $t-\mathrm{sech}\,t\tanh t$ of the first harmonic has a cubic stationary
point at $t=0$, and this harmonic decays only like $\Lambda^{-1/3}$. When the rotor
frequency varies along the homoclinic orbit, the phase of the harmonic of
order $k$ is $k\int\nu_\vartheta\,\mathrm dt$, and the ratio at the hyperbolic orbit does
not fix the decay; along the EKL separatrix at $i_0=65^\circ$ the ratio grows from
$0.21$ to $2.63$ (Section~\ref{sec:ekl}). The ratio $|\nu_h|\lambda^{-1}$ at the
hyperbolic orbit (Table~\ref{tab:slowfast}) is therefore used only as a
diagnostic of scale separation: a ratio of order one leaves the first-order
prediction open to a direct numerical test (Section~\ref{sec:ekl}), and a
large ratio indicates exponentially small splitting only where the two
hypotheses hold; no such analysis is carried out here. For the
equatorial $J_2$ separatrix the rotor frequency $\Theta r^{-2}$ decreases along the
homoclinic orbit by the factor $(u_a/u_s)^2$ (Appendix~\ref{app:separatrices}),
and the exact Melnikov function \eqref{eq:M-j2} replaces the estimate. For the
critical inclination the nodal rate $-\frac32J_2HG^{-5}L^{-3}$ of the secular $J_2$
Hamiltonian ($\mu=R_e=1$) depends on $G$ only, and along the separatrix of the
pendulum model of the libration in $(g,G)$ it varies by $1\%$ of $\nu_h$ or less
for the two orbits of Table~\ref{tab:slowfast} inside the domain of that model.
A relative variation of $1\%$ of the rotor frequency shifts the phase of the
harmonic of order $k$ by up to $0.01\,k|\nu_h|\lambda^{-1}$ over the time $\lambda^{-1}$,
which is $2.2k$ and $6.3k$ for the ratios $224$ and $627$ of
Table~\ref{tab:slowfast}, so that the first hypothesis does not hold to the
accuracy that the exponential estimate requires. The domain of that model requires
the half-excursion $2|b/a|^{1/2}$ of $G$ on its separatrix, with $a$ and $b$ the
coefficients of the quadratic and of the $\cos2g$ terms, to stay below the
distance $L-G$ to circular orbits, where $(g,G)$ is singular; the check fails for
$a=1.1$ and $e=0.01$, with $8.7\times10^{-5}$ against $5.2\times10^{-5}$, and holds for
$e>0.0165$, $0.0121$ and $0.0043$ at $a=1.1$, $1.5$ and $4.2$. Since the
half-excursion is of order $J_2^{1/2}e$ and the distance $L-G$ of order $e^2$, the
check fails for every eccentricity below a limit proportional to $J_2^{1/2}$
at fixed $a$: near-circular orbits lie outside the pendulum model, which
describes the libration in $(g,G)$ only away from circular orbits, and its
ratio for $a=1.1$ and $e=0.01$ in Table~\ref{tab:slowfast} is a formal value.

\begin{table}[ht]
\centering
\caption{Ratio of the rotor frequency to the separatrix exponent for three
separatrices. EKL: test-particle quadrupole problem at initial inclination
$i_0$. Equatorial $J_2$: the radial separatrix of Section~\ref{sec:j2}, where the
ratio is $(2\beta)^{-1}$, at apocentre parameter $u_a$. All ratios are evaluated
at the hyperbolic orbit. Critical inclination:
secular $J_2$ Hamiltonian of the Earth with the long-period terms of $J_4$ and
$J_2^2$, eq.~\eqref{eq:critical}, $\mu=R_e=1$, semimajor axis $a$ and
eccentricity $e$, with the pericentre above the surface; $^\dagger$formal value,
outside the domain of the pendulum model, whose half-excursion in $G$ exceeds
the distance $L-G$ to circular orbits.}
\label{tab:slowfast}
\begin{tabular}{@{}lr@{}}
\toprule
Separatrix & $|\nu_h|/\lambda$\\
\midrule
EKL, $i_0=50^\circ$ & $0.47$\\
EKL, $i_0=65^\circ$ & $0.21$\\
EKL, $i_0=85^\circ$ & $0.036$\\
Equatorial $J_2$, $u_a=0.1$ & $1.50$\\
Equatorial $J_2$, $u_a=0.5$ & $2.24$\\
Critical inclination, $a=1.1$, $e=0.01$ & $5.4\times10^{3\,\dagger}$\\
Critical inclination, $a=1.5$, $e=0.3$ & $224$\\
Critical inclination, $a=4.2$, $e=0.3$ & $627$\\
\bottomrule
\end{tabular}
\end{table}

For the critical inclination the secular $J_2$ Hamiltonian is
$k_2\eta^3p^{-3}(1-3c^2)$, and the libration in $(g,G)$ is driven by
\eqref{eq:critical} \citep{CoffeyDepritMiller1986,Jupp1969}. Its exponent is of
order $n\,e\,(J_2|J_4+J_2^2|)^{1/2}=O(nJ_2^{3/2}e)$ for $J_4=O(J_2^2)$, whereas the
nodal rate is of order $nJ_2$, so the ratio grows like $J_2^{-1/2}e^{-1}$ as $e$
decreases to the limit of the domain of the model, and this growth does not
extend to near-circular orbits; with
the $J_2^2$ term it is larger by a factor of $1.9$ than with the $J_4$ term
alone, because of the partial cancellation of Section~\ref{sec:second}. For a
perturbation that depends on the node, such as the lunisolar terms, a
first-order splitting of this separatrix for a low orbit is therefore
difficult to justify; the lunisolar resonances of medium Earth orbits, which
produce chaotic transport \citep{Rosengren2015,Daquin2016,Gkolias2016}, are not
treated here.

\subsection{The eccentric Kozai--Lidov problem}
\label{sec:ekl}

A test particle on an inner orbit perturbed by a distant body on an
eccentric outer orbit is the simplest hierarchical three-body problem
\citep{Kozai1962,Lidov1962,Naoz2016}. After double averaging and expansion to
octupole order, the averaged potential is $F=F_{\mathrm{qu}}+\varepsilon F_{\mathrm{oc}}$ in
the eccentricity $e$, the inclination $i$, the argument of pericentre $\omega$ and
the node $\Omega$, with the octupole strength $\varepsilon$
\citep[eqs.~(8)--(11)]{LithwickNaoz2011}. In the chart
$x=\sqrt{2(1-J)}\cos\omega$, $y=\sqrt{2(1-J)}\sin\omega$, with $J=\sqrt{1-e^2}$ and
$J_z=J\cos i$, which is regular at circular orbits as the Poincar\'e pair of
Section~\ref{sec:charts} is regular at equatorial orbits, the quadrupole
Hamiltonian $H_0=-F_{\mathrm{qu}}$ has for $0<J_z^2<\frac35$ a hyperbolic equilibrium at
the circular orbit with
\begin{equation}
 \lambda=\sqrt{8(3-5J_z^2)},\qquad \nu_h=\frac{\partial H_0}{\partial J_z}\Big|_{x=y=0}=-2J_z,
\end{equation}
whose two homoclinic loops, smooth in the chart $(x,y)$ for $J_z\ne0$, bound the
libration lobes centred at $\omega=90^\circ$ and $270^\circ$ \citep{Naoz2016}; as $J_z\to0$
the eccentric branch reaches the boundary $e=1$ of the non-collision domain. The quadrupole problem belongs to the
rotor class with $(q,p)=(x,y)$, rotor angle $\Omega$ and action $J_z$, and the
octupole potential $V=-F_{\mathrm{oc}}$ vanishes at $e=0$, so that (F3) applies
with the Melnikov potential. The rotor frequency does not vanish along the
separatrices for $i_0\in[42^\circ,89^\circ]$, and at the saddle $|\nu_h|<\lambda$ if and only if
$4J_z^2<8(3-5J_z^2)$, that is, $\cos^2i_0<\frac{6}{11}$ or $i_0>42.392^\circ$. Along the
homoclinic orbit the ratio grows: at $i_0=65^\circ$ it rises from $0.21$ at the saddle
to $2.63$ at maximum eccentricity.

\begin{figure}[ht]
\centering
\includegraphics[width=\textwidth]{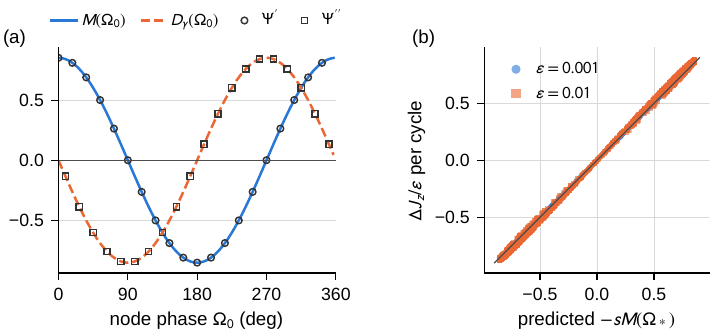}
\caption{EKL problem at $i_0=65^\circ$. (a) Melnikov function $M$, integrated
defect $D_\gamma$ and the first and second derivatives of the Melnikov potential
$\Psi$, versus the node phase. (b) Changes of $J_z$ over finite Kozai cycles
versus the first-order kick, for $\varepsilon=10^{-3}$ and $10^{-2}$; the line is the
diagonal}
\label{fig:ekl}
\end{figure}

At $i_0=65^\circ$, where the maximal eccentricity on the separatrix is $0.838$,
the Melnikov function $M$, the integrated defect $D_\gamma$ and the Melnikov
potential are pure first harmonics of the node phase, because every term of
$F_{\mathrm{oc}}$ is a first harmonic in $\Omega$ \citep{LithwickNaoz2011}; the
homoclinic orbit and the quadratures are described in
Appendix~\ref{app:separatrices}, and the identities (F3) hold to the accuracy of
the quadratures (Figure~\ref{fig:ekl}(a)). Every zero of $M$ is simple, so the
separatrices split transversally at first order \citep{GuckenheimerHolmes}.

The numerical test of this section is a finite-cycle kick test. It compares
the change of $J_z$ over one Kozai cycle with the first-order kick
$-\varepsilon sM(\Omega_*)$, with $\Omega_*$ the node at maximum eccentricity and $s=\pm1$ the
lobe; it does not measure the distance between the stable and unstable
manifolds, which would require their intersections with a common phase
section at several strengths. A cycle runs between two consecutive minima of
the eccentricity, at the distance $d$ from the saddle in the regular plane
$(x,y)$. On the energy surface the physical action $J_z$ and the reduced
unperturbed action $I_0$, defined by $H_0(x,y,I_0)=E_h$, differ at first order by
$-\varepsilon V\nu_\Omega^{-1}$, and the changes of both over a cycle carry endpoint terms of
order $\varepsilon d$, which vanish only in the limit $d\to0$ that defines the
Melnikov kick. Table~\ref{tab:ekl} gives $d$ and the errors for integrations
of the octupole system started on the energy surface of the unperturbed
separatrix from six initial nodes (Appendix~\ref{app:separatrices}). The ratio
of the median error of the change of $J_z$ to the median of $d$ is $0.42$, $0.32$
and $0.44$ at the three strengths, consistent with endpoint terms of order $\varepsilon d$;
the distance $d$ grows by the factors $3.2$ and $3.7$ per decade of $\varepsilon$, since the
orbits follow levels at an offset of order $\varepsilon$ from the separatrix, and the
errors relative to the kick decrease as $\varepsilon$ decreases (Figure~\ref{fig:ekl}(b)). The change of $I_0$
is farther from the kick than that of $J_z$, by a factor between $1.6$ and $2.5$
in the medians.

\begin{table}[ht]
\centering
\caption{Finite-cycle kick test in the test-particle octupole problem at
$i_0=65^\circ$: changes over one Kozai cycle against the first-order (Melnikov)
approximation $-\varepsilon sM(\Omega_*)$ of the kick, with $M$ evaluated on the
unperturbed separatrix at the action $J_h$ and $\Omega_*$ the node at maximum
eccentricity of the cycle. Columns 3 and 4: median over all cycles of the
endpoint distance $d$, the larger of the distances of the two ends of a cycle
from the saddle in the plane $(x,y)$, and its range over the six initial
nodes. Columns 5--7: median error $|\Delta J_z+\varepsilon sM(\Omega_*)|$ of the
first-order approximation, with its $90$th percentile, its range over the six
initial nodes and the ratio of the median error, in units of $\varepsilon\max|M|$,
to the median of $d$. Column 8: median error $|\Delta I_0+\varepsilon sM(\Omega_*)|$, with
its $90$th percentile. Errors in units of $\varepsilon\max|M|$.}
\label{tab:ekl}
\footnotesize\setlength{\tabcolsep}{3.5pt}
\begin{tabular}{@{}lrrrrrrr@{}}
\toprule
& & \multicolumn{2}{c}{endpoint distance $d$} & \multicolumn{3}{c}{error of first-order $\Delta J_z$} & error of $\Delta I_0$\\
\cmidrule(lr){3-4}\cmidrule(lr){5-7}\cmidrule(lr){8-8}
$\varepsilon$ & cycles & median & range & median (90\%) & range & ratio & median (90\%)\\
\midrule
$10^{-4}$ & $751$ & $0.0051$ & $0.0049$--$0.0056$ & $0.0021$ ($0.0046$) & $0.0017$--$0.0024$ & $0.42$ & $0.0052$ ($0.016$)\\
$10^{-3}$ & $921$ & $0.0163$ & $0.0153$--$0.0175$ & $0.0053$ ($0.0133$) & $0.0045$--$0.0058$ & $0.32$ & $0.0125$ ($0.041$)\\
$10^{-2}$ & $1245$ & $0.0601$ & $0.0551$--$0.0725$ & $0.0263$ ($0.0504$) & $0.0250$--$0.0295$ & $0.44$ & $0.0429$ ($0.104$)\\
\bottomrule
\end{tabular}
\end{table}

The integrated defect also gives a heuristic half-width of the chaotic layer
through the separatrix map \citep{Chirikov1979},
\begin{equation}
 w=\frac{\varepsilon|\nu_h|}{\lambda}\,\max|D_\gamma|
 \label{eq:layer-width}
\end{equation}
\citep[eq.~(4.19)]{MartinusiPD}, which is $0.176\,\varepsilon$ in $J_z$ at $i_0=65^\circ$.
Finite-time Lyapunov exponents at two node phases and three strengths
(Appendix~\ref{app:separatrices} and Table~\ref{tab:ekl-layer}) bracket the edge
of the layer on the libration side between $2.46w$ and $3.14w$ for $\varepsilon$ between
$3\times10^{-3}$ and $3\times10^{-2}$, with the same brackets over $500$ and $1000$ time
units, so that \eqref{eq:layer-width} gives the scale of the layer in this
range; the scan does not establish a law for the width. On the circulating
side the brackets, also the same at both horizons, move outward in units of
$w$ as $\varepsilon$ grows, from $5$--$6w$ at $\varepsilon=3\times10^{-3}$ to $8$--$13w$ at
$\varepsilon=3\times10^{-2}$, the wider bracket at the largest strength because the
offset $10w$ is chaotic at one node phase only (Table~\ref{tab:ekl-layer}). The
one-passage estimate does not explain this growth;
circulating orbits pass near both lobes in each cycle, and the octupole
resonances of the circulating region \citep{Li2014} may contribute. Per-cycle maps
for librating cycles at high eccentricity \citep{Klein2026}, averaging over
Kozai cycles \citep{KatzDongMalhotra2011} and resonance-overlap studies
\citep{Li2014} address other parts of the phase space.
\citet{FejozGuardia2016} computed the homoclinic solution of the first-order
secular spatial three-body problem with a distant third body, its complex
singularities and the Melnikov potential, and proved the existence of a
horseshoe; their model and asymptotic regime differ from the test-particle
octupole problem, and the additions of this section are restricted to the
operator form (F3) of the derivatives of the Melnikov potential and to the
finite-cycle test of the first-order kick.

\subsection{The equatorial \texorpdfstring{$J_2$}{J2} separatrix}
\label{sec:j2}

The intermediary with $A_{30}=-2$ and all other coefficients zero, which
coincides with the efficient Cid--Lahulla choice $A_{30}=1-3c^2$ on prograde
equatorial orbits, has the radial Hamiltonian
\begin{equation}
 H_0=\frac12R^2+\frac{\Theta^2}{2r^2}-\frac{\mu}{r}-\frac{\mu J}{2r^3},\qquad J=J_2R_e^2,
\end{equation}
and the remainder of the $J_2$ problem is
$V=H_{10}-H_{01}=\frac32\mu J r^{-3}\sin^2\varphi$, with $\sin\varphi$ given by
\eqref{eq:sinphi}. The attractive $r^{-3}$ term creates an unstable circular
orbit at $u_s=r_s^{-1}$. By axisymmetry, $N=\Theta-\frac12(\xi^2+\zeta^2)$ is an integral
of $H_0+\varepsilon V$, and for $s\ne0$ the system reduced at fixed $N$ is the
two-degree-of-freedom system in the polar-nodal pairs $(r,R)$ and
$(\theta,\Theta)$, with $s=(1-N^2\Theta^{-2})^{1/2}$; in it, $H_0$ belongs to the rotor
class with $(q,p)=(r,R)$, rotor angle $\theta$, action $\Theta$ and rotor frequency
$\Theta r^{-2}$, so that (F3) applies. The construction degenerates at the
invariant equatorial plane $\xi=\zeta=0$, on which $V$ vanishes. With $\mu=J=1$, the
homoclinic orbit in $u=r^{-1}$ is $u(\Phi)=u_s-D\,\mathrm{sech}^2(\beta\Phi)$, where $\Phi$
is the angle swept along the orbit, $u_a$ the value of $u$ at apocentre,
$D=u_s-u_a$, $A=\Theta^{-2}$ and $4\beta^2=AD$, and since $\sin^2\varphi=s^2\sin^2\theta$, the
Melnikov function of the reduced system is
\begin{equation}
 M(\theta_0)=-\frac{3\pi s^2D}{\Theta\,\beta^2\sinh(\pi/\beta)}\,\sin2\theta_0,
 \label{eq:M-j2}
\end{equation}
which agrees with a direct quadrature (Appendix~\ref{app:separatrices}). By (F3),
$D_\gamma=M'$ is proportional to $\cos2\theta_0$, and its integrand is the
radial defect \eqref{eq:Lrpsi} of $V$ along the homoclinic orbit, weighted by
$\nu_\vartheta^{-2}$ with $\nu_\vartheta=\Theta r^{-2}$. The zeros are simple for $s\ne0$, so on
each reduced level set with $s\ne0$ the family $H_0+\varepsilon V$ splits
transversally for sufficiently small $\varepsilon\ne0$.

The factor $\sinh(\pi\beta^{-1})^{-1}$ decays like $\exp(-2\pi|\nu_h|\lambda^{-1})$, with the
ratio $|\nu_h|\lambda^{-1}=(2\beta)^{-1}$ of Table~\ref{tab:slowfast}, at the rate fixed by
the poles of $\mathrm{sech}^2(\beta\Phi)$ at $\beta\Phi=\pm\mathrm i\pi/2$. The $J_2$ problem
itself is the member $\varepsilon=1$, which the first-order theory does not reach, and
its non-integrability is proved by other methods
\citep{IrigoyenSimo1993,CellettiNegrini1995}. The hyperbolic circular orbit lies
at $r_s\approx(0.71$--$1.22)R_e\sqrt{J_2}$, inside the planet for every real body, so
that this example illustrates (F3) with a closed-form Melnikov function and
has no exterior counterpart.

\section{Event surfaces}
\label{sec:event}

\subsection{Transverse zero sets and the zonal problem}

Numerical propagators and semi-analytical theories use event surfaces such
as the passage through a node or the pericentre \citep{Lara2021}. A zero set
of a defect $\sigma=\B_{a,b;c,d}[H_*]$ supplies such a surface for the motion
generated by a Hamiltonian $K$, possibly different from $H_*$, near a point $z_*$
with $\sigma(z_*)=0$ and $\{\sigma,K\}(z_*)\ne0$: there $\dd\sigma\ne0$, the implicit function
theorem gives a hypersurface crossed transversally by the flow, and its
intersection with the energy surface has codimension two, since $\dd\sigma$ and
$\dd K$ are independent where their Poisson pairing does not vanish.

For the zonal problem in the modified chart, the zero set of the radial
defect \eqref{eq:torque-form} contains the apsidal surface $R=0$ and the zeros of
the weighted torque, which for $J_2$ alone are the surfaces of extreme latitude
and the nodal surface $\sin\varphi=0$, the classical event surfaces of satellite
theory, smooth up to the equatorial plane in this chart; with $J_3$ the zero
near a node moves to the height $R_eJ_3J_2^{-1}$, and the geometric node is the
zero of the degree-two term \eqref{eq:zonal-defect} on the full motion
(Section~\ref{sec:propagated}). The coupling \eqref{eq:axisym-xi} vanishes to
second order on $F_\varsigma=0$, where the transversality condition does not apply,
and the crossing of the force equator, near the centre $z=\frac12R_eJ_3J_2^{-1}$ of
the shifted oblate structure, is detected by a simple root of its factor
$F_\varsigma$.

\subsection{Solar radiation pressure in resonant canonical variables}
\label{sec:rcv}

The resonant canonical variables of \citet{Martinusi2023} are action--angle
variables of the planar Kepler problem lifted by the Levi-Civita
transformation \citep{LeviCivita1920}. With polar coordinates $(r,\theta)$ in the
plane, radial velocity $R$ and angular momentum $\Theta$, the Levi-Civita
coordinates and momenta are
\begin{equation}
 \alpha=\sqrt{2r}\cos\frac\theta2,\quad \beta=\sqrt{2r}\sin\frac\theta2,\quad
 A=R\,\alpha-\frac{\Theta\beta}{r},\quad B=R\,\beta+\frac{\Theta\alpha}{r},
 \label{eq:lc-vars}
\end{equation}
so that $r=\frac12(\alpha^2+\beta^2)$ and $x=r\cos\theta=\frac12(\alpha^2-\beta^2)$, and $r\pm x$ are
the parabolic coordinates aligned with the $x$ axis. For a frequency $\omega>0$,
\begin{equation}
 \begin{gathered}
 U=\tfrac12(\omega^2\alpha^2+A^2),\qquad W=\tfrac12(\omega^2\beta^2+B^2),\\
 u=\frac1\omega\operatorname{atan2}(\omega\alpha,A),\qquad v=\frac1\omega\operatorname{atan2}(\omega\beta,B),
 \end{gathered}
 \label{eq:rcv}
\end{equation}
and the pairs $(u,U)$ and $(v,W)$ are canonical, with
\begin{equation}
 r\Bigl(H_K+\frac{\omega^2}{2}\Bigr)=\frac{U+W}{2}-\mu
 \label{eq:rcv-identity}
\end{equation}
for the Kepler Hamiltonian $H_K$, identically \citep{Martinusi2023}. For a perturbation $V$ and
an energy $h$, the Poincar\'e--Sundman Hamiltonian
$\mathcal K=r(H_K+V-h)=\frac12(U+W)-\mu+\Phi$, with
$\Phi=rV-(h+\frac12\omega^2)r$, generates the motion on the level $\mathcal K=0$ in the
fictitious time $\tau$ with $\dd t=r\,\dd\tau$. When $rV$ is a sum of a function of $\alpha$
and a function of $\beta$, as for the Kepler problem and the Stark term, the
numerator is separated and $\mathcal K$ is of the form \eqref{eq:liouville} in the
Levi-Civita variables \citep{LeviCivita1920}. A radial term $F(r)$ with
$r=\frac12(\alpha^2+\beta^2)$ is of this form only when it is affine, $F''=0$, and
otherwise it is the source of the defect in the following statement.

In the cannonball model of solar radiation pressure, with constant
area-to-mass ratio and reflectivity, a fixed Sun direction and distance and
no eclipses \citep{MontenbruckGill2000}, the perturbation is the constant
acceleration $-\varpi\,\hat{\mathbf x}$ along the Sun direction, derived from the
potential $\varpi x$, and the perturbed Kepler problem is the Stark problem
\citep{LantoineRussell2011}. In the planar problem below the Sun direction lies
in the equatorial plane, which is invariant under the $J_2$ term.

\begin{proposition}[Planar perturbations in resonant variables]
\label{prop:rcv}
Let $\Phi=\phi_1(\alpha^2)+\phi_2(\beta^2)+F(r)$ with smooth functions $\phi_1$, $\phi_2$ and $F$,
and $\mathcal K=\frac12(U+W)-\mu+\Phi$. Then the following hold.
\begin{enumerate}[label=(\alph*)]
 \item For every $\omega>0$, exactly and to all orders in $\Phi$,
 \begin{equation}
  b_{uv}[\mathcal K]=-\frac14\,F''(r)\,\alpha A\,\beta B .
  \label{eq:rcv-defect}
 \end{equation}
 \item The Stark potential $V=\varpi x$ and the constant $-(h+\frac12\omega^2)r$ are of the
 form $\phi_1+\phi_2$, so that the Stark problem has $b_{uv}=0$ identically for every
 $\omega$.
 \item Along a motion on the level $\mathcal K=0$, $\alpha A=r\,\frac{\dd}{\dd t}(r+x)$ and
 $\beta B=r\,\frac{\dd}{\dd t}(r-x)$, so that
 \begin{equation}
  b_{uv}[\mathcal K]=-\frac{r^2}{4}\,F''(r)\,\bigl(\dot r^2-\dot x^2\bigr),
  \label{eq:rcv-physical}
 \end{equation}
 independently of $\omega$. For the equatorial $J_2$ term, $V=-\frac12\mu J_2R_e^2r^{-3}$,
 $F=rV$ and \eqref{eq:rcv-physical} becomes
 $\frac34\mu J_2R_e^2\,r^{-2}(\dot r^2-\dot x^2)$.
 \item Where $F''(r)\ne0$, the zero set of \eqref{eq:rcv-physical} consists of the
 states at which one of the parabolic coordinates $r+x$ and $r-x$ is
 stationary. A simple root in time of one factor of $\dot r^2-\dot x^2=(\dot r+\dot x)(\dot r-\dot x)$
 at which the other factor does not vanish is a transverse event; both factors
 vanish at a crossing of the $x$ axis with $\dot x=0$.
\end{enumerate}
\end{proposition}

\begin{proof}
\emph{(a)} Put $r_1=\frac12\alpha^2$, a function of $(u,U)$, and $r_2=\frac12\beta^2$, a function of
$(v,W)$, so that $r=r_1+r_2$. The only term of $\mathcal K$ that depends on both
factors is $F(r_1+r_2)$, so every mixed second derivative of $\mathcal K$ is
$F''\,\partial_xr_1\,\partial_yr_2$, with $x\in\{u,U\}$ and $y\in\{v,W\}$. Substitution in
\eqref{eq:LC} gives the factorisation
\[
 b_{uv}[\mathcal K]=F''\,\bigl(\mathcal K_u\,\partial_Ur_1-\mathcal K_U\,\partial_ur_1\bigr)
 \bigl(\mathcal K_W\,\partial_vr_2-\mathcal K_v\,\partial_Wr_2\bigr).
\]
In the first factor, $\mathcal K_u=c_1\partial_ur_1$ and $\mathcal K_U=\frac12+c_1\partial_Ur_1$ with
$c_1=2\phi_1'+F'$, so the factor is $-\frac12\partial_ur_1$; similarly the second factor is
$\frac12\partial_vr_2$. From \eqref{eq:rcv}, $\alpha=\omega^{-1}\sqrt{2U}\sin\omega u$ and
$A=\sqrt{2U}\cos\omega u$, so $\partial_u\alpha=A$, $\partial_ur_1=\alpha A$, and likewise $\partial_vr_2=\beta B$.

\emph{(b)} The two terms are $\varpi rx=\frac14\varpi(\alpha^4-\beta^4)$ and
$-(h+\frac12\omega^2)r=-\frac12(h+\frac12\omega^2)(\alpha^2+\beta^2)$, and (a) applies with $F=0$.

\emph{(c)} In the Cartesian form of the first pair, $\mathcal K$ depends on $A$ only
through $\frac12U$, so $\dd\alpha/\dd\tau=\partial\mathcal K/\partial A=\frac12A$, and $\alpha A=\dd(\alpha^2)/\dd\tau=r\,\dd(r+x)/\dd t$;
the second pair is analogous. Then $\alpha A\,\beta B=r^2(\dot r+\dot x)(\dot r-\dot x)$.
For the $J_2$ term, $F=-\frac12\mu J_2R_e^2r^{-2}$ and $F''=-3\mu J_2R_e^2r^{-4}$.

\emph{(d)} By (c), the defect vanishes where $F''\ne0$ if and only if
$\dot r+\dot x=0$ or $\dot r-\dot x=0$. At a simple root of one factor where the other
factor and $F''$ do not vanish, the time derivative of \eqref{eq:rcv-physical}
is the product of nonzero terms, so the defect changes sign and the event is
transverse. On the $x$ axis $r=|x|$, so that one factor vanishes at every
crossing, and the other vanishes as well when $\dot x=0$ there.
\end{proof}

\begin{figure}[ht]
\centering
\includegraphics[width=\textwidth]{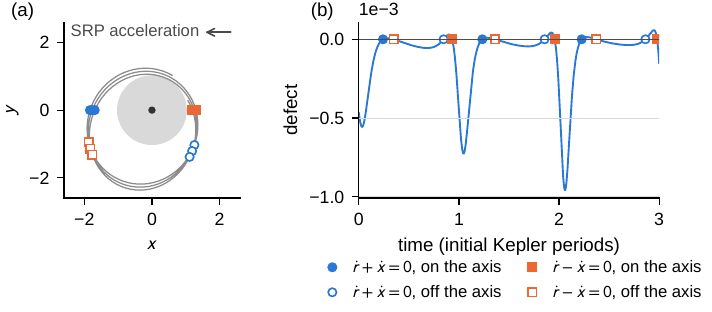}
\caption{Planar Kepler problem with solar radiation pressure in the cannonball
model and the equatorial $J_2$ term, initial $a=1.7\,R_e$, $e=0.3$, argument of
pericentre $60^\circ$, launched at the osculating pericentre,
$J_2=1.08263\times10^{-3}$, $\varpi=2\times10^{-3}$, $\mu=R_e=1$, three initial Kepler periods; the
minimum radius is $1.031\,R_e$, and the value of $\varpi$ is much larger than the
radiation pressure on a typical satellite and makes the precession visible.
(a) Orbit and planet, with the Sun along the positive $x$ axis, so that the
acceleration points along the negative $x$ axis, and
the events of Proposition~\ref{prop:rcv}(d), located as simple roots on the
dense solution: stationary points of $r+x$ (circles) and of $r-x$ (squares),
filled on the $x$ axis and open off it. (b) Defect \eqref{eq:rcv-physical} versus
time in initial Kepler periods, with the same events}
\label{fig:rcv}
\end{figure}

Statement (b) is the operator form of the classical separation of the Stark
problem in parabolic coordinates \citep{LantoineRussell2011,Arnold}, and by
(a) a radial potential with $F''\ne0$, such as the equatorial $J_2$ term, is the
only source of a defect in this class, with a value independent of $\omega$. Since
$\alpha^2=r+x$ vanishes on the negative $x$ axis and $\beta^2=r-x$ on the positive
one, the transverse events of (d) on an orbit that encircles the planet, the
simple roots of one factor at which the other factor does not vanish, are
crossings of the line of the Sun and points where the velocity is tangent to a
parabola with focus at the planet and axis along the Sun direction. On the
orbit of Figure~\ref{fig:rcv}, the roots located on the dense solution are
twelve simple roots in three revolutions, six of them on the Sun axis, each
with the other factor nonzero. On the sum and difference pairs of
\citet{Martinusi2023}, with coordinates $u\pm v$ and momenta $\frac12(U\pm W)$, the
unperturbed part of $\mathcal K$ is linear in the momentum of the sum pair and
does not depend on the difference pair, and the first-order defect of every
perturbation vanishes, as for the Poincar\'e pair of the modified chart
(Proposition~\ref{prop:structure}(a) and Remark~\ref{rem:degenerate}), so that these events
are specific to the pairs $(u,U)$ and $(v,W)$.

\section{Limitations}
\label{sec:limitations}

The verdicts of Section~\ref{sec:intermediaries} refer to the modified and the
polar-nodal charts, to the phase-space transcriptions \eqref{eq:general} and to
the originals \eqref{eq:originals}; a nonzero defect does not exclude
separation in other variables, and in the modified chart vanishing defects do
not imply separation near the prograde equatorial orbits
(Remark~\ref{rem:degenerate}). The interaction structure of
Proposition~\ref{prop:axisym} concerns axisymmetric potentials, and the
interactions of tesseral and third-body terms have not been computed. The
numerical tests of Section~\ref{sec:generator} use one orbit with $J_2$, $J_3$ and
$J_4$ and three with $J_2$ only. Proposition~\ref{prop:second-mean} concerns the
$J_2$ problem and means between passages through the osculating pericentre, the
terms of $J_3$ and $J_4$ enter its remainder, and the bound on the remainder holds
at fixed $e$. The statements of Section~\ref{sec:second} hold in the gauge of
\eqref{eq:W1}, up to the regular axially symmetric changes of gauge discussed
there, and the second-order generator $W_2$ has not been computed.
Propositions~\ref{prop:vinti-response} and \ref{prop:shifted} concern the
residuals and the defects of the oblate structures, and the response of
Proposition~\ref{prop:shifted}(c) is established to leading order in $J_2$; the
test of Table~\ref{tab:drift} measures the approximate conservation of $I_3$ on
$36$ orbits of two planets, and it is not a measure of the error of orbit
propagation. The first-order Melnikov theory applies to the family
$H_0+\varepsilon V$ for small $\varepsilon$, whereas the physical problems correspond to fixed
values of the small parameters; the ratios of Table~\ref{tab:slowfast} are
diagnostics that do not give the size of the splitting, the EKL test concerns
finite-cycle kicks and not the manifolds themselves, the layer brackets of
Table~\ref{tab:ekl-layer} come from a sampled finite-time classifier at two
node phases and three strengths, no splitting has been computed for the
critical inclination, and the lunisolar regime of medium Earth orbits has not
been treated. Proposition~\ref{prop:rcv} concerns the planar cannonball model
with a fixed Sun direction and without eclipses.

\section{Conclusions}
\label{sec:conclusions}

The five-input operator of \citet{MartinusiPD} has been followed through
satellite theory as one scalar on one family of product structures, and
several statements obtained classically by separate arguments appear as its
properties. On the intermediaries the linearised defects of the general form
vanish if and only if the integrability ansatz holds, which in polar-nodal
variables is also the condition for additive first-order Hamilton--Jacobi
separation; the members with terms in the argument of latitude have explicit
second-order defects, and the coupling with the nodal pair is produced by the
dependence on the physical latitude, common to the zonal term, to Aksnes's
intermediary and to the originals of Sterne and Garfinkel. On the zonal
problem the radial defect is exactly linear, with a torque form whose zeros
separate the geometric node, the force equator and the zero of the weighted
torque, and the first-order defect is the operator response to the first-order
Lie generator. Along propagated orbits the same generator gives the mean
radial defect over a revolution at second order in $J_2$, and in its gauge the
long-period part of the second-order average is common to the intermediaries
without latitude terms.

The combination $J_4+J_2^2$ governs both the resonant term at the critical
inclination, which is known since \citet{Brouwer1959} and which no regular
axially symmetric change of gauge alters, and the coordinate response that selects
Vinti's focal radius. With the third harmonic, the exact separation by displaced
oblate structures reduces to the three-term recurrence of Vinti's translated
potential, and along propagated orbits the displaced structures improve the
approximate conservation of the third integral by median factors between $23$
and $77$.
On secular separatrices the integrated defect is the derivative of the
Melnikov function, the first-order kick is supported by finite-cycle tests in
the EKL problem, and transverse zero sets of the operator give event surfaces.
The second-order generator, the manifold splitting itself, the tesseral and
lunisolar perturbations of medium Earth orbits and the perturbation theory in
resonant canonical variables are the natural continuations.


\section*{Statements and Declarations}

\begin{itemize}
\item \textbf{Funding.} No funding was received for this work.
\item \textbf{Competing interests.} The author declares no competing interests.
\item \textbf{Ethics approval and consent to participate.} Not applicable.
\item \textbf{Consent for publication.} Not applicable.
\item \textbf{Data availability.} All numerical results of this paper, the data
behind every table and figure and the outputs of the codes, are archived with
the codes \citep{CodeCMDA}; no external data sets were used beyond the published
zonal coefficients cited in the text.
\item \textbf{Materials availability.} Not applicable.
\item \textbf{Code availability.} The codes that reproduce every number and
table and produce every figure of this paper are archived in a public
repository \citep{CodeCMDA}. They use SymPy \citep{SymPy2017} for the averages over the
mean anomaly and for the exact zero tests of Table~\ref{tab:intermediaries},
NumPy \citep{NumPy2020} and SciPy \citep{SciPy2020} for the quadratures and
integrations, and Matplotlib \citep{Matplotlib2007} for the figures. A single
driver regenerates all results from an empty output directory, the scans of
Table~\ref{tab:ekl-layer} included, and records the package versions.
\item \textbf{Author contributions.} The author is the sole contributor to this work.
\item \textbf{Use of generative AI.} As stated in Section~\ref{sec:verification}, the
author used Claude (Anthropic) to check derivations, write and run the numerical
code, and edit the language of the manuscript, reviewed and edited the content,
and takes full responsibility for the content of the article.
\end{itemize}

\begin{appendices}
\setcounter{table}{0}
\renewcommand{\thetable}{A\arabic{table}}
\renewcommand{\theHtable}{A\arabic{table}}

\section{Numerical procedures and diagnostics}
\label{app:numerics}

This appendix describes the numerical procedures of the paper and collects
the diagnostics that support the statements of the main text. All
computations use the units $\mu=R_e=1$ and the zonal coefficients of
\citet{Pavlis2012} for the Earth, and every number is reproduced by the codes
of \citet{CodeCMDA}.

\subsection{Exact zero tests (Section~\ref{sec:intermediaries})}
\label{app:zero}

The defects of Table~\ref{tab:intermediaries} were computed symbolically in the
modified chart from \eqref{eq:LC}, and each coefficient was tested for identical
vanishing by exact algebra. After the substitutions $\cos2\psi\mapsto C$,
$\sin2\psi\mapsto S$ and $(-2H_{00})^{1/2}\mapsto u\,r^{-1}$, which make the expression rational in
$(r,R,\Theta,\xi,\zeta,u,C,S)$, its numerator was reduced modulo the ideal generated
by $u^2-(2r-R^2r^2-\Theta^2)$ and $C^2+S^2-1$ ($\mu=1$). The two generators have coprime
leading monomials and therefore form a Gr\"obner basis
\citep[Chap.~2, \S9]{CoxLittleOShea2015}, the ideal is prime, and the real
points of the domain are Zariski dense in its zero set, so that the
coefficient vanishes identically on the domain if and only if the remainder
is zero. Evaluations at random points serve as a cross-check in both
directions: every coefficient declared zero vanishes at all samples to
rounding, and every coefficient declared nonzero is nonzero at some sample.
The originals \eqref{eq:originals} are evaluated with $\sin\varphi$ of
\eqref{eq:sinphi} at the exact reference values $c_0=\frac35$, $p_0=\frac{11}{10}$ and
$\eta_0=\frac{19}{20}$, and the second-order coefficient of $b_{\psi\xi}$ minus
\eqref{eq:originals-xi} reduces to zero; the column on the polar-nodal chart
applies Proposition~\ref{prop:radial-angular}(d), with
$\sin^2\varphi=(1-N^2\Theta^{-2})\sin^2\theta$, and the same exact test.

\subsection{Propagated orbits (Sections~\ref{sec:propagated} and \ref{sec:second-mean})}
\label{app:orbit}

The orbit of Figure~\ref{fig:propagated}, with $a=1.2$, $e=0.1$, $i=50^\circ$ and
initial argument of pericentre $30^\circ$, was integrated under the $J_2$, $J_3$ and
$J_4$ terms with the eighth-order Dormand--Prince method at relative tolerance
$10^{-12}$ over one period of the argument of pericentre, which comprises 1664
revolutions. A revolution runs between two consecutive passages through the
osculating pericentre; the initial state is a pericentre by construction and
is inserted as the first passage, so that the first revolution does not
depend on the rounding sign of the initial radial velocity. The time
integrals of $b_{r\psi}$ and of the Delaunay elements over each revolution are
integrated together with the equations of motion, and every integration is
checked for successful termination. The zeros of $b_{r\psi}$ in
Figure~\ref{fig:propagated}(a) are located by root finding on the dense
solution, and each is attributed to the factor $R$ or to the weighted torque
that vanishes there. The difference between the mean of $b_{r\psi}$ over each revolution and
the closed-form averages \eqref{eq:brpsi-avg}, with the $J_3$ and $J_4$ terms,
at the elements averaged over the same revolution was fitted by a constant
and a harmonic in $2g$. The zero of the
latitudinal force near the first ascending node was located on the
integrated orbit. Table~\ref{tab:orbit-diag} lists the results.

\begin{table}[ht]
\centering
\caption{Diagnostics of the propagated orbit of Figure~\ref{fig:propagated}.
Percentages refer to the amplitude of the closed-form average.}
\label{tab:orbit-diag}
\begin{tabular}{@{}lr@{}}
\toprule
Quantity & Value\\
\midrule
Largest difference of the means from the first-order averages & $1.64\%$\\
Constant part of the difference & $-1.21\%$\\
Amplitude of the harmonic in $2g$ of the difference & $0.43\%$\\
Rms of the difference after the two-term fit & $0.003\%$\\
Latitude of the zero of the latitudinal force at $r=1.094$ & $-0.0611^\circ$\\
Linear approximation $\frac12J_3J_2^{-1}r^{-1}$ & $-0.0613^\circ$\\
Height $z=r\sin\varphi$ of this zero & $-7.44$~km\\
Centre $\frac12R_eJ_3J_2^{-1}$ of the shifted structure & $-7.46$~km\\
Height of the weighted-torque zero next to each node & $-14.8$~km\\
$|b_{r\psi}|$ at the geometric nodes, relative to its maximum & $1.2$--$4.1\times10^{-3}$\\
Zeros of $b_{r\psi}$ in three revolutions (apsides, torque zeros) & $18$ ($6$, $12$)\\
\bottomrule
\end{tabular}
\end{table}

The integrations of the $J_2$ problem from the same initial state, with $J_2$,
with $J_2/2$ and with $e=0.05$, cover 1664, 3328 and 1689 revolutions. Halving $J_2$
multiplies the constant and the harmonic in $2g$ of the difference, relative to
the amplitude of \eqref{eq:brpsi-avg}, by $0.498$ and $0.500$. The closed form
\eqref{eq:second-mean-cf} was evaluated at the elements averaged over each
revolution and compared with the integrated means (Table~\ref{tab:second-check}).
An evaluation of \eqref{eq:second-mean} by quadrature over $\ell$ with $4096$
points and central differences reproduces \eqref{eq:second-mean-cf} at four
states to the relative difference $10^{-7}$, the truncation error of the
differences. For the orbit of Figure~\ref{fig:propagated}, with $J_3$ and $J_4$,
the largest misfit is $0.0098\%$ of the amplitude, against the largest
difference $1.64\%$.

\begin{table}[ht]
\centering
\caption{Second-order mean of the radial defect (Proposition~\ref{prop:second-mean})
against integrations of the $J_2$ problem with $a=1.2$ and $i=50^\circ$. Columns 3--6:
constant and coefficient of $\cos2g$ of $\bar b-\langle b_{r\psi}\rangle_\ell$, in
percent of the amplitude of \eqref{eq:brpsi-avg}, fitted to the integration and
given by \eqref{eq:second-mean-cf}; the integrated coefficient of $\cos2g$ comes
from a joint fit with $\cos2g$ and $\sin2g$, whose coefficient of $\sin2g$ is below
$5\times10^{-5}$ in the same units. Column 7: largest difference between the
integrated values and \eqref{eq:second-mean-cf} over all revolutions, relative to
the largest value of $\bar b-\langle b_{r\psi}\rangle_\ell$.}
\label{tab:second-check}
\begin{tabular}{@{}llrrrrr@{}}
\toprule
& & \multicolumn{2}{c}{integration} & \multicolumn{2}{c}{closed form} &\\
\cmidrule(lr){3-4}\cmidrule(lr){5-6}
Field & $e$ & constant & $\cos2g$ & constant & $\cos2g$ & misfit\\
\midrule
$J_2$ & $0.1$ & $-1.2227$ & $0.4423$ & $-1.2229$ & $0.4443$ & $1.7\times10^{-3}$\\
$J_2/2$ & $0.1$ & $-0.6085$ & $0.2210$ & $-0.6085$ & $0.2215$ & $8.3\times10^{-4}$\\
$J_2$ & $0.05$ & $-4.5995$ & $0.8415$ & $-4.6007$ & $0.8495$ & $2.0\times10^{-3}$\\
\bottomrule
\end{tabular}
\end{table}

From $e=0.1$ to $e=0.05$, with $J_2$ only, the amplitude of the first-order
average decreases by the factor $4.2$ and the constant part of the difference
by the factor $1.1$, as the growth $e^{-2}$ of the relative constant part
requires. At $e=0.1$ the part $\Gamma_2^{\mathrm d}$ gives $0.435\%$ of the
coefficient $0.444\%$ of $\cos2g$. For this orbit, the two limits of the
Delaunay derivation discussed after Proposition~\ref{prop:second-mean} are
reached at $e=0.0274$, where $J_2e^{-2}G^{-4}=1$, and at $e=0.0105$, where the
constant part of \eqref{eq:second-mean-cf} equals the first-order amplitude, and
the definition of the revolution breaks down at
$e\approx J_2s^2G^{-4}=4.5\times10^{-4}$. At $e=0.05$, where $J_2e^{-2}G^{-4}=0.31$, the
misfit of Table~\ref{tab:second-check} is close to its value at $e=0.1$.

\subsection{Second-order averages (Section~\ref{sec:second})}
\label{app:second}

The averages over the mean anomaly of the $J_2^2$ bracket of
$\langle Q_2\rangle_\ell$ were computed by the trapezoidal rule in the true anomaly
with $4001$ points and the weight $\dd\ell/\dd f$, at eight values of $g$. The
Poisson brackets are formed after the inclination-dependent coefficients of
the intermediaries are inserted, so that their chain-rule terms are included
pointwise. For the intermediaries of Brouwer, Cid and Lahulla and Deprit at
four states, the coefficients of $\cos2g$ agree to all printed digits. On the
$45$ states with $a\in\{1.2,2,4.2\}$, $e\in\{0.05,0.2,0.5\}$ and five inclinations, the
quadrature agrees with \eqref{eq:J22lp} to $3.7\times10^{-13}$ of the $J_4$ coefficient
of \eqref{eq:J4avg}, and on nine states at the critical inclination the total
coefficient agrees with \eqref{eq:critical} to $7\times10^{-13}$. The cosine term of
\citet[eq.~(12)]{Lara2021redux}, multiplied by $\frac12J_2^2$ and written in these units,
minus \eqref{eq:J22lp} simplifies to zero, and so does the $J_3$ term of
$\langle Q_2\rangle_\ell$ after the addition of \eqref{eq:J3gauge} to $W_1$. Figure~\ref{fig:second}
uses $a=1.3$, whose pericentre at $e=0.2$ lies above the surface. For the Earth,
$J_4+J_2^2=-4.5\times10^{-7}$.

\subsection{Oblate structures (Section~\ref{sec:vinti})}
\label{app:vinti}

The residual objectives of Propositions~\ref{prop:vinti-response} and
\ref{prop:shifted} were evaluated on a grid of $120$ radii in
$1.05\le r\le3$ and $161$ latitudes in $5^\circ\le\varphi\le85^\circ$, and in
$-85^\circ\le\varphi\le-5^\circ$ for Table~\ref{tab:shifted}, with uniform weight,
and minimised by linear least squares. The local estimate of
Table~\ref{tab:planets} and Figure~\ref{fig:vinti} is evaluated at $r=1.05$. The
defect objective $\mathcal E_B$ of Proposition~\ref{prop:vinti-response}(d) was
evaluated with isotropic Gaussian momenta of variance $\sigma^2=(4r)^{-1}$ per
component, a standard deviation equal to half the local circular speed, which
are integrated out analytically: the numerator of \eqref{eq:Qlambda} has the
mean square $\sigma^4\bigl(4R^2z^2+(z^2-R^2+\lambda)^2\bigr)$, and the remaining spatial
integral is evaluated on the uniform grid. For the Earth the minimiser is
$\lambda_B=0.9996022\,J_2R_e^2$, the minimiser of the residual objective with the
weight $wQ_0^2$ is $0.9996010\,J_2R_e^2$, and \eqref{eq:lambda-B} gives
$0.9995974\,J_2R_e^2$; for Saturn the three values are $0.96848$, $0.96663$ and
$0.95994$, in units of $J_2R_e^2$. The centre and focal radius of
Table~\ref{tab:shifted} follow from \eqref{eq:vinti1966} with the equatorial
radii $6378.137$~km for the Earth and $3396.0$~km for Mars.

For Table~\ref{tab:drift}, the initial state of each orbit is the osculating
pericentre of the stated elements, with the speed $(1+e)p^{-1/2}$, and the
osculating $a$, $e$ and $i$ of every initial state are checked to $10^{-12}$. The
orbits were integrated in Cartesian coordinates under the $J_2$, $J_3$ and $J_4$
terms over $300$ initial Kepler periods with the eighth-order Dormand--Prince
method at relative tolerance $10^{-12}$ and absolute tolerance $10^{-13}$, with
$12001$ samples, and $I_3$ of \eqref{eq:I3} was evaluated at each sample, with $E$
the initial zonal energy, after the transformation to the spheroidal
coordinates of each structure. The relative energy error stays below
$1.5\times10^{-10}$, a run at relative tolerance $10^{-13}$ changes the per-orbit
maxima by at most $0.2\%$, and $48001$ samples change them by at most $3.6\%$. As a
check, $I_3$ is conserved to $1.3\times10^{-12}$ of $G_0^2$ over $20$ periods for the
displaced potential of Proposition~\ref{prop:shifted}(b) with $a=0.002$ and
$\delta=0.008$.

The test of Section~\ref{sec:vinti} with $J_4$ replaced by $J_3^2J_2^{-1}-J_2^2$ uses
the same initial states, method, tolerances and samples. Over $30$ periods the
median per-orbit factors for S$_1$ are $31.7$ (Earth) and $47.4$ (Mars), and between
$30$ and $300$ periods the median excursion of S$_1$ in this field grows by the
factors $5.80$ and $4.46$, against $1.70$ and $1.46$ in the field of
Table~\ref{tab:drift}. The relative energy error stays below $1.5\times10^{-10}$, a
run at relative tolerance $10^{-13}$ changes the per-orbit maxima by at most $2.7\%$,
and $48001$ samples change them by at most $2.9\%$. In the displaced potential of
Proposition~\ref{prop:shifted}(b) with the parameters of S$_1$, for which $I_3$ of
S$_1$ is an exact integral, the same integration gives excursions of at most
$1.1\times10^{-10}$, at most $3.1\%$ of the value with $J_4=J_3^2J_2^{-1}-J_2^2$ on the
same orbit.

\subsection{Separatrices (Section~\ref{sec:separatrices})}
\label{app:separatrices}

For the equatorial $J_2$ separatrix, the factor $(u_a/u_s)^2$ by which the rotor
frequency decreases along the homoclinic orbit is $0.006$ and $0.25$ for the
two cases of Table~\ref{tab:slowfast}, and the Melnikov function
\eqref{eq:M-j2} agrees with a direct quadrature to $1.3\times10^{-5}$ of its
amplitude at $u_a=0.3$. For the critical inclination, the nodal rate stays
within $1.0\%$ and $0.36\%$ of $\nu_h$ along the pendulum separatrix for the two
orbits of Table~\ref{tab:slowfast} inside the domain of the model, with
$e=0.3$ at $a=1.5$ and $4.2$.

For the EKL problem, the homoclinic orbit is obtained by integrating the
quadrupole flow forward and backward in time from the point $(x,y)=(0,y_m)$ of
maximum eccentricity until $(x^2+y^2)^{1/2}=\rho_c$, and $M$ and $D_\gamma$ are
trapezoidal quadratures over the truncated orbit, on $4\times10^4$ nodes
($8\times10^4$ for $\rho_c=10^{-6}$), at 120 node phases. Since $V$ vanishes at $e=0$, the truncated tails are of the order
of $\rho_c/\lambda$; for $\rho_c=10^{-3}$, $10^{-4}$, $10^{-5}$ and $10^{-6}$ the amplitude of
$M$ at $i_0=65^\circ$ is $0.85388$, $0.85487$, $0.85482$ and $0.85480$, quadrupling the
number of nodes at $\rho_c=10^{-4}$ leaves it unchanged to seven digits, and the
computations use $\rho_c=10^{-4}$, whose relative error of $8\times10^{-5}$ is negligible
against the errors of Table~\ref{tab:ekl}. The spectral derivative of $M$
deviates from $D_\gamma$ by $5\times10^{-14}$ of its amplitude, and $M$ and $D_\gamma$
deviate from the derivatives of the truncated potential $\Psi_T$ by
$4.2\times10^{-6}$, the endpoint term $[V\nu_\vartheta^{-1}]$ of the integration by parts
over the truncated orbit.

For the kick test of Table~\ref{tab:ekl}, the initial states lie on the energy
surface $H_0+\varepsilon V=E_h$ of the unperturbed separatrix, with $x=0$, $y$ equal
to its value at maximum eccentricity on the unperturbed separatrix of the
lobe $s=+1$, the six nodes $\Omega=0^\circ,60^\circ,\dots,300^\circ$ and $J_z$ fixed by the
energy, and each state was integrated over 400 time units at relative
tolerance $10^{-11}$, with a check of successful termination and of the domain
$|J_z|\le J\le1$ on a uniform grid of $4\times10^5$ points of the dense solution. A Kozai
cycle runs between two consecutive minima of $x^2+y^2$ on this grid; its lobe $s$ is
the sign of $y$ at the intermediate maximum, $\Omega_*$ is the node there, the
measured change of $J_z$ is the difference between the two minima, and $d$ is
the larger of the two values of $(x^2+y^2)^{1/2}$ at the minima. The function $M$ is
evaluated on the unperturbed separatrix at the action $J_h$, and only cycles
whose maximum of $(x^2+y^2)^{1/2}$ exceeds $0.9$ of its value on the separatrix are
used, which holds for all cycles of Table~\ref{tab:ekl}. The results for each
initial node are in the archive \citep{CodeCMDA}.

For the layer of Table~\ref{tab:ekl-layer}, an offset $h$ of the action, given
in the table in units of the half-width $w$ of \eqref{eq:layer-width}, defines the initial state with $x=0$, $\Omega\in\{0,\pi\}$
and the value of $y$ for which the unperturbed quadrupole level through
$(0,y)$ at the separatrix energy $E_h$ has the action $J_h+h$, that is,
$I_0(0,y)=J_h+h$ with $I_0(x,y)$ defined by $H_0(x,y,0,I_0)=E_h$, and $J_z$
follows from $H_0+\varepsilon V=E_h$. The finite-time Lyapunov exponent is
$T^{-1}\sum_k\ln(d_k/d_0)$, where a separation of length $d_0=10^{-9}$ in
$(x,y,\Omega,J_z)$, with the initial direction $(1,0.3,0.2,0.1)$, is integrated
with the state and renormalised to $d_0$ every 20 time units, $d_k$ being its
length before the $k$-th renormalisation; the integrator is DOP853 with
relative and absolute tolerances $10^{-11}$ and $10^{-13}$, and $w=0.175997\,\varepsilon$ is
the unrounded value of \eqref{eq:layer-width}, by which the exact offset
$I_0(0,y)-J_h$ of each launch is divided. An offset is classified as chaotic
when the exponent of either initial node exceeds $0.06$, and the table gives,
for $T=500$ and $T=1000$, the bracket between the outermost chaotic offset and
the next sampled one. With the larger of the two node exponents, the chaotic
offsets have exponents between $0.069$ and $0.27$ at $T=500$ and between $0.12$
and $0.21$ at $T=1000$, and the regular ones below $0.036$ and $0.030$. The
value closest to the threshold is $0.069$, at $\varepsilon=3\times10^{-2}$ and $10w$ on the
circulating side for $T=500$, where the node $\pi$ gives $0.028$; over $T=1000$ the
node $0$ gives $0.117$ there, and the bracket $[10,13]$ holds at both horizons.
The libration side is sampled with spacing $w/6$ and the
circulating side with spacings of $w$ to $3w$ near the edge. Every trajectory
of the three scans ($318$ in all) reached its end time inside the physical
domain $|J_z|\le J\le1$, which is checked at every renormalisation. The
exponents of both nodes next to each edge are in the archive \citep{CodeCMDA}.

\begin{table}[ht]
\centering
\caption{Edge of the chaotic layer of the EKL problem at $i_0=65^\circ$, from a
sampled finite-time classifier, in units of the half-width $w=0.175997\,\varepsilon$ of
\eqref{eq:layer-width}: brackets between the outermost chaotic offset and the
next sampled one, from finite-time Lyapunov exponents over the time $T$ at the
node phases $0$ and $\pi$ (threshold $0.06$; DOP853, tolerances $10^{-11}$ and
$10^{-13}$, separation $10^{-9}$ renormalised every $20$ time units).
Sensitivity: the brackets do not change for any threshold between $0.0355$ and
$0.0685$ at $T=500$, and between $0.0298$ and $0.1171$ at $T=1000$; with one node
phase alone, the circulating brackets become $[5.05,6.00]$ at $\varepsilon=10^{-2}$
(phase $0$) and $[8.00,10.00]$ at $\varepsilon=3\times10^{-2}$ (phase $\pi$), and the
libration brackets do not depend on the phase. Single trajectories next to
an edge can stay near regular regions for long times, and their exponents
can cross the threshold under small changes of the launch or of the
tolerances; the circulating edge at $\varepsilon=3\times10^{-2}$ is therefore taken as
$[8,13]$ in the text.}
\label{tab:ekl-layer}
\begin{tabular}{@{}lcccc@{}}
\toprule
& \multicolumn{2}{c}{libration} & \multicolumn{2}{c}{circulating}\\
\cmidrule(lr){2-3}\cmidrule(lr){4-5}
$\varepsilon$ & $T=500$ & $T=1000$ & $T=500$ & $T=1000$\\
\midrule
$3\times10^{-3}$ & $[2.66,2.83]$ & $[2.66,2.83]$ & $[5.02,6.00]$ & $[5.02,6.00]$\\
$10^{-2}$ & $[2.98,3.14]$ & $[2.98,3.14]$ & $[6.00,8.00]$ & $[6.00,8.00]$\\
$3\times10^{-2}$ & $[2.46,2.62]$ & $[2.46,2.62]$ & $[10.00,13.00]$ & $[10.00,13.00]$\\
\bottomrule
\end{tabular}
\end{table}

\end{appendices}

\FloatBarrier

\bibliographystyle{unsrtnat}
\bibliography{references}

\end{document}